\documentclass[times,final]{elsarticle}

\usepackage{jcomp}
\usepackage{framed,multirow}

\usepackage{amssymb,amsmath,mathtools}
\usepackage{latexsym}
\newcommand{\RR}{\mathit{Re}}
\newcommand{\We}{\mathit{We}}

\newcommand{\bs}{\boldsymbol}
\newcommand{\f}{\frac}
\newcommand{\ov}{\overline}

\usepackage{multirow}
\usepackage{array}
\usepackage{pifont}
\newcommand{\cmark}{\ding{51}}
\newcommand{\xmark}{\ding{56}}

\usepackage[ruled,vlined]{algorithm2e}
\SetKw{KwBy}{by}

\usepackage{url}
\usepackage{xcolor}
\definecolor{newcolor}{rgb}{.8,.349,.1}

\usepackage[colorlinks=true,allcolors=blue]{hyperref}
\usepackage[export]{adjustbox} 

\journal{Journal of Computational Physics}

\begin{document}

\verso{W.H.R. Chan \textit{et al.}}

\begin{frontmatter}

\title{Identifying and tracking bubbles and drops in simulations: a toolbox for obtaining sizes, lineages, and breakup and coalescence statistics}

\author[1]{Wai Hong Ronald \snm{Chan}\corref{cor1}}
\ead{whrchan@stanford.edu}
\cortext[cor1]{Corresponding author.}
\author[1,2]{Michael S. \snm{Dodd}}
\ead{mchl.dodd@gmail.com}
\author[1,3]{Perry L. \snm{Johnson}}
\ead{perry.johnson@uci.edu}
\author[1]{Parviz \snm{Moin}}
\ead{moin@stanford.edu}

\address[1]{Center for Turbulence Research (CTR), Stanford University, Stanford, CA 94305, USA}
\address[2]{Bellevue, WA 98005, USA}
\address[3]{The Henry Samueli School of Engineering, University of California, Irvine, Irvine, CA 92697, USA}

\received{TBC}
\finalform{TBC}
\accepted{TBC}
\availableonline{TBC}

\begin{abstract}
Knowledge of bubble and drop size distributions in two-phase flows is important for characterizing a wide range of phenomena, including combustor ignition, sonar communication, and cloud formation. The physical mechanisms driving the background flow also drive the time evolution of these distributions. Accurate and robust identification and tracking algorithms for the dispersed phase are necessary to reliably measure this evolution and thereby quantify the underlying mechanisms in interface-resolving flow simulations. The identification of individual bubbles and drops traditionally relies on an algorithm used to identify connected regions. This traditional algorithm can be sensitive to the presence of spurious structures. A cost-effective refinement is proposed to maximize volume accuracy while minimizing the identification of spurious bubbles and drops. An accurate identification scheme is crucial for distinguishing bubble and drop pairs with large size ratios. The identified bubbles and drops need to be tracked in time to obtain breakup and coalescence statistics that characterize the evolution of the size distribution, including breakup and coalescence frequencies, and the probability distributions of parent and child bubble and drop sizes. An algorithm based on mass conservation is proposed to construct bubble and drop lineages using simulation snapshots that are not necessarily from consecutive time-steps. These lineages are then used to detect breakup and coalescence events, and obtain the desired statistics. Accurate identification of large-size-ratio bubble and drop pairs enables accurate detection of breakup and coalescence events over a large size range. Accurate detection of successive breakup and coalescence events requires that the snapshot interval be an order of magnitude smaller than the characteristic breakup and coalescence times to capture these successive events while minimizing the identification of repeated confounding events. Together, these algorithms enable insights into the mechanisms behind bubble and drop formation and evolution in flows of practical importance.
\end{abstract}


\end{frontmatter}



\section{Introduction}\label{sec:intro}

Bubble and drop size distributions are of interest in various physical processes of practical importance, including liquid atomization in combustors and ocean sprays~\citep[and references therein]{Lin1,Lasheras2,Villermaux1,Gorokhovski1,Theofanous1,Veron1}, liquid agglomeration in clouds~\citep[and references therein]{Shaw1,Grabowski1,Pumir1}, bubble formation in oceans and reactors~\citep[and references therein]{Veron1,Melville1,Reed1,Kulkarni1,Kiger1,Risso1,Chu1}, and foam generation~\citep[and references therein]{Pugh1,Drenckhan1,Wang3,Hill1}. Knowledge of size distributions enables the quantification of metrics like combustor ignition probabilities, radiation scattering coefficients, and mass transfer rates, which are typically strong functions of bubble and drop sizes. These distributions evolve with the background flow, which causes the bubbles and drops to break up and coalesce. Breakup and coalescence mechanisms may be characterized by statistics such as breakup and coalescence frequencies, the probability distribution of child bubble and drop sizes in breakup events, and the analogous distribution of parent sizes in coalescence events. Accurate measurements of breakup and coalescence events in detailed interface-resolving two-phase flow simulations enable computation of breakup and coalescence statistics that help to build physical intuition of the underlying mechanisms~\citep[and references therein]{Lasheras1,Liao1,MartinezBazan3,Solsvik2}. For example, these statistics may shed light on the nature of mass transfer from large to small bubble sizes in breaking waves~\citep{Garrett1}, or the influence of turbulence on the nature of drop breakup~\citep{Dodd2}. The macroscopic effects of breakup and coalescence events on the evolution of the size distributions may be modeled by population balance equations~\citep{Smoluchowski1,Smoluchowski2,Melzak1,Williams2,Friedlander1,Friedlander2,Valentas2,Valentas1}. The analysis of the relevant statistics aids in evaluating and improving kernels in these model equations, and enhancing modeling efforts for dispersed-phase dynamics in a variety of flows.

The identification of bubbles and drops in a flow simulation is a natural intermediate step towards the computation of bubble and drop sizes. These sizes may be used to generate the size distribution and total interfacial area. This can be challenging in interface-resolving simulations where the bubbles and drops may deform and are not necessarily spherical. One way to identify them is to search for connected regions of computational nodes or cells corresponding to individual bubbles and drops. One can then integrate the dispersed-phase volume fraction over each of these regions to obtain the total volume of each individual bubble and drop. Note that this process may be applied to a variety of interface-advection schemes where an equivalent volume-conserving volume fraction field may be constructed, including volume-of-fluid (VoF), level-set, diffuse-interface, and front-tracking methods, among others. The identification of connected regions has been a recurring theme in digital image processing~\citep{Rosenfeld1,He1} and computer graphics~\citep{Glassner1,Smith1,Shoup1}, including the ``bucket-fill" operation in modern raster graphics editors, which is carried out by the well-known flood-fill algorithm. In the context of computational fluid dynamics, this identification process was first documented in Refs.~\citep{Hebert1,Herrmann2,Tomar1} for the purpose of identifying underresolved Eulerian bubbles and drops and transforming them to Lagrangian point particles in a hybrid Eulerian--Lagrangian framework. The process includes all computational nodes or cells with nonzero dispersed-phase volume fraction. Preliminary work~\citep{Chan3,Chan4} has suggested that this traditional algorithm may be sensitive to the presence of spurious structures that may occur due to interface-advection errors or energetic surface collisions. Spurious structures may contaminate the bubble and drop size distributions. In this work, a refinement is proposed to exclude these spurious bubbles and drops, while ensuring that the volumes of the other identified structures remain accurate. This refinement is designed in a lightweight manner such that limited additional computational cost is incurred.

Tracking the motion of the identified bubbles and drops in time is crucial for computing breakup and coalescence statistics, which characterize the evolution of the bubble and drop size distributions. Bubble and drop tracking enables the construction of lineages that help determine the frequency and nature of breakup and coalescence events. Tracking algorithms have been developed in the experimental fluids community~\citep{RodriguezRodriguez2,RodriguezRodriguez1,Honkanen1} using standard image processing techniques to identify bubbles and drops, and suitable matching criteria to link bubbles and drops between successive image frames. Additional matching criteria are used to detect breakup and coalescence events. In the cell and particle tracking community~\citep[and references therein]{Li1,Jaqaman1,Meijering1,Meijering2}, probabilistic tools like local filtering schemes and the global multiple-hypothesis tracking method have been employed to augment the linking process between successive frames. These tools are useful in regions where the cell or particle loading is high, and in datasets where the successive frames are significantly spaced apart in time. With access to the three-dimensional information of bubble and drop shapes and locations, flow simulations have the potential to carry out this linking process with fewer assumptions. 

In the level-set context, Refs.~\citep{Fang1,Fang2} developed a linking scheme for bubbles using tagging information in each bubble and its surrounding liquid shell from the current and preceding time-steps. However, performance of this scheme at high void fractions, where the separation between bubbles is comparable to the grid spacing, has not been demonstrated, and the scheme also displays significant sensitivity to the extent of this liquid shell. A similar scheme involving the overlap of tagging information between consecutive time-steps was developed in the VoF context by Ref.~\citep{Langlois1} with the understanding that the Courant number of the simulation is typically small. In a VoF-based method, the linking process between successive frames may also be performed deterministically by relying on mass conservation properties in the underlying solver. This was carried out implicitly by Ref.~\citep{Rubel1}, where one may ride on the underlying advection process for the volume fraction field to advect bubble and drop identification tags every time-step. It turns out that when the simulation time-step is small compared to the integral time-scale, bubbles and drops may not separate quickly enough after breaking up. This may result in the sustained detection of repeated breakup and coalescence events, when the children bubbles and drops remain close to each other with insufficient resolution for their separation. An analogous situation may be conceived for slowly coalescing bubbles and drops. This peculiarity was observed in Ref.~\citep{Rubel1}, but is expected in the other schemes discussed above as well. In light of this, it may be advantageous to have the flexibility to carry out the tracking algorithm between simulation snapshots and not necessarily using consecutive time-steps, so as to skip over these confounding breakup and coalescence events arising from the underresolution of momentarily occurring thin features. In this work, the mass conservation properties of the underlying solver are explicitly used to track bubbles and drops between successive simulation snapshots using only the volume and location of each identified bubble and drop in every snapshot. The proposed algorithm differs from the existing algorithms above in that the aforementioned algorithms need to be executed over consecutive time-steps, whereas the successive simulation snapshots required by the proposed algorithm may be spaced several time-steps apart. Also, unlike in the algorithms introduced by Refs.~\citep{Fang1,Fang2,Langlois1}, where comparisons of level sets or volume fraction fields from consecutive time-steps need to be made, the proposed algorithm does not necessarily require the direct input of the underlying volume fraction field. Finally, while the algorithm introduced by Ref.~\citep{Rubel1} needs to be implemented in a flow solver in order to directly advect and compare identification tags over consecutive time-steps, the proposed algorithm can be executed either as a post-processing procedure, as long as the saved snapshots resolve all relevant physical time-scales, or as a runtime routine in tandem with the rest of the simulation.

Together, these identification and tracking algorithms form a toolbox for two-phase flow simulations where dispersed-phase statistics relating to breakup and coalescence are sought. This paper introduces this toolbox with the following organization of material. In \S~\ref{sec:ident}, the bubble and drop identification algorithm is introduced, and several considerations on minimizing volume error are discussed. A number of test cases for this algorithm are provided in \S~\ref{sec:ident-test}, and a demonstration case for the algorithm involving an energetic two-phase flow with large nonspherical structures is presented in \S~\ref{sec:ident-demo}. In \S~\ref{sec:track}, the bubble and drop tracking algorithm is introduced. The volume accuracy of the scheme introduced in \S~\ref{sec:ident} is revisited vis-\`{a}-vis the accuracy of the tracking algorithm. A number of test and demonstration cases for this algorithm are provided in \S~\ref{sec:track-test}. Finally, conclusions are drawn in \S~\ref{sec:concl}.

\section{Description of the identification algorithm}\label{sec:ident}

\begin{figure}
  \centerline{
(a)
\includegraphics[width=0.375\linewidth,valign=t]{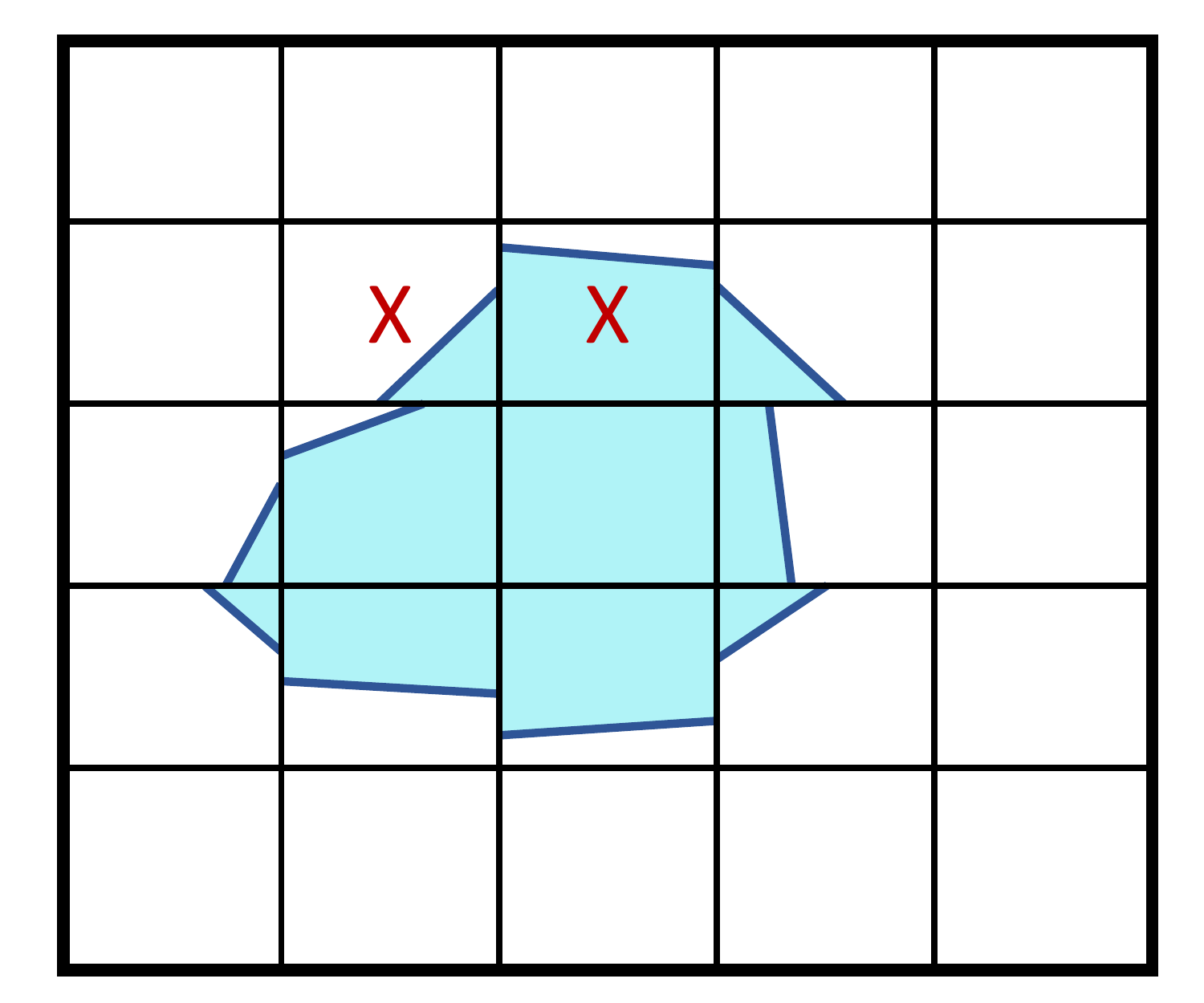}
\quad
(b)
\includegraphics[width=0.375\linewidth,valign=t]{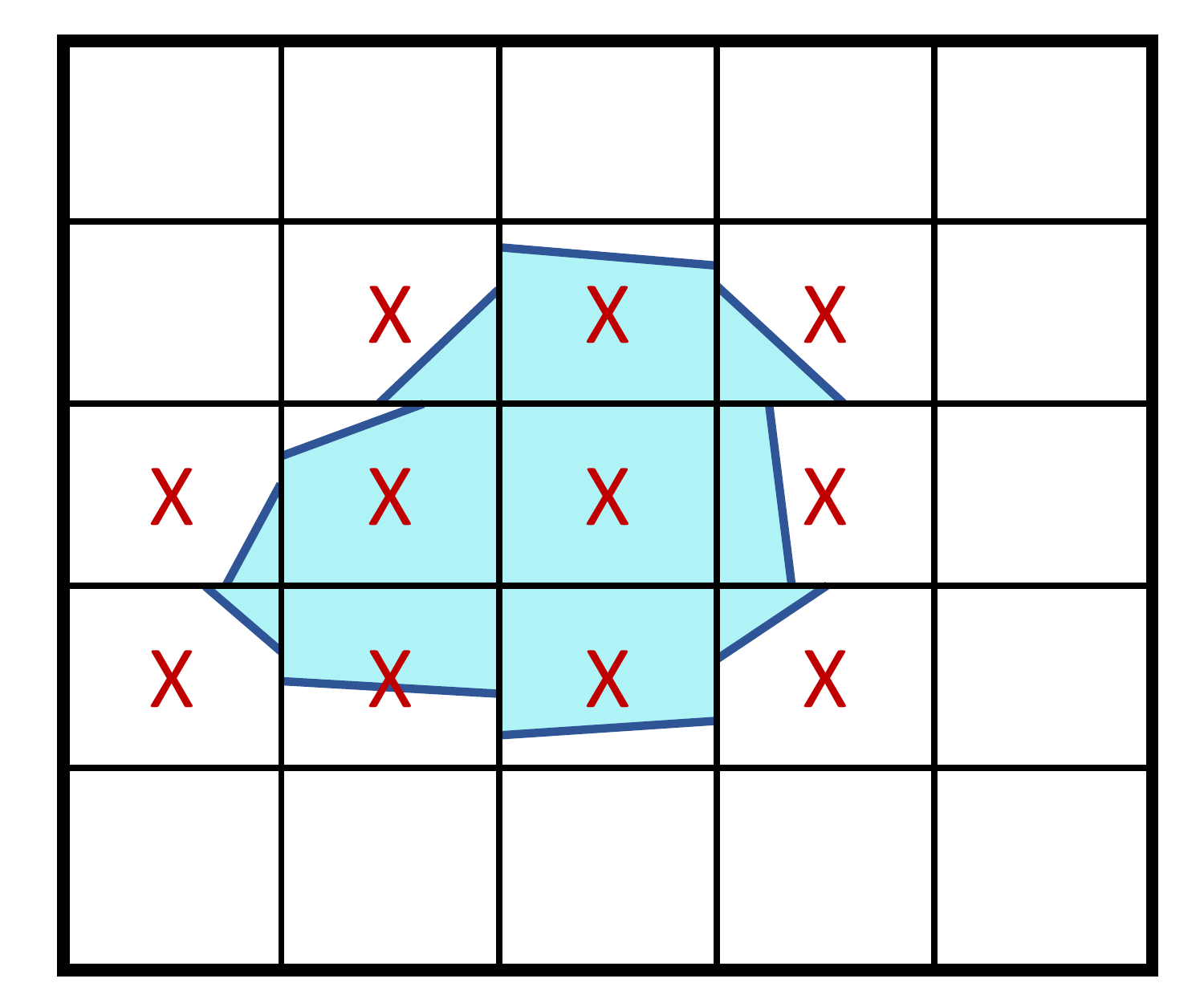}
}
  \caption{Schematics illustrating the dispersed-phase identification algorithm in a two-dimensional system simulated by a VoF-based two-phase solver. Each box in each grid corresponds to a single computational cell. In the case of a node-centered code, this specifically refers to the median dual cell of each node. Each of the lines traversing some of the boxes corresponds to a numerical phase interface representation compatible with the conventional piecewise-linear interface calculation (PLIC) scheme. Suppose that the shaded region contains liquid and the other region contains gas. (a) The pair of crosses highlights a pair of neighboring cells that satisfies the grouping criterion of the employed identification algorithm. (b) The crosses highlight all the cells that are grouped together after the identification algorithm is executed. This contiguous collection of cells will be associated with a single drop.}
\label{fig:flood}
\end{figure}

The traditional method for identifying bubbles and drops is to identify sets of connected computational nodes or cells corresponding to individual bubbles and drops. This algorithm may be applied to interface-advection schemes with a well-defined phase interface such that the volume fraction in each node or cell may be reconstructed. This includes geometric VoF, level-set, and front-tracking methods. Diffuse-interface and algebraic VoF methods are also eligible, provided care is taken to sharpen the diffuse interface in a volume-conserving manner before the algorithm is executed, or to analytically reconstruct any missing mass that remains outside identified bubbles and drops after the algorithm is executed. Each of these connected sets is built by linking cell pairs that satisfy a particular grouping criterion related to the dispersed-phase volume fraction, $\phi$. Without loss of generality and for brevity, assume that the dispersed phase comprises liquid drops. The traditional grouping criterion is as follows: if two neighboring cells $i$ and $j$ have $\phi=\phi_i$ and $\phi=\phi_j$, respectively, where $0 < \phi_i,\phi_j \leq 1$, then $\phi_i$ and $\phi_j$ should each exceed some threshold volume fraction $\phi_c \geq 0$. The most general choice for $\phi_c$ is $\phi_c = 0$, which involves no arbitrary assumptions and guarantees that the sum of the volumes of all identified drops equals the total liquid volume. Irrespective of the criterion, the general grouping process is illustrated in Fig.~\ref{fig:flood} for a VoF-based scheme.

\begin{figure}
  \centerline{
(a)
\includegraphics[trim={0pt 300pt 20pt 0pt},clip=true,width=0.65\linewidth,valign=t]{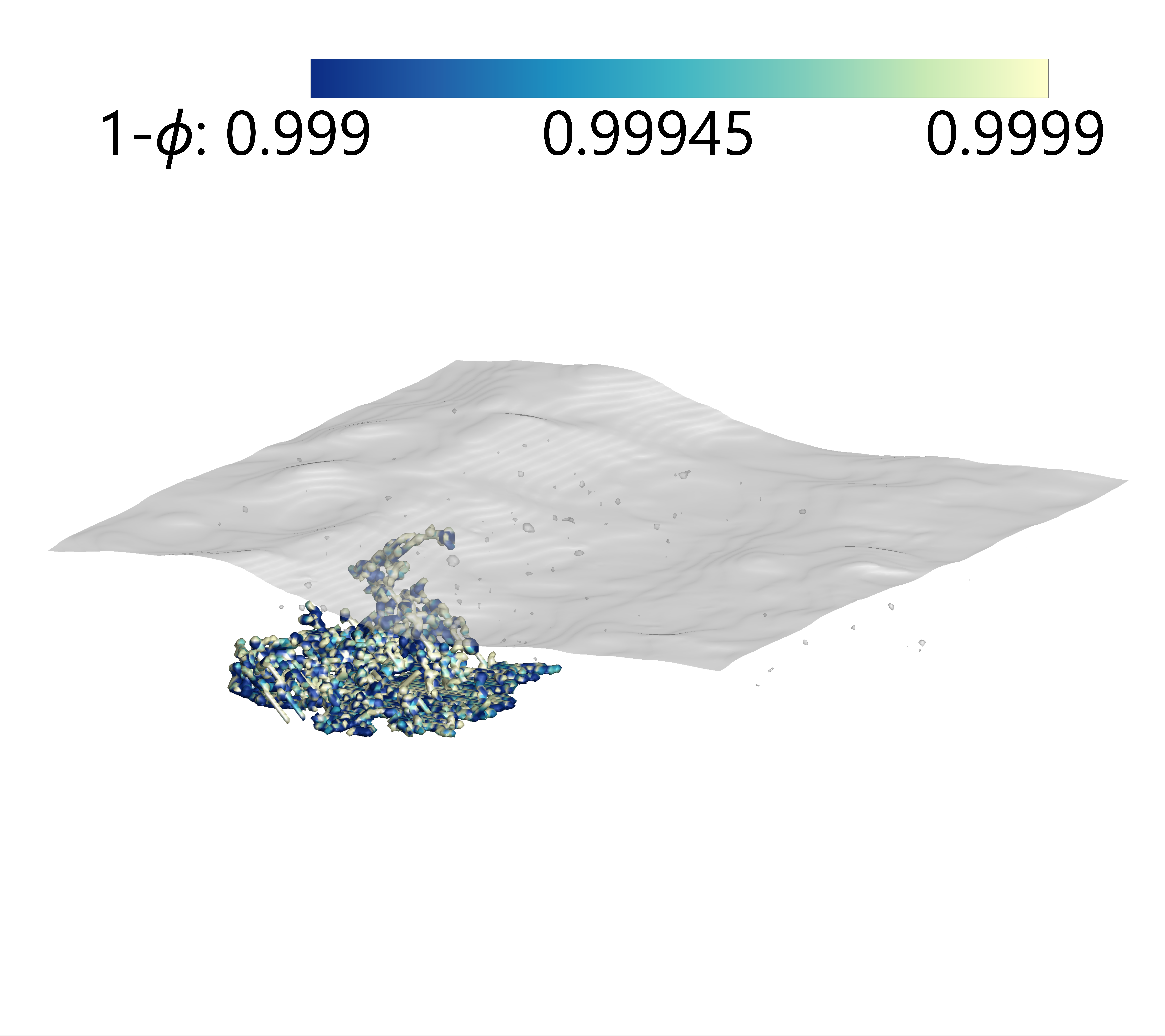}
}
\quad\\
  \centerline{
(b)
\includegraphics[trim={0pt 300pt 20pt 0pt},clip=true,width=0.65\linewidth,valign=t]{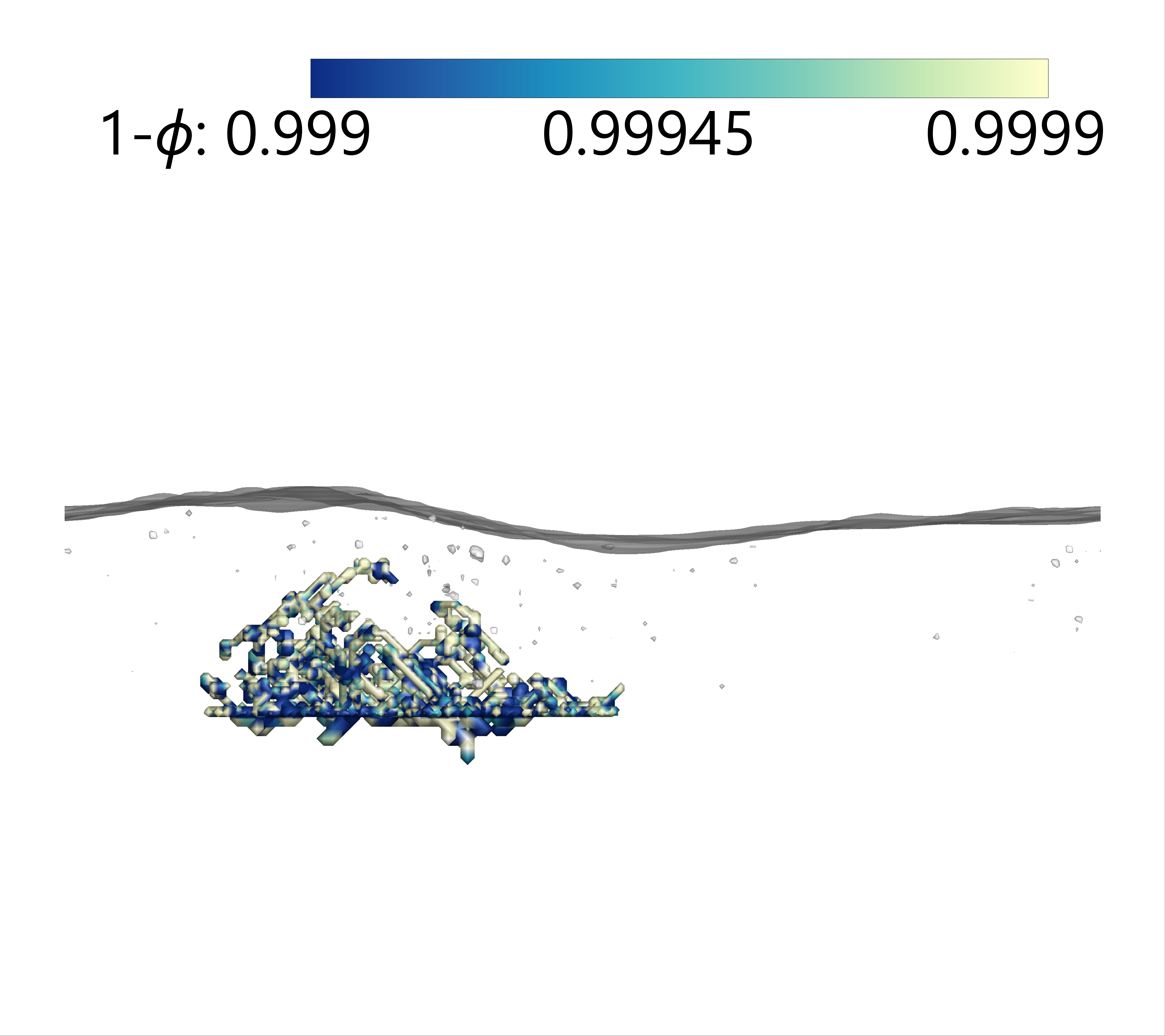}
}
  \caption{A bubble of spurious numerical origin identified in a breaking-wave simulation using the traditional identification algorithm with $\phi_c=0$. In this simulation snapshot, the wave has already been broken for a relatively long time, and large bubbles are not expected in the flow during this late stage of the wave-breaking process. The spurious bubble is made up of wisps of air in numerous neighboring cells, each having a small air volume fraction, which were grouped together by the traditional algorithm. A similar scenario is depicted in the fourth row of Table~\ref{tab:groupcase}. The figure depicts the spurious bubble with (a) an axonometric projection from above the wave and (b) a spanwise cross-section of the translucent $\phi=0.5$ isosurface, where $\phi$ is the dispersed-phase volume fraction, or the gaseous volume fraction in this case. These isosurfaces were obtained from a flow snapshot about 4.7 wave periods after the wave was initialized, and about 4.2 periods after it breaks. The perimeter nodes of the spurious bubble are shaded by the liquid volume fraction, $1-\phi$, in each node. Observe that $\phi$ in these nodes is of the order $O(10^{-3}\text{--}10^{-4})$, i.e., $0.999 < 1-\phi < 0.9999$.}
\label{fig:spurious}
\end{figure}

The drawback of the threshold choice $\phi_c = 0$ is that drops of spurious numerical origin, which could arise due to the presence of wisps, may contaminate the identification process. Wisps are very small dispersed-phase fragments detached from a main body, corresponding to cells with very small $\phi$, that might be numerically generated by the inadequate resolution of pinch-off events due to energetic surface impacts and turbulent eddies, or by errors in the interface-advection scheme employed. If many of these wisps are closely spaced in a region, which is likely in the presence of energetic collisions, then they could be falsely grouped as a contiguous set of liquid cells. If these wisps are not excluded, then large spurious drops may appear in the size distribution at resolvable sizes. The spurious drops may also have considerable surface area that may contribute erroneously to the total surface energy. An example of such a spurious structure from a breaking-wave simulation is depicted in Fig.~\ref{fig:spurious}. Here, a 27-cm-long deep water wave at atmospheric conditions was simulated with integral-scale Weber and Reynolds numbers $\We_L = 1.6\times10^3$ and $\RR_L = 1.8\times10^5$, respectively. For more simulation details, refer to~\citet{Chan3,Chan4,Chan5}, as well as \S~\ref{sec:ident-test-wave}. Making $\phi_c$ sufficiently large eliminates the contributions of these wisps. For example, a value of $\phi_c$ larger than $10^{-3}$ can eliminate the spurious bubble depicted in Fig.~\ref{fig:spurious}. Ref.~\citep{Yu3} uses $\phi_c=0.05$, while Ref.~\citep{Pepiot1} uses $\phi_c=0.85$. Generally speaking, the more energetic the flow, the larger the likely magnitude of the wisps, and the larger the value of $\phi_c$ required to eliminate these wisps. However, making $\phi_c$ large also decreases the computed volumes of the resolved drops, especially those of sizes close to the mesh resolution, since cells with small $\phi$ are then indiscriminately clipped throughout the computational domain regardless of their origin. Hence, accurate reporting of dispersed-phase size distributions and surface energies in two-phase flow simulations, especially of flows with energetic impacts and turbulent eddies, demands the exclusion of these spurious dispersed-phase structures through a more introspective modification of the grouping criterion in the identification algorithm, as described below.

In order to strike a balance between neglecting wisps and not unnecessarily clipping cells with small $\phi$, the following grouping criterion is proposed. In order for a pair of liquid-containing cells $i$ and $j$ to be eligible for grouping, at least one of the cells should have a liquid volume fraction above some threshold value $\phi_{c,m}$. In other words, small wisps of liquid are only considered for grouping if they are attached to a large liquid mass. This prevents closely spaced wisps from being grouped to form drops of spurious numerical origin, while reducing the mass clipped from cells with small $\phi$ on the perimeter of resolved drops. Note that the presence of these cells is inherent in volume discretization: if the drop surface does not conform to the underlying mesh geometry, then cut cells with a broad distribution of $\phi$ are almost certain to be generated where the surface intersects the mesh. It is thus desirable to not unnecessarily exclude these cells in the computation of drop volumes in a general fashion. The proposed grouping criterion retains these cells whenever possible by recognizing that cells with small $\phi$ are likely to contain liquid that is part of a resolved drop if they reside next to cells with large $\phi$, as the former would likely be near the drop boundary and the latter near the drop interior. Unlike the method proposed by~\citet{Hendrickson1,Hendrickson2}, this grouping criterion only requires a single pass over the computational domain to accurately capture all dispersed-phase structures since it does not involve multiple thresholds. Table~\ref{tab:groupcase} illustrates the cell pairs that are permissible for grouping based on the grouping criteria discussed above.

\begin{table}
\begin{center}
\begin{tabular}{c | >{\centering\arraybackslash} m{0.12\textwidth} >{\centering\arraybackslash} m{0.12\textwidth} >{\centering\arraybackslash} m{0.12\textwidth} >{\centering\arraybackslash} m{0.12\textwidth}}
\multirow{3}{*}{Case} & $\phi_c=0$ & $\phi_c>0$ & $\phi_c=0$\\
& $\phi_{c,m}=0$ & $\phi_{c,m}=0$ & $\phi_{c,m}>0$\\
& (Criterion A) & (Criterion B) & (Criterion C)\\
\raisebox{-.5\height}{\includegraphics[width=0.3\textwidth]{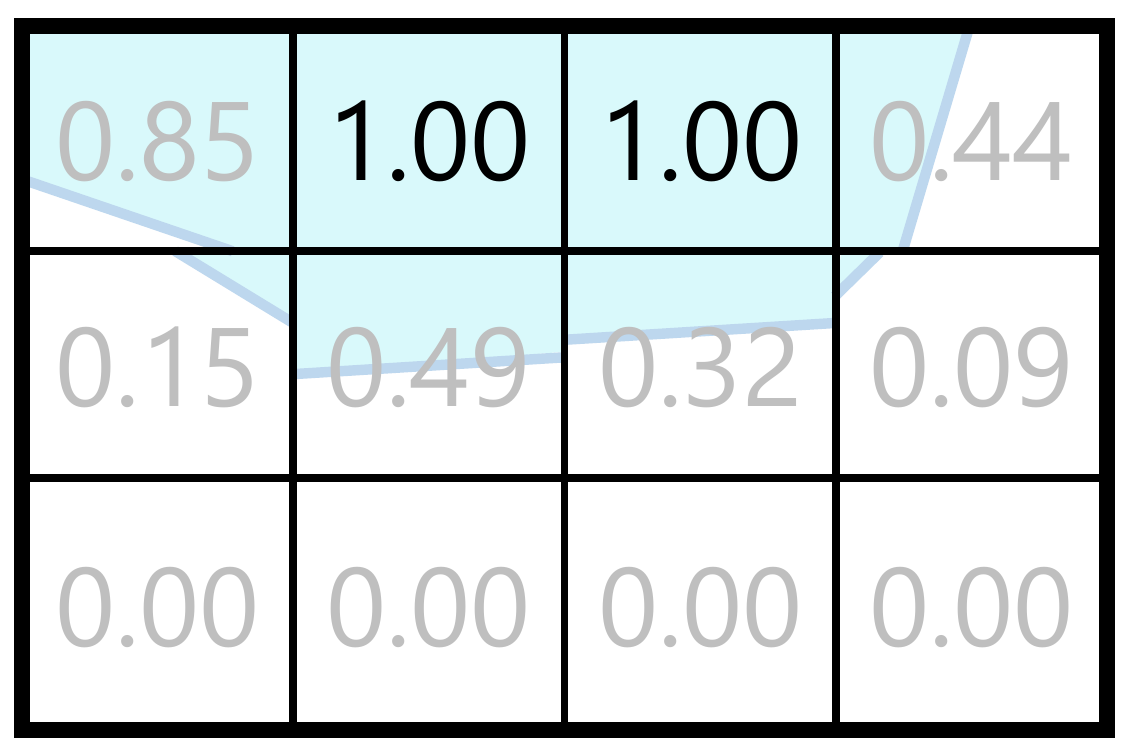}} & \Huge\cmark & \Huge\cmark & \Huge\cmark \\
\raisebox{-.5\height}{\includegraphics[width=0.3\textwidth]{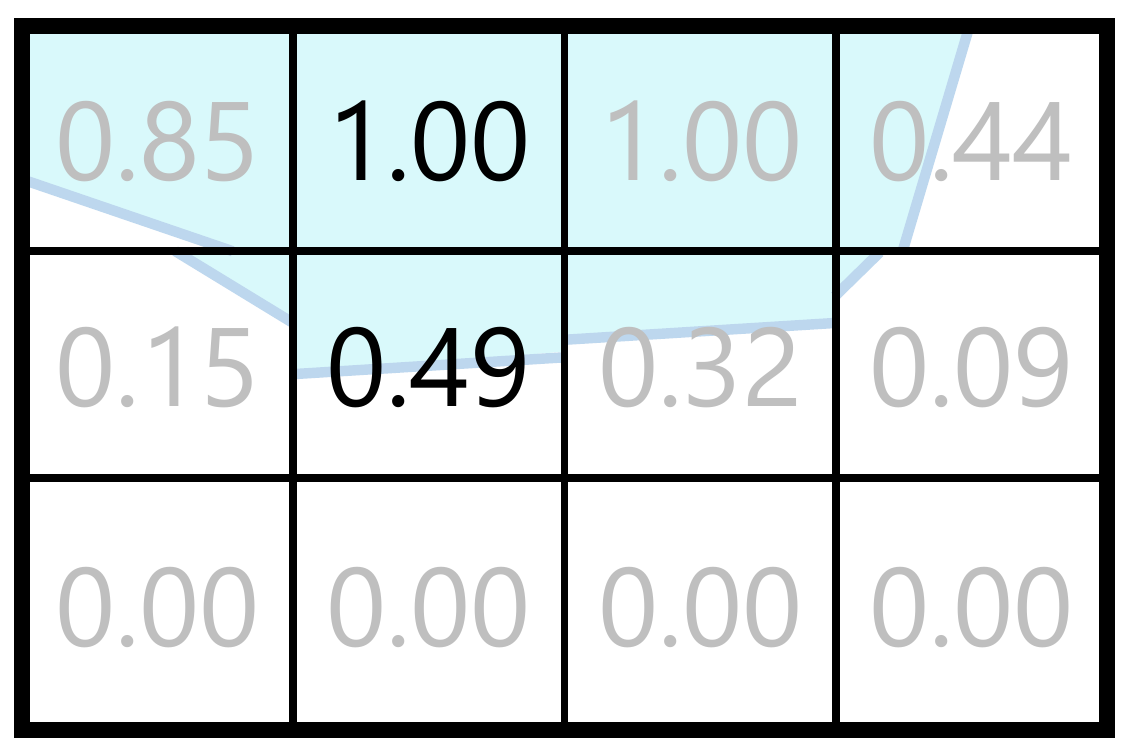}} & \Huge\cmark & \Huge\xmark & \Huge\cmark \\
\raisebox{-.5\height}{\includegraphics[width=0.3\textwidth]{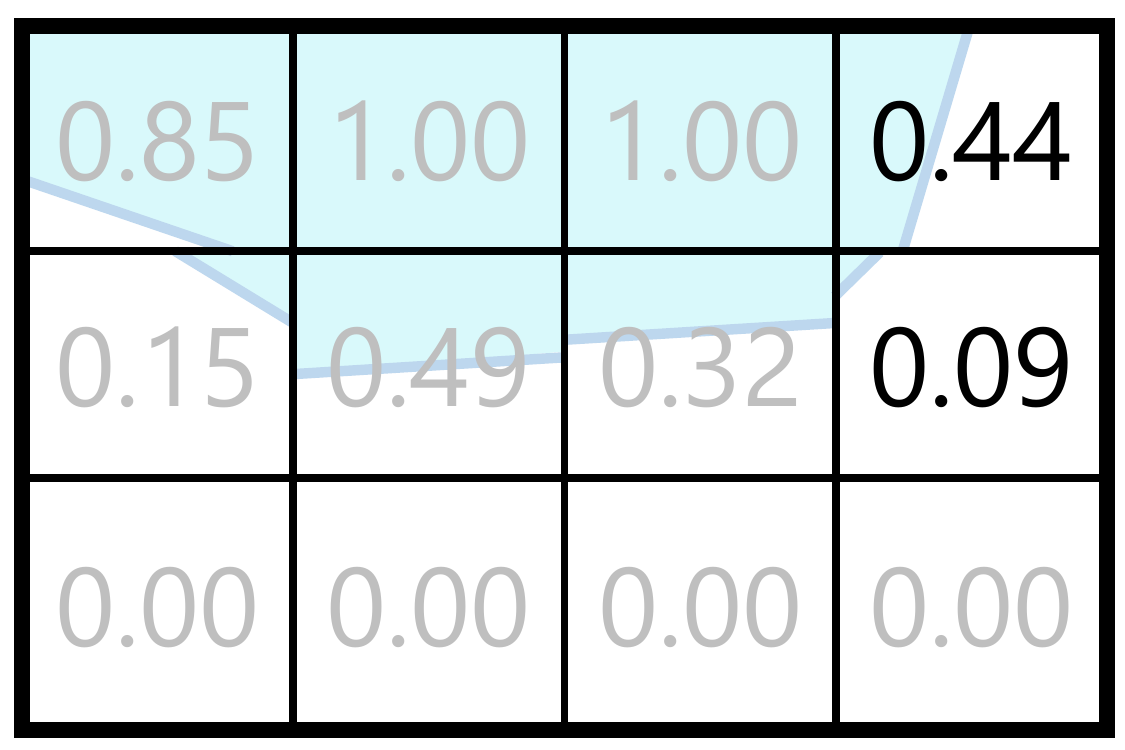}} & \Huge\cmark & \Huge\xmark & \Huge\xmark \\
\raisebox{-.5\height}{\includegraphics[width=0.3\textwidth]{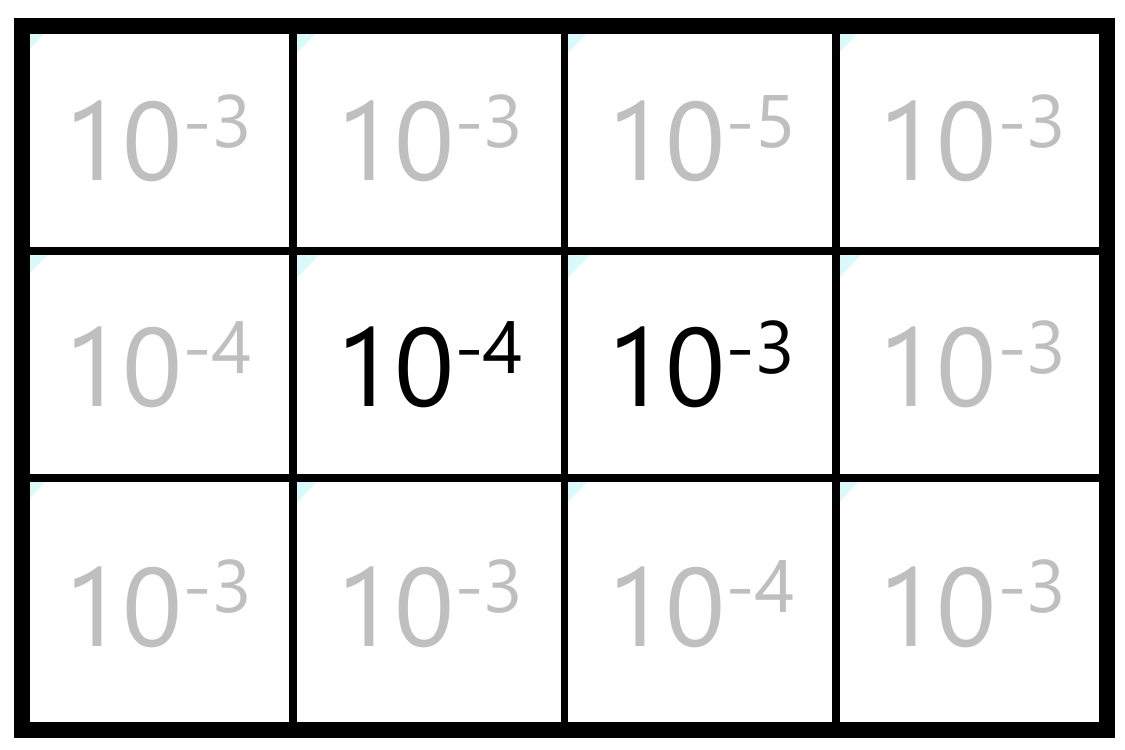}} & \Huge\cmark & \Huge\xmark & \Huge\xmark \\
\raisebox{-.5\height}{\includegraphics[width=0.3\textwidth]{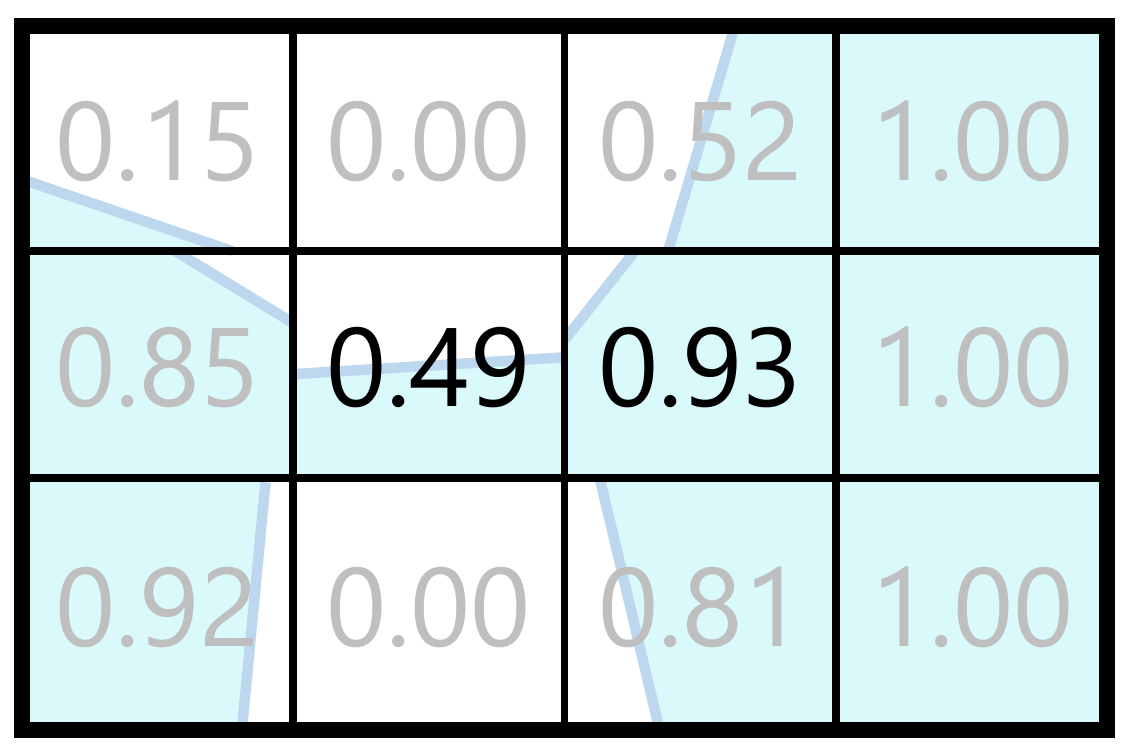}} & \Huge\cmark & \Huge\xmark & \Huge\cmark \\
\normalsize
\end{tabular}
\caption{Cell pairs (in bold text) that are permitted (\cmark) or not permitted (\xmark) for grouping in the identification algorithm subject to the following grouping criteria: traditional grouping criterion with $\phi_c=0$ and $\phi_{c,m}=0$ (Criterion A), clipping of small $\phi$ such that $\phi_c>0$ while $\phi_{c,m}=0$ (Criterion B), and the proposed grouping criterion with $\phi_c=0$ and $\phi_{c,m}>0$ (Criterion C). For illustration, the thresholds $\phi_c=0.5$ and $\phi_{c,m}=0.5$ are chosen for the second and third criteria, respectively. Each row corresponds to a particular pair of cells, depicted together with some of its neighbors, in the first column. For an explanation of the grid, cell shading, and lines in each subfigure in the first column, refer to the caption of Fig.~\ref{fig:flood}. In particular, the sloped lines depict the numerical representation of the phase interface. The numbers in each cell correspond to $\phi$. The $\phi$ fields in the first three rows of the table are identical, with different cell pairs highlighted in each row in bold, and represent the corner of a drop. The fourth row illustrates a region filled with wisps, such as one that gave rise to the spurious structure depicted in Fig.~\ref{fig:spurious}. The last row illustrates the corner case discussed at the end of \S~\ref{sec:ident}.}
\label{tab:groupcase}
\end{center}
\end{table}

This section is closed with a discussion of the corner case that afflicts many identification schemes: distinguishing two drops spaced a grid cell apart from a dumbbell-shaped drop where two liquid masses are connected by a thin liquid bridge with the dimensions of a grid cell. Note that the occurrence of this corner case does not necessarily suggest a deficiency in any of these schemes. Instead, it is representative of an inherent limitation of a discrete volume fraction field with finite numerical resolution: when a thin liquid bridge is numerically indistinguishable from a small underresolved drop or a small liquid protrusion on the surface of a larger drop, none of the geometries necessarily represent reality more accurately in the absence of additional information. Consequently, any decision by any scheme to favor any of the geometries is essentially arbitrary to a certain degree. That said, the outcomes of various identification schemes have not been convergent. For example, the scheme by~\citet{Hendrickson1,Hendrickson2} will always link the liquid bridge to one of the large drops, but the decision of which drop to be attached to depends on the threshold selected in the scheme. As illustrated in the last row of Table~\ref{tab:groupcase}, the grouping criterion introduced in this work collectively identifies the liquid masses and the liquid bridge as a single drop if the bridge is connected to cells in both liquid masses with sufficiently large $\phi$, e.g., $\phi > \phi_{c,m} = 0.5$ in the referenced example. Otherwise, it returns two large drops and a small underresolved drop. This corner case is revisited in the context of the test cases in \S~\ref{sec:ident-test}.

\section{Test cases for the identification algorithm}\label{sec:ident-test}

In the following test and demonstration cases, five grouping criteria will be referenced: the traditional criterion (Criterion A), clipping of small $\phi$ with two choices for the threshold $\phi_c$ (Criteria B1 and B2), and the proposed criterion with two choices for the threshold $\phi_{c,m}$ (Criteria C1 and C2). These criteria are summarized in Table~\ref{tab:criteria}. Note that the values of $\phi_c$ examined in Criteria B1 and B2 are in the ballpark of the values considered by Refs.~\citep{Yu3,Pepiot1}, and are representative of values applicable to realistic flow solvers and conditions of physical and engineering interest.

\begin{table}
\begin{center}
\begin{tabular} { c | >{\centering\arraybackslash} m{0.3\textwidth} || >{\centering\arraybackslash} m{0.1\textwidth} | >{\centering\arraybackslash} m{0.1\textwidth} }
Criterion & Description & $\phi_c$ & $\phi_{c,m}$ \\
&&&\\
A & Traditional criterion & 0 & 0 \\
B1 & Clipping of small $\phi$ & 0.5 & 0 \\
B2 & Clipping of small $\phi$ & 0.1 & 0 \\
C1 & Proposed criterion & 0 & 0.5\\
C2 & Proposed criterion & 0 & 0.1 \\
\end{tabular}
\end{center}
\caption{Summary of the grouping criteria to be compared in the cases in \S~\ref{sec:ident-test} and \S~\ref{sec:ident-demo}.}
\label{tab:criteria}
\end{table}

\subsection{Advection of a single drop}\label{sec:ident-test-single}

In this test case, errors in the volume and centroid of a three-dimensional water drop of diameter $D$ moving in quiescent air are computed for the criteria listed in Table~\ref{tab:criteria}. The drop is advected with a prescribed velocity without deformation across a uniform Cartesian mesh of size $(5D)^3$ with grid spacing $\Delta x$, and the phase interface is reconstructed using the VoF-based solver described in Refs.~\citep{Kim2,Ham2,Bravo4}. This effectively provides samples of the volume and centroid at different drop positions relative to the mesh. The errors incurred by the criteria, averaged over 200 of these samples, are listed in Tables~\ref{tab:errtrans1a} and \ref{tab:errtrans1b} for two different drop resolutions. Here, the traditional criterion (Criterion A) provides the ground truth since there is only one drop and no wisps, so all computed errors are relative to the quantities yielded with this criterion. The errors introduced by the proposed criterion (Criteria C1/C2) are consistently smaller than those introduced by clipping (Criteria B1/B2). Note also that the centroid errors are orders of magnitude smaller than the grid spacing for all criteria considered. Identical tests were also performed for an air bubble moving in quiescent water with comparable errors (not shown here).

\begin{table}
\begin{center}
\begin{tabular} { c || >{\centering\arraybackslash} m{0.2\textwidth} | >{\centering\arraybackslash} m{0.2\textwidth} | >{\centering\arraybackslash} m{0.3\textwidth}  }
Criterion & Volume error/$(\Delta x)^3$ & Centroid error/$(\Delta x)$ & Area-normalized volume error [First column ${} \div \left\{\pi D^2 / (\Delta x)^2\right\}$] \\
&&&\\
B1 & $26.5 \pm 0.3$ & $(2.2 \pm 0.2) \times 10^{-2}$ &  $(1.32 \pm 0.01) \times 10^{-1}$\\
B2 & $2.44 \pm 0.04$ & $(4.6 \pm 0.3) \times 10^{-3}$ & $(1.21 \pm 0.02) \times 10^{-2}$\\
C1 & $(4.1 \pm 0.4) \times 10^{-2}$ & $(3.0 \pm 0.3) \times 10^{-4}$ & $(2.0 \pm 0.2) \times 10^{-4}$ \\
C2 & Less than m.p. & Less than m.p. & Less than m.p.\\
\end{tabular}
\end{center}
\caption{Nondimensional volume and centroid errors for the drop advected with a prescribed velocity in \S~\ref{sec:ident-test-single} with the drop resolution $D/(\Delta x) = 8$. In each entry, the first value refers to the mean absolute deviation from the ground truth obtained using Criterion A over 200 samples, while the second value denotes twice the standard error over the 200 samples. For descriptions of the tested criteria, refer to Table~\ref{tab:criteria}. Here, m.p. denotes machine precision, and ``less than m.p." indicates that the computed error is associated with a value less than the standard double-precision machine epsilon $\sim 10^{-16}$.}
\label{tab:errtrans1a}
\end{table}

\begin{table}
\begin{center}
\begin{tabular} { c || >{\centering\arraybackslash} m{0.2\textwidth} | >{\centering\arraybackslash} m{0.2\textwidth} | >{\centering\arraybackslash} m{0.3\textwidth} }
Criterion & Volume error/$(\Delta x)^3$ & Centroid error/$(\Delta x)$ & Area-normalized volume error [First column ${} \div \left\{\pi D^2 / (\Delta x)^2\right\}$] \\
&&&\\
B1 & $100.8 \pm 0.6$ & $(1.3 \pm 0.1) \times 10^{-2}$ & $(1.254 \pm 0.007) \times 10^{-1}$\\
B2 & $8.80 \pm 0.07$ & $(2.3 \pm 0.1) \times 10^{-3}$ & $(1.094 \pm 0.009) \times 10^{-2}$\\
C1 & $(7.7 \pm 0.5) \times 10^{-2}$ & $(1.3 \pm 0.1) \times 10^{-4}$ & $(9.6 \pm 0.6) \times 10^{-5}$\\
C2 & Less than m.p. & Less than m.p. & Less than m.p.\\
\end{tabular}
\end{center}
\caption{The same quantities in Table~\ref{tab:errtrans1a} with the drop resolution $D/(\Delta x) = 16$.}
\label{tab:errtrans1b}
\end{table}

\subsection{Large drop--small drop pair}\label{sec:ident-test-pair}

\begin{figure}
  \centerline{
(a)
\includegraphics[width=0.275\linewidth,valign=t]{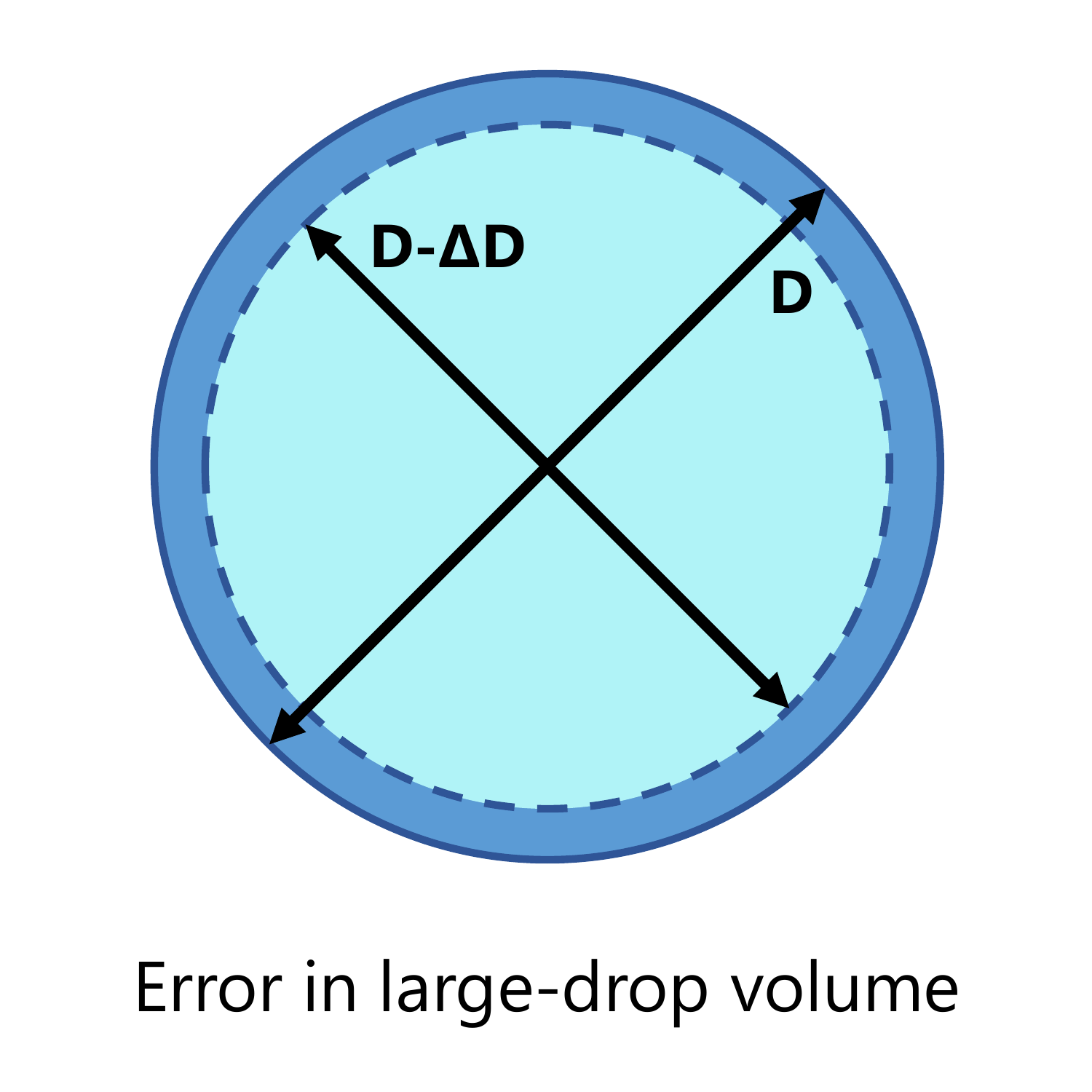}
\quad
(b)
\includegraphics[width=0.275\linewidth,valign=t]{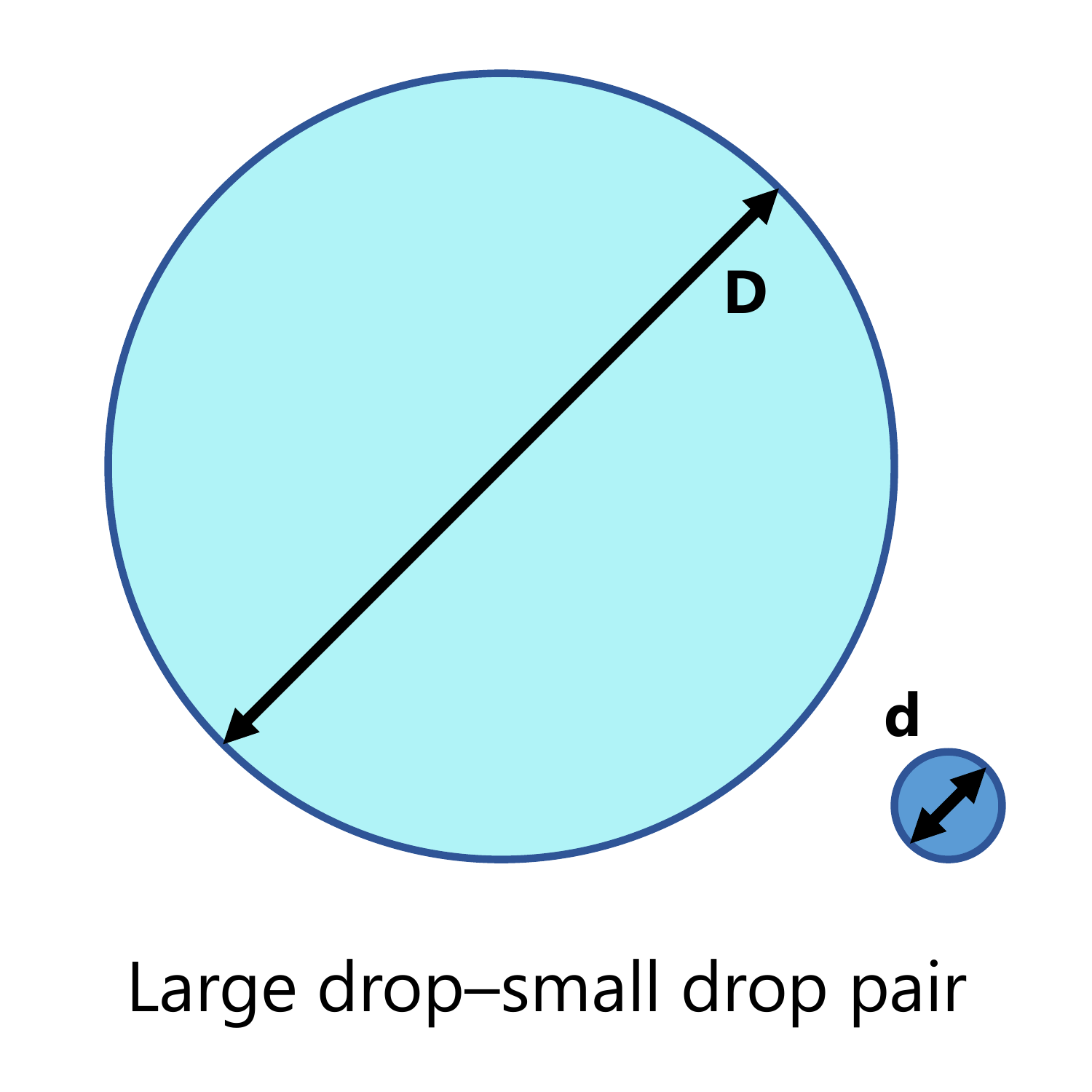}
}
  \caption{(a) Schematic depicting variations in the volume of a large drop of diameter $D$. The circle with a dashed outline represents a slightly shrunken drop of diameter $D - \Delta D$. (b) If these variations in the large-drop volume are comparable to the volume of a small drop of diameter $d$, then the presence of the large drop--small drop pair cannot be distinguished from fluctuations in the large-drop volume.}
\label{fig:droppair}
\end{figure}

The volume error due to the advection of a single drop may be generalized by considering if a pair of drops comprising one large drop of diameter $D$ and one small drop of diameter $d$ may be distinguished from fluctuations in the volume of the large drop, recalling from the end of \S~\ref{sec:ident} that a drop pair with limited separation may be difficult to distinguish from a single drop due to finite numerical resolution, and from \S~\ref{sec:ident-test-single} that the alignment of a drop with the underlying mesh results in variations in the computed drop volume between time-steps. First, consider the volume variation $\Delta \mathcal{V}_\text{err}$ of a sufficiently large spherical drop of diameter $D$, as illustrated in Fig.~\ref{fig:droppair}(a). Tables~\ref{tab:errtrans1a} and \ref{tab:errtrans1b} suggest that the nondimensional volume variation, $\Delta \mathcal{V}_\text{err}/(\Delta x)^3$, is proportional to the nondimensional drop surface area, $\pi D^2/(\Delta x)^2$, and that the constant of proportionality, $M$, may be approximated as a function of only the grouping criterion used. One may also surmise this by assuming that the only source of volume error is the culling of several cut cells at the drop surface by the grouping criterion---see, e.g., the first three cases in Table \ref{tab:groupcase}. The volume error may be approximated as being proportional to the difference between the drop volume, $\mathcal{V}_1 = \pi D^3 / 6$, and the volume of a slightly shrunken drop, $\mathcal{V}_2 = \pi (D-\Delta D)^3 / 6$, where $\Delta D \sim \Delta x$. Assuming $\Delta D \ll D$, this volume difference may be written as $\delta \mathcal{V} = \mathcal{V}_1 - \mathcal{V}_2 \approx \pi D^2 \Delta D/2$. Correspondingly, $\Delta \mathcal{V}_\text{err} \propto \delta \mathcal{V}$ may be expressed as
\begin{equation}
\f{\Delta \mathcal{V}_\text{err}}{(\Delta x)^3} = M \left[ \pi \left(\f{D}{\Delta x}\right)^2 \right].
\label{eqn:volprop}
\end{equation}
It may further be shown that the corresponding fractional diameter error $\Delta D_\text{err}/D$ may be estimated as 
\begin{equation}
\f{\Delta D_\text{err}}{D} \approx 2M\f{\Delta x}{D}.
\label{eqn:nondimvolerror}
\end{equation}
In particular, one may obtain, from \eqref{eqn:volprop}, $\Delta \mathcal{V}_\text{err} = M \pi D^2 \Delta x \approx \pi D^2 \Delta D_\text{err} / 2$. Next, consider the scenario where the volume of the small drop, $\pi d^3/6$, in the large drop--small drop pair is comparable to $\Delta \mathcal{V}_\text{err}$, as illustrated in Fig.~\ref{fig:droppair}(b). Then, one may write
\begin{equation}
\f{\pi\left(\f{d}{\Delta x}\right)^3}{6} = \f{\pi N^3}{6} \sim M \left[ \pi \left(\f{D}{d}\f{d}{\Delta x}\right)^2 \right] = M \left[ \pi \left(\f{D}{d}N\right)^2 \right],
\label{eqn:volpropratio}
\end{equation}
or 
\begin{equation}
r = \f{D}{d} \sim \sqrt{\f{N}{6M}},
\label{eqn:critratio}
\end{equation}
where $r$ is the critical size ratio corresponding to this scenario, and $N \equiv d/(\Delta x)$ is the number of cells across the small drop. Observe that $r/\sqrt{N}$ is a function of only $M$. Thus, as the resolution of the small drop ($N$) increases, the critical size ratio ($r$) between a large drop--small drop pair that the algorithm is able to correctly identify also increases. A decrease in the volume error ($M$) committed by the algorithm in \S~\ref{sec:ident} results in an increase in $r$ as well. Returning to the single-drop test case in \S~\ref{sec:ident-test-single}, Table \ref{tab:errtrans2} re-expresses the volume errors in Tables \ref{tab:errtrans1a} and \ref{tab:errtrans1b} in terms of $M$ and $r/\sqrt{N}$. Note that these quantities do not seem to be too sensitive to the drop resolution, $D/(\Delta x)$. These observations support the hypothesis that $M$ and thus $r/\sqrt{N}$ are functions of only the grouping criterion for sufficiently well-resolved drops. 

The critical size ratio, $r$, is a crucial parameter driving the accurate detection of breakup and coalescence events and will be revisited in \S~\ref{sec:track} and \S~\ref{sec:track-test}. For example, consider a population of fragmenting drops of sizes between 10 \textmu m and 1 mm. If $r \sim O(10^1)$, then events involving a 1-mm drop breaking up into drops of sizes smaller than 100 \textmu m, as well as events involving 100-\textmu m drops and/or smaller drops fragmenting near a 1-mm drop, may not be distinguishable from volume fluctuations in the 1-mm drop. Increasing $r$ to $O(10^2\text{--}10^3)$ through a higher mesh resolution and/or a more accurate identification algorithm enables the detection of these events.

\begin{table}
\begin{center}
\begin{tabular} { c || >{\centering\arraybackslash} m{0.16\textwidth} | >{\centering\arraybackslash} m{0.16\textwidth} || >{\centering\arraybackslash} m{0.16\textwidth} | >{\centering\arraybackslash} m{0.16\textwidth} }
\multirow{2}{*}{Criterion} & $M$ & $r/\sqrt{N}$ & $M$ & $r/\sqrt{N}$ \\
& [$D/(\Delta x) = 8$] & [$D/(\Delta x) = 8$] & [$D/(\Delta x) = 16$] & [$D/(\Delta x) = 16$] \\
&&&&\\
B1 & $0.1$ & $1$ & $0.1$ & $1$ \\
B2 & $0.01$ & $4$ & $0.01$ & $4$ \\
C1 & $2 \times 10^{-4}$ & $30$ & $1 \times 10^{-4}$ & $40$ \\
C2 & Less than m.p. & 1/(Less than m.p.) & Less than m.p.  & 1/(Less than m.p.) \\
\end{tabular}
\end{center}
\caption{Values of $M$ and $r$ as defined in \eqref{eqn:volprop} and \eqref{eqn:critratio}, respectively, for the drop in \S~\ref{sec:ident-test-single}. Refer to Table \ref{tab:criteria} for descriptions of the criteria.}
\label{tab:errtrans2}
\end{table}

\subsection{A known drop size distribution}\label{sec:ident-test-dist}

\begin{figure}
\begin{center}
\includegraphics[width=0.55\textwidth]{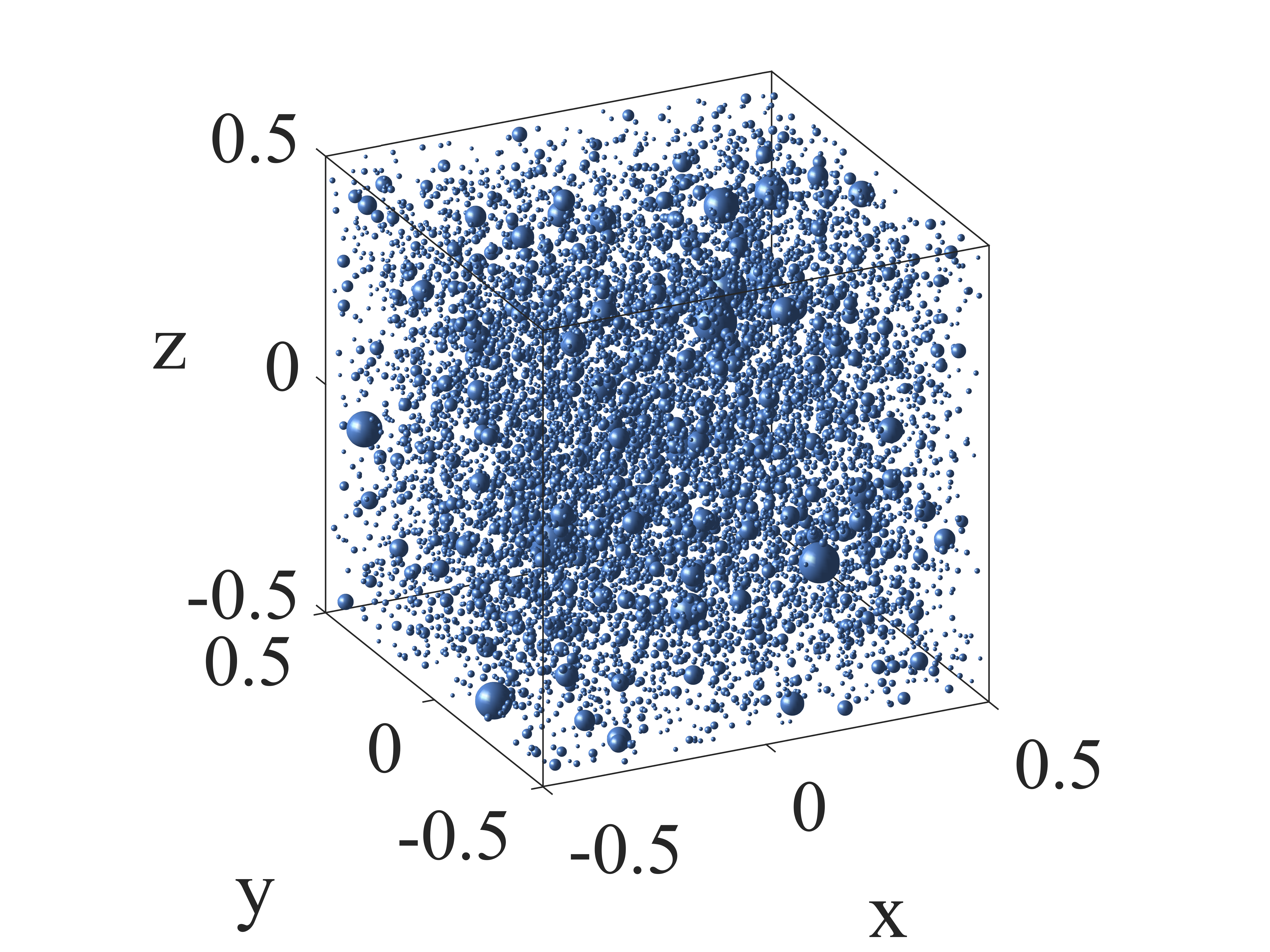}
\caption{An illustration of one possible realization of the 10{,}000-drop system described in \S~\ref{sec:ident-test-dist}. Here, $L=1$.}
\label{fig:drops}
\end{center}
\end{figure}

In this test case, the criteria in Table~\ref{tab:criteria} are evaluated on an analytically prescribed drop size distribution. 10{,}000 drops are randomly seeded following a power-law distribution with exponent $-10/3$ in a domain of size $L^3 = (500\Delta x)^3$. Seeded drops have radii $R \in [2\Delta x,200\Delta x]$ and are separated by at least two grid cells. A sample seeding pattern is depicted in Fig.~\ref{fig:drops}. Note that this two-grid-cell steric restriction has the effect of reducing the number of large drops eligible for seeding and depresses the large-drop distribution away from the idealized $-10/3$ power-law scaling. This seeding yields a global packing fraction of 2.2\%, i.e., 2.2\% of the domain volume is occupied by drops. The seeding procedure is implemented in the solver referenced in \S~\ref{sec:ident-test-single} so that the identification algorithm is executed on the corresponding VoF field. Since the drops are identified immediately after seeding, no wisps are generated, so the traditional criterion (Criterion A) provides the ground truth once again and is used to supply the reference distribution. 

\begin{figure}
  \centerline{
\includegraphics[width=0.42\linewidth,valign=t]{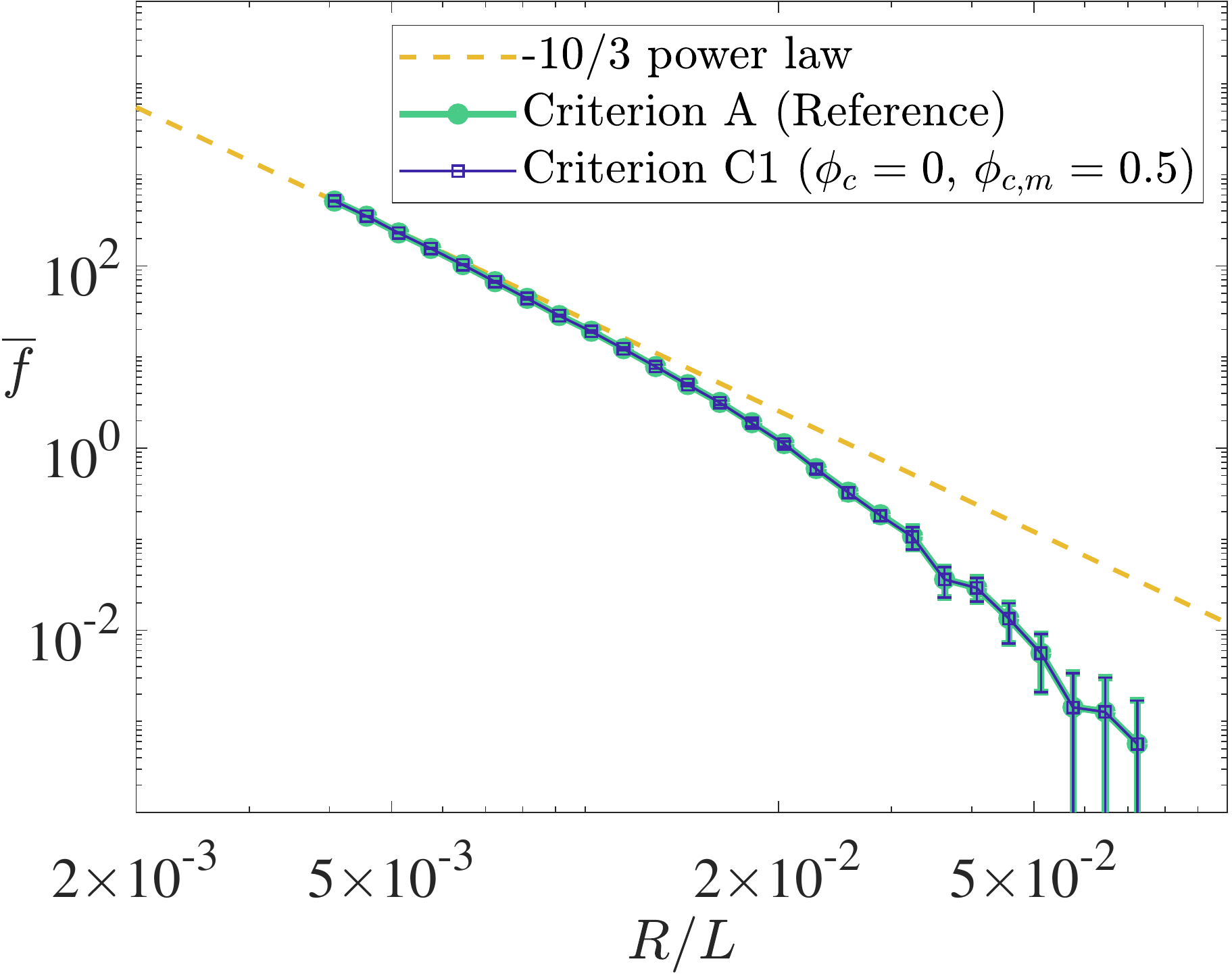}
}
 \caption{The drop size distribution, $\ov{f}$, averaged over 20 statistically independent realizations of the 10{,}000-drop system using Criterion C1. Refer to Table \ref{tab:criteria} for a description of the criterion. The distribution was computed using histogram bins of equal logarithmic spacing, and is normalized such that $\sum_j \ov{f}_j \Delta(R_j/L) = 1$. The smallest employed bin is larger than the radius error \eqref{eqn:nondimvolerror} incurred. The error bars denote twice the standard error over the 20 realizations.}
\label{fig:dist}
\end{figure}

\begin{figure}
  \centerline{
(a)
\includegraphics[width=0.42\linewidth,valign=t]{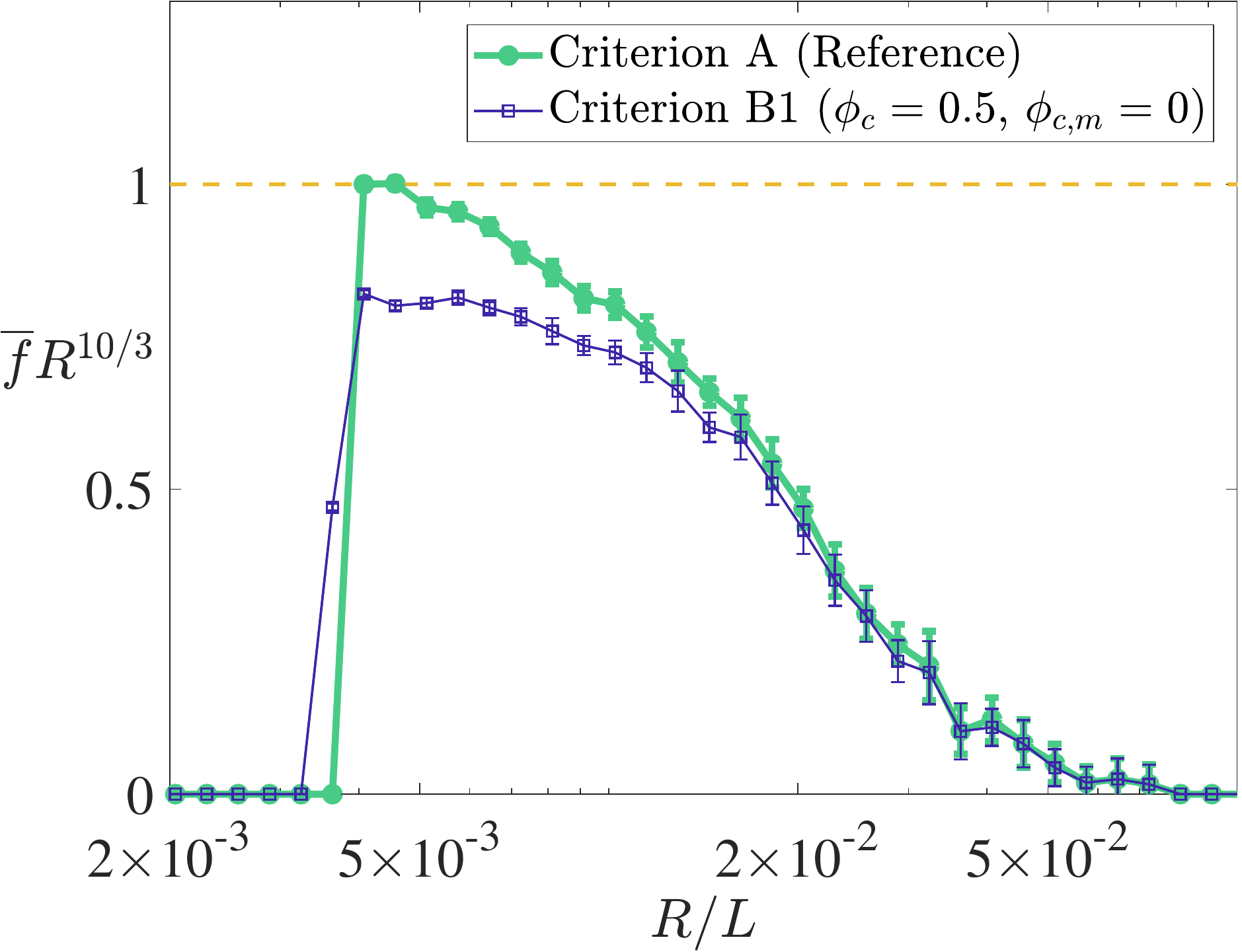}
\quad
(b)
\includegraphics[width=0.42\linewidth,valign=t]{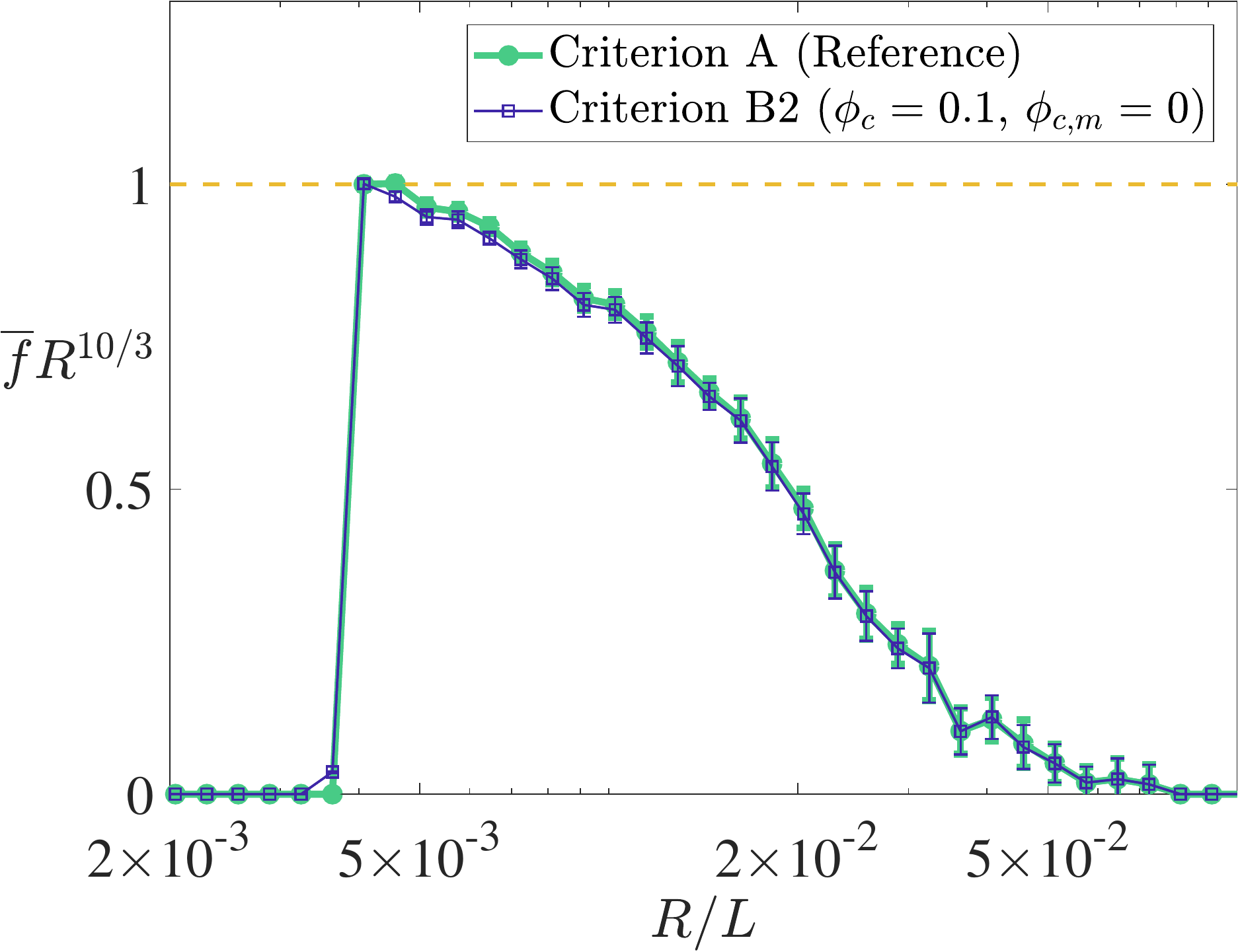}
}
  \centerline{
(c)
\includegraphics[width=0.42\linewidth,valign=t]{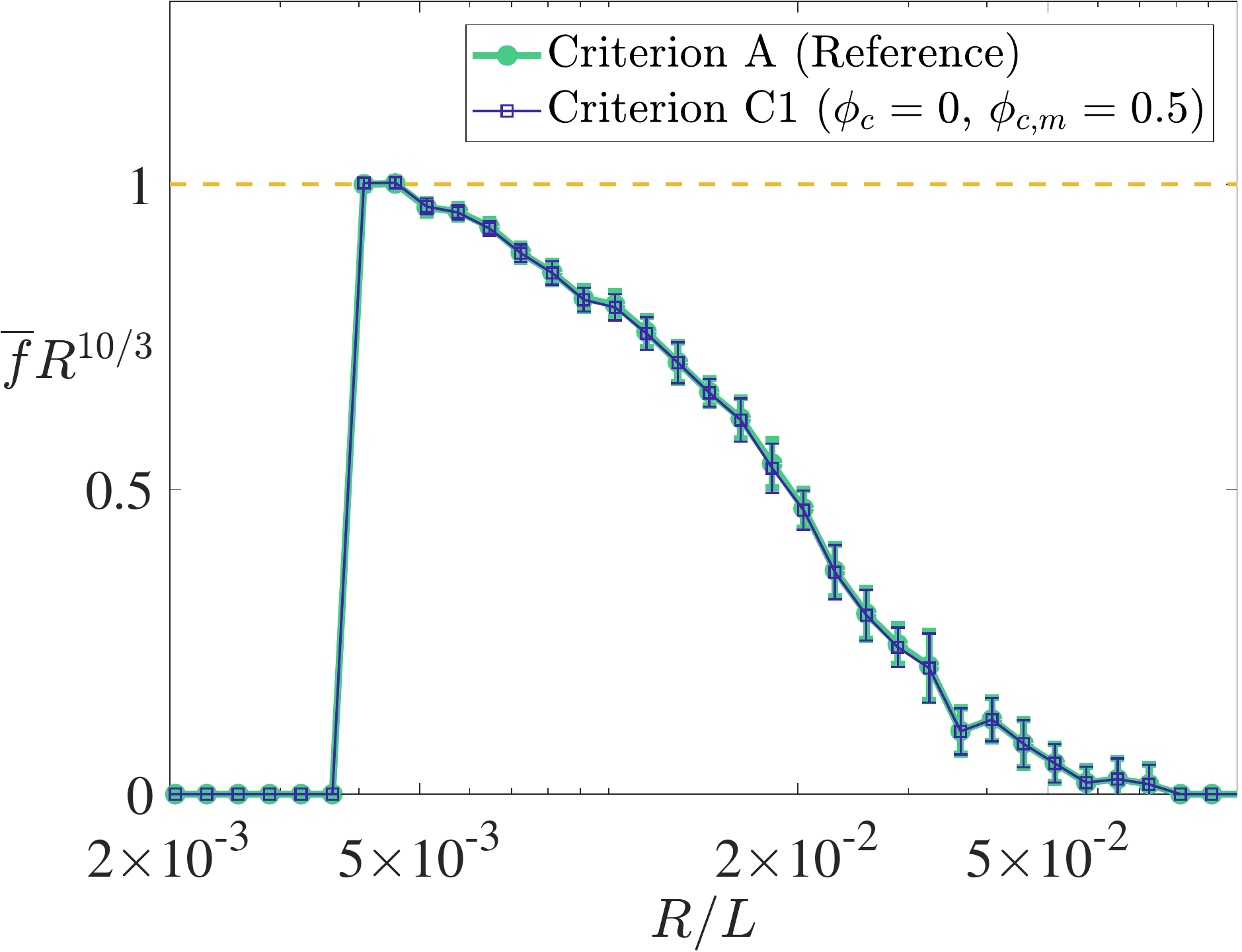}
\quad
(d)
\includegraphics[width=0.42\linewidth,valign=t]{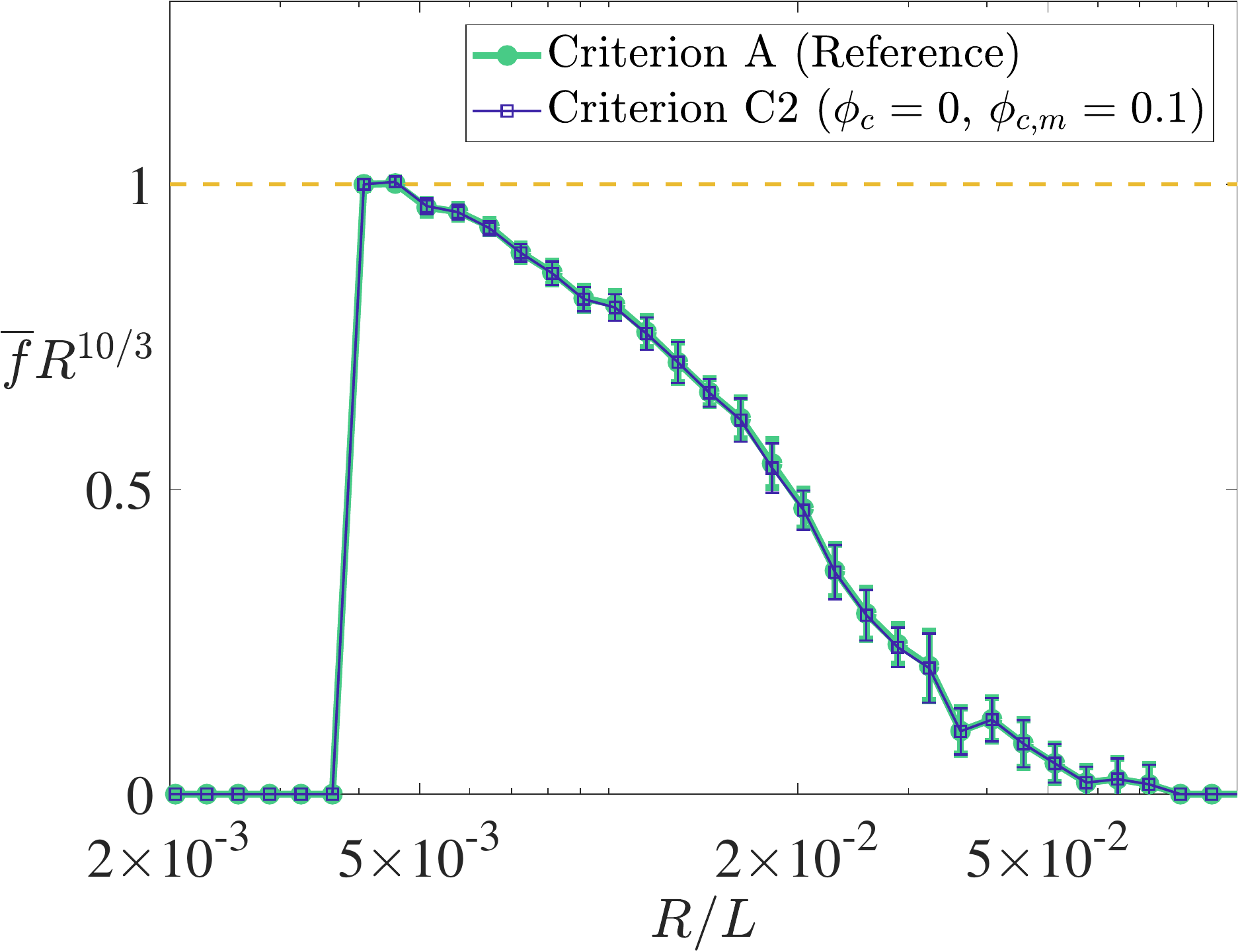}
}
  \caption{The compensated drop size distribution, $\ov{f}R^{10/3}$, premultiplied by the inverse of the idealized $R^{-10/3}$ scaling, using (a) Criterion B1, (b) Criterion B2, (c) Criterion C1, and (d) Criterion C2. Refer to Table \ref{tab:criteria} for descriptions of the criteria. The compensated distribution is normalized by the value of the reference distribution from Criterion A for the drop radius $R/L = 4.07\times10^{-3}$. For a description of the histogram bins and error bars, refer to the caption of Fig.~\ref{fig:dist}.}
\label{fig:comp}
\end{figure}

The distribution $\ov{f}$ obtained using Criterion C1 is plotted on logarithmic axes in Fig.~\ref{fig:dist} after averaging over 20 statistically independent realizations. The other criteria yield visually similar distributions. Deviations from the reference distribution may be visualized more clearly in the compensated distributions in Fig.~\ref{fig:comp}, which plots the same distribution premultiplied by the inverse of the idealized power-law scaling ($\ov{f}R^{10/3}$) on logarithmic--linear axes for the various criteria. These compensated distributions reveal that clipping (Criteria B1/B2) reduces the computed volume of small drops and influences the shape of the resulting size distribution. In addition, it appears that there is more sensitivity in the shape of the size distribution to the selected threshold in the case of clipping ($\phi_c$, Criteria B1/B2) as compared to the proposed criterion ($\phi_{c,m}$, Criteria C1/C2).

Two other diagnostics for the grouping criteria are the global packing fraction (volume fraction) and the number of identified drops. Table~\ref{tab:dropvoidnum} lists the global packing fraction for the aforediscussed criteria. Its accuracy is reduced by clipping (Criteria B1/B2), but is effectively maintained by the proposed criterion (Criteria C1/C2) since cells with small $\phi$ are not indiscriminately excluded. As remarked earlier, Fig.~\ref{fig:comp} suggests that volume accuracy is critical to the accuracy of the shape of the size distribution. Since the closest drops are spaced two grid cells apart, some of these drop pairs may be merged in the corresponding VoF field since the separation between the drops is not well resolved, thus decreasing the number of identified drops. This is in fact the onset of the corner case discussed at the end of \S~\ref{sec:ident}, and the minimum drop spacing is deliberately selected in this test case to highlight this issue. It is reiterated that the presence of this corner case does not necessarily signify a deficiency in the algorithm per se, but rather highlights the limitations of a discrete volume fraction field with finite numerical resolution. As alluded to by the case considered in the final row of Table~\ref{tab:criteria}, clipping (Criteria B1/B2) eliminates this merging. The proposed criterion also suppresses this merging with the choice of a moderate threshold ($\phi_{c,m} = 0.5$, Criterion C1), but an average of 1 merger in $10{,}000$ drops occurs in each ensemble realization with the choice of a more aggressive threshold ($\phi_{c,m} = 0.1$, Criterion C2). This difference in the number of mergers is reflective of the difference in approach taken by each of these criteria to deal with the limitations in the discrete volume fraction field introduced by finite numerical resolution.

The results of Fig.~\ref{fig:comp} and Table~\ref{tab:dropvoidnum} demonstrate that when the proposed Criterion C is employed, most global quantities of interest, such as the size distribution and the global packing fraction, are insensitive to the choice of $\phi_{c,m}$ between the values $\phi_{c,m}=0.1$ and $\phi_{c,m}=0.5$ in the absence of wisps. In other words, the choice of $\phi_{c,m}$ should not significantly influence these global quantities of interest within this threshold range on the basis of the accuracy of the algorithm. In light of the minimal sensitivity to this threshold in this test case, it is suggested that the proposed criterion generally be used with a moderate threshold, e.g., $\phi_{c,m}=0.5$, to achieve a balance between mass conservation, accuracy in the shape of the size distribution, and minimization of the merger of closely spaced dispersed-phase structures. In particular, the choice of a moderate threshold ameliorates the sensitivity of structure mergers to the threshold. In this case, the sensitivity was found to be weak and on the order of $0.01\%$. Such mergers could possibly be further minimized by applying conservative sharpening filters on the $\phi$ field if necessary. Note that the presence of structure mergers did not have an observable influence on the global quantities of interest. It is emphasized that the decision on whether to promote or suppress mergers in no way improves the inherent accuracy of the algorithm, and should instead be informed by whether these mergers are beneficial or detrimental to the eventual applications of interest, such as the computation of absorption and scattering coefficients, or subgrid-scale modeling.

\begin{table}
\begin{center}
\begin{tabular} { c || >{\centering\arraybackslash} m{0.28\textwidth} | >{\centering\arraybackslash} m{0.28\textwidth}}
Criterion & Global packing fraction & Error with reference to A\\
&\\
A & $(2.23\pm0.05) \times 10^{-2}$ & N/A\\
B1 & $(2.07\pm0.05) \times 10^{-2}$ & $1.6 \times10^{-3}$\\
B2 & $(2.21\pm0.05) \times 10^{-2}$ & $1.5 \times10^{-4}$\\
C1 & $(2.23\pm0.05) \times 10^{-2}$ & $3.2 \times10^{-6}$\\
C2 & $(2.23\pm0.05) \times 10^{-2}$ & $4.5 \times10^{-13}$\\
\end{tabular}
\end{center}
\caption{The global packing fraction identified by various criteria, averaged over the 20 realizations of the 10{,}000-drop system, and the absolute error in the mean fraction relative to the mean fraction of Criterion A. Refer to Table \ref{tab:criteria} for descriptions of the criteria. The uncertainties denote twice the standard error over the 20 realizations.}
\label{tab:dropvoidnum}
\end{table}

\section{Demonstration case for the identification algorithm}\label{sec:ident-demo}

The performance of the identification algorithm using the grouping criteria referenced in Table~\ref{tab:criteria} is now demonstrated on an energetic two-phase flow with large nonspherical bubbles, specifically the generation of bubbles in oceanic breaking waves. Large nonspherical bubbles near the wave surface complicate the computation of the bubble size distribution since they may have multiple points of approach or connection with the atmosphere. These approaches and connections may occur in tandem with spurious connections due to wisps. The key distinction between this demonstration case and the test cases of \S~\ref{sec:ident-test} is that no obvious ground truth is available for the bubble statistics in a breaking wave due to the presence of these approaches and connections, except for a theoretical power-law scaling expected in the bubble size distribution at intermediate sizes in the early wave-breaking stages~\citep{Garrett1,Chan7}. While a reference size distribution is not accessible in the demonstration case, the complexity of the flow provides a valuable opportunity to discriminate the size distributions resulting from the various grouping criteria, in order to inform the selection of the grouping criterion in general cases. The results from the test cases in \S~\ref{sec:ident-test} are crucial to provide this discrimination and meaningfully interpret the size distributions in the ensuing demonstration case.\\

\subsection{Bubble size distribution in a breaking wave}\label{sec:ident-test-wave}

\begin{figure}
  \centerline{
\includegraphics[width=0.75\linewidth]{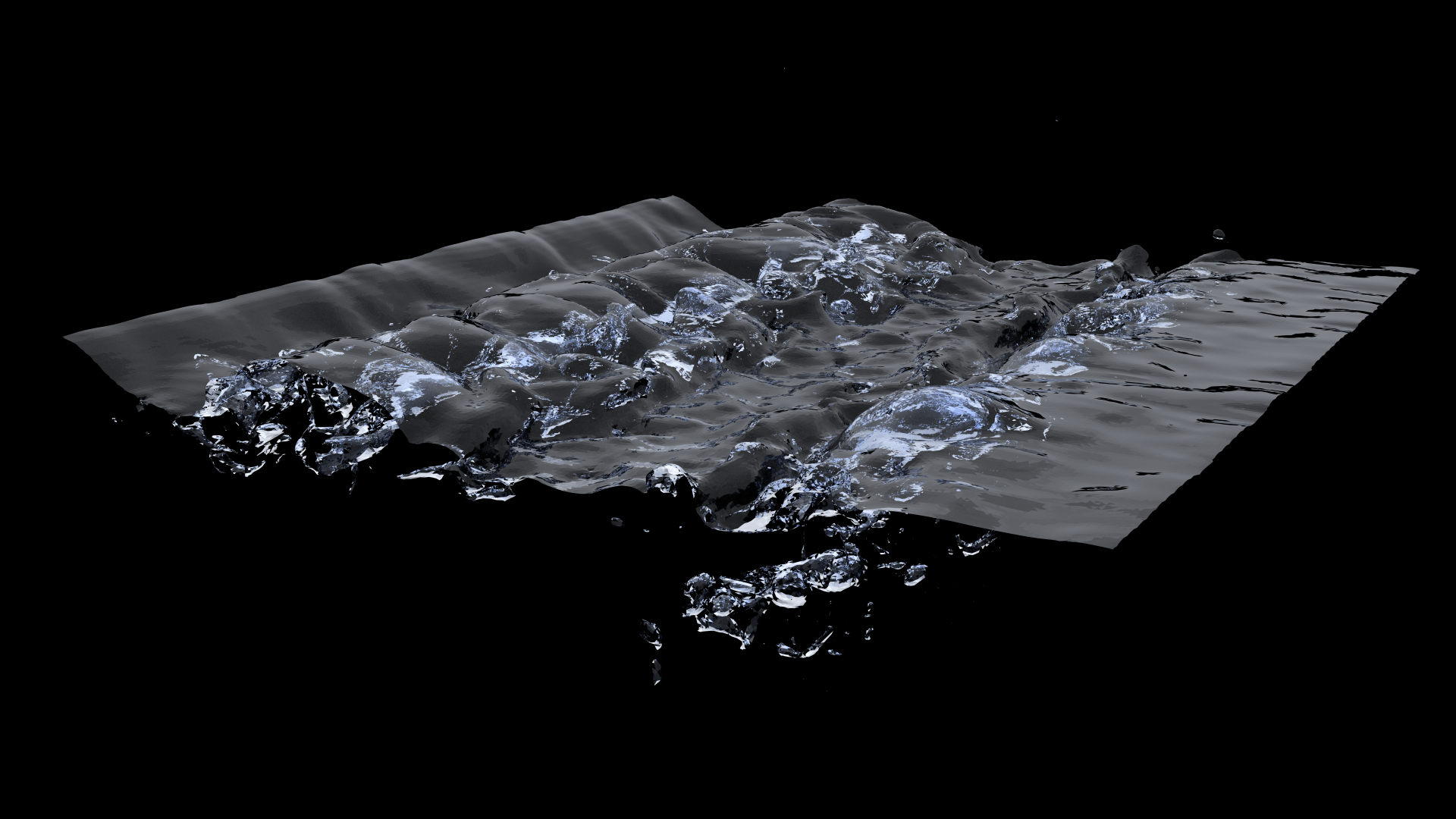}
}
  \caption{An axonometric rendering of the $\phi=0.5$ isosurface from a single realization of the breaking-wave simulation ensemble described in \S~\ref{sec:ident} and \S~\ref{sec:ident-test-wave}. The snapshot was obtained about 1.3 wave periods after the wave was initialized, and about 0.9 periods after it breaks.}
\label{fig:wave}
\end{figure}

\begin{figure}
  \centerline{
(a)
\includegraphics[width=0.42\linewidth,valign=t]{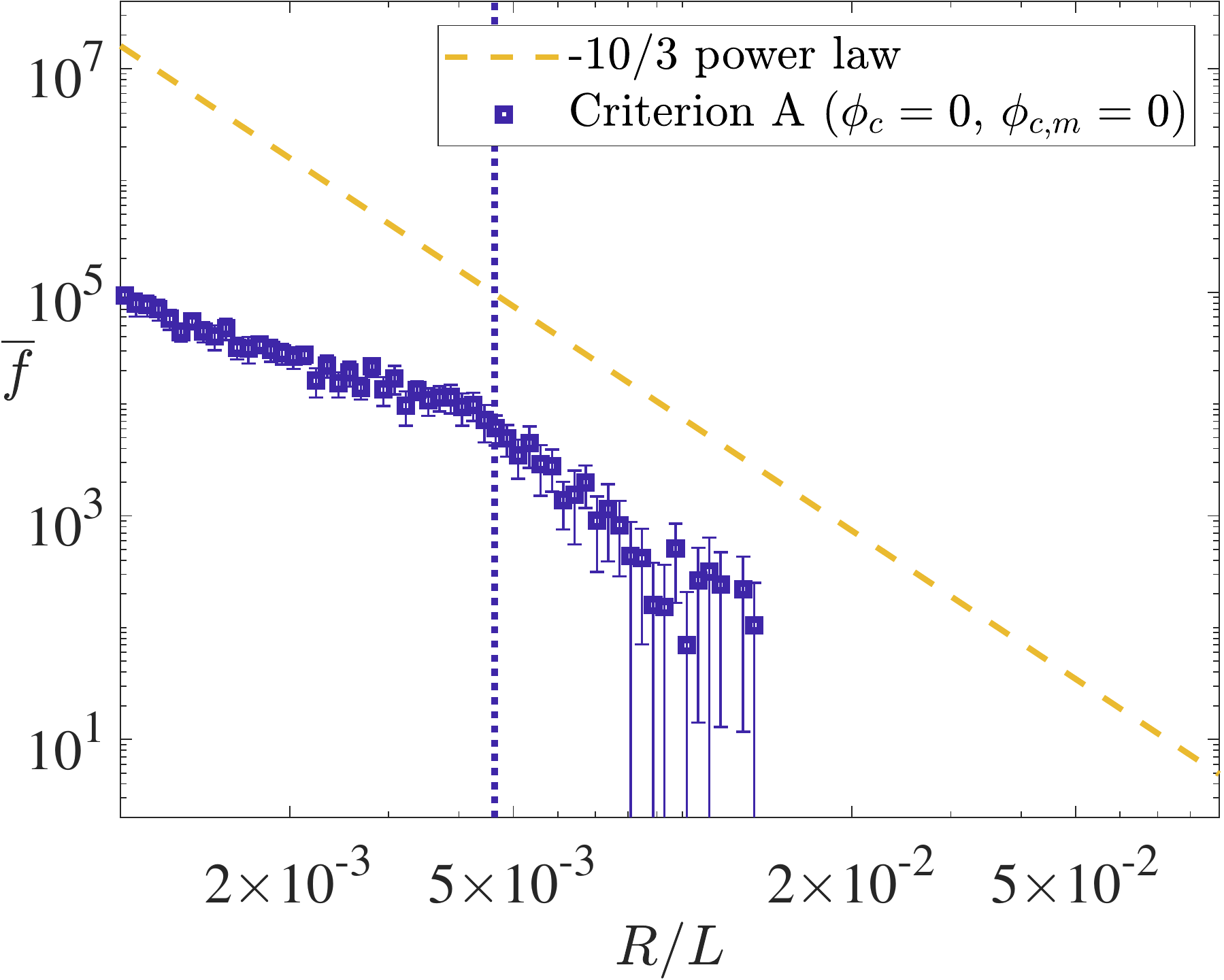}
}
  \centerline{
(b)
\includegraphics[width=0.42\linewidth,valign=t]{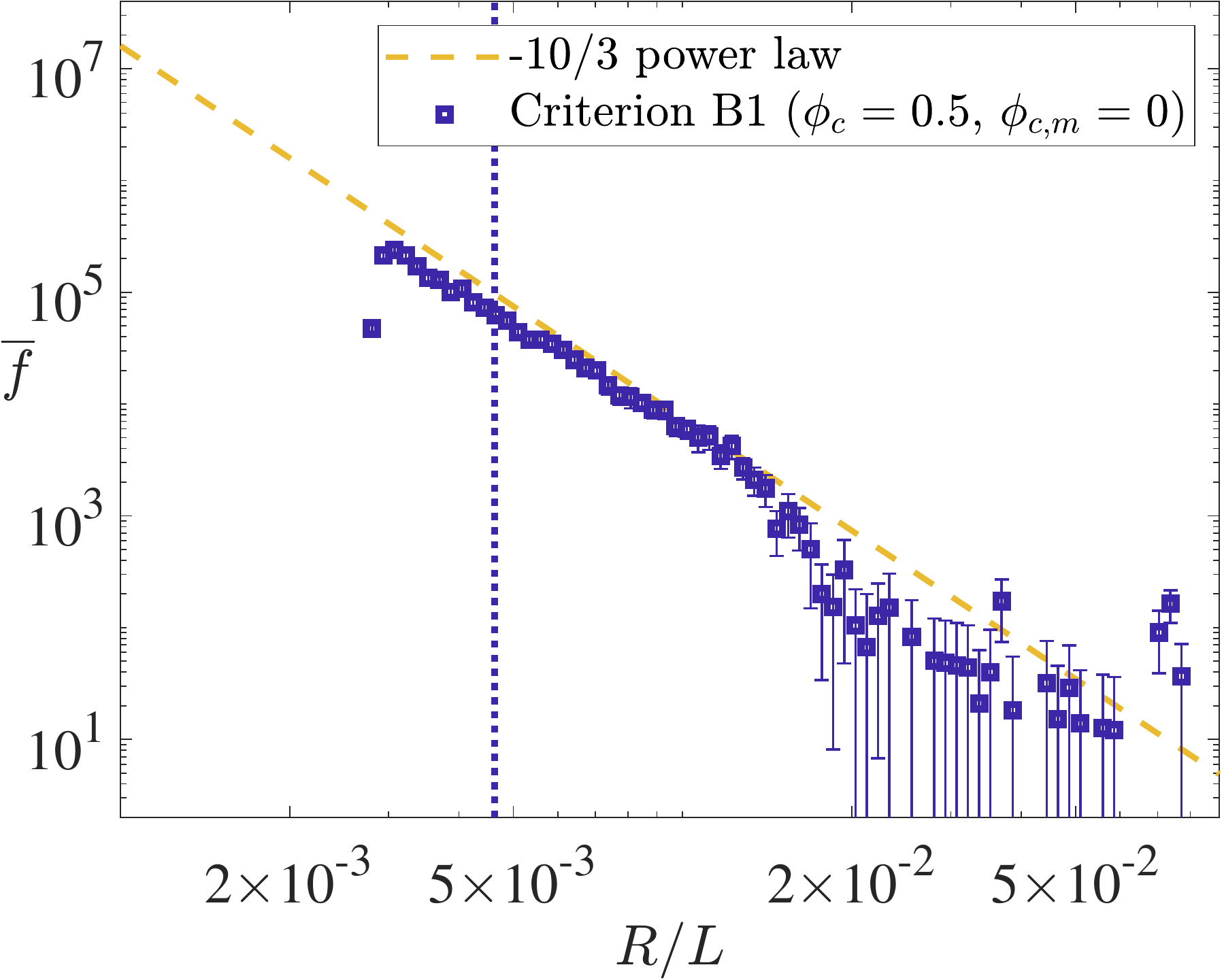}
\quad
(c)
\includegraphics[width=0.42\linewidth,valign=t]{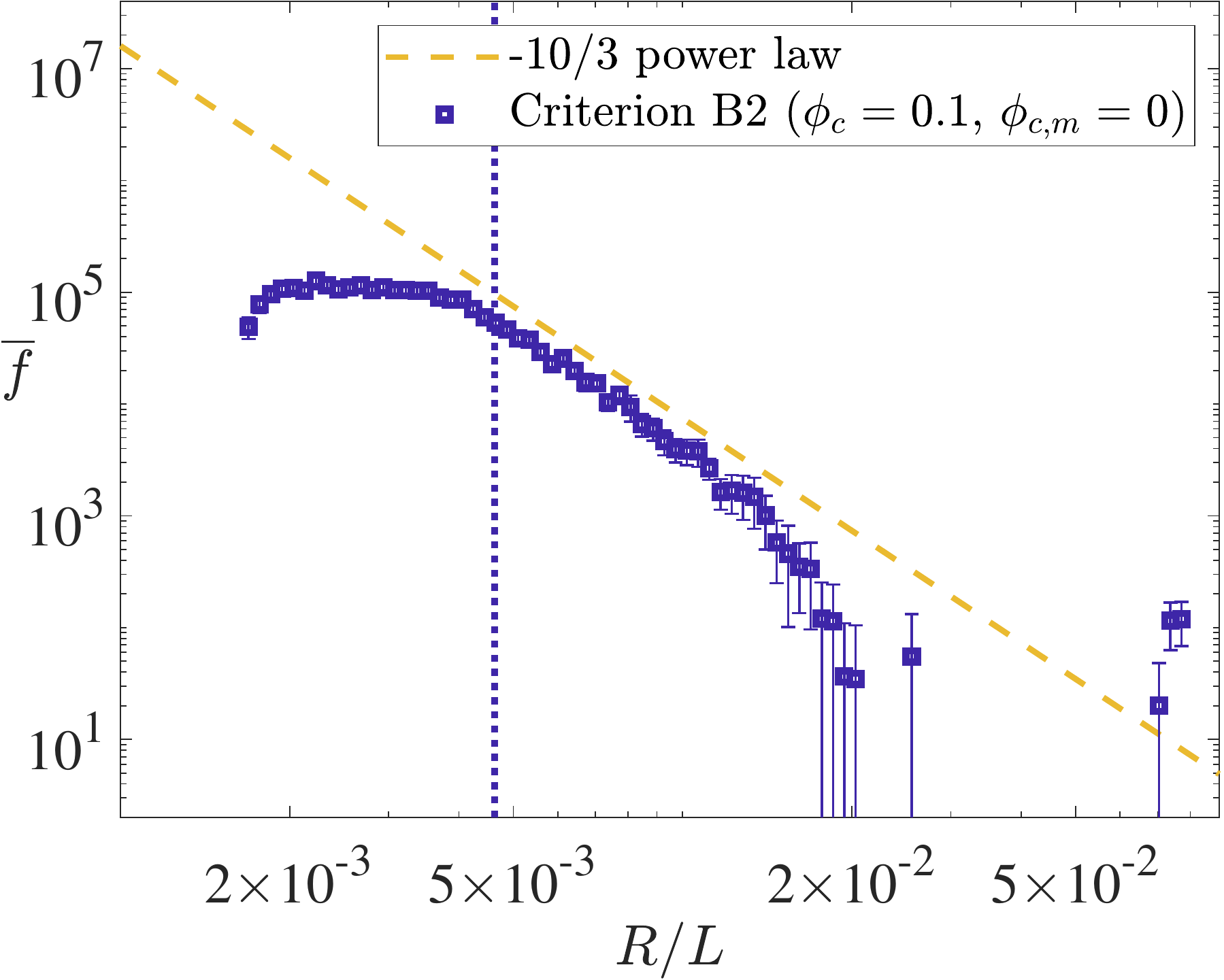}
}
  \centerline{
(d)
\includegraphics[width=0.42\linewidth,valign=t]{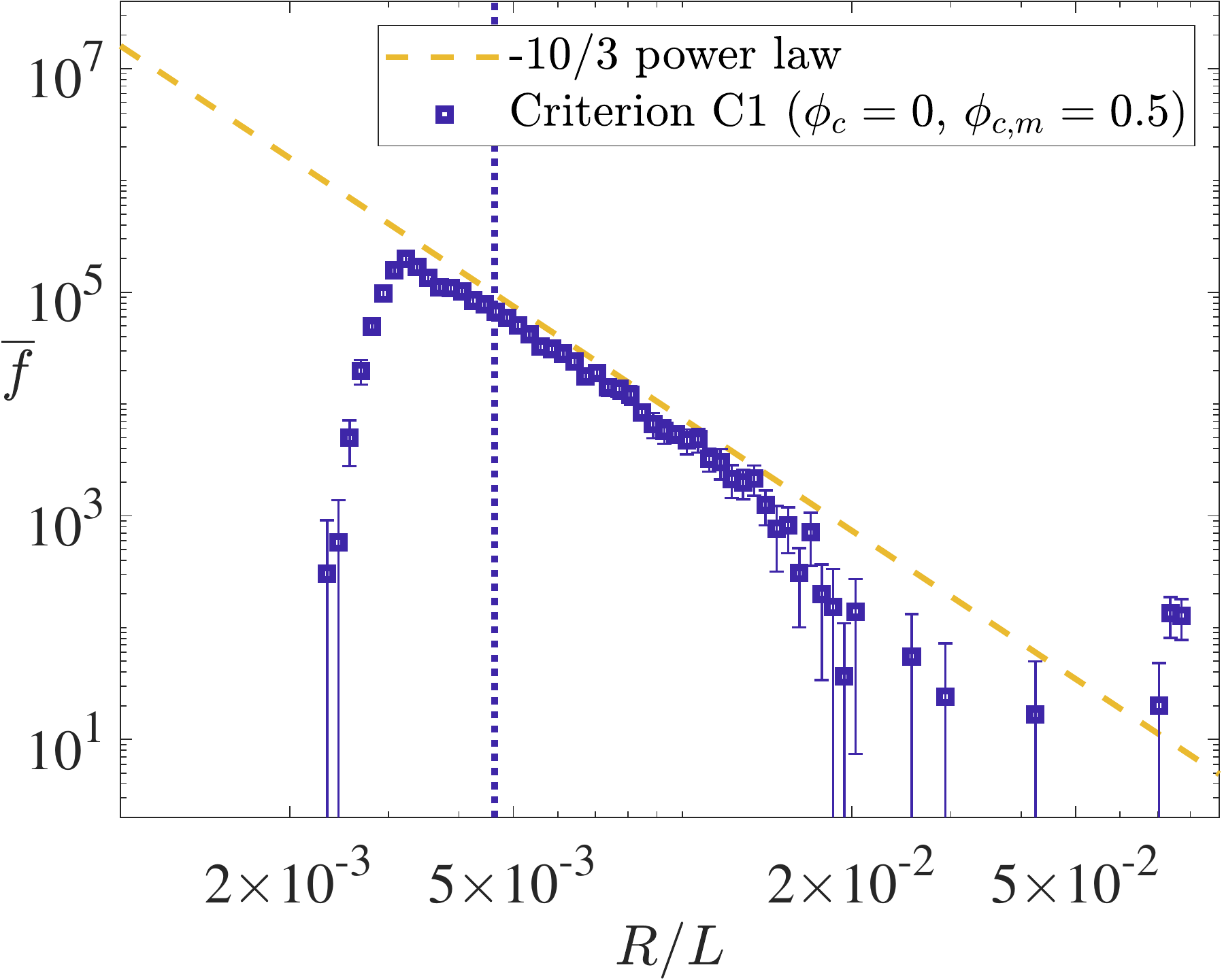}
\quad
(e)
\includegraphics[width=0.42\linewidth,valign=t]{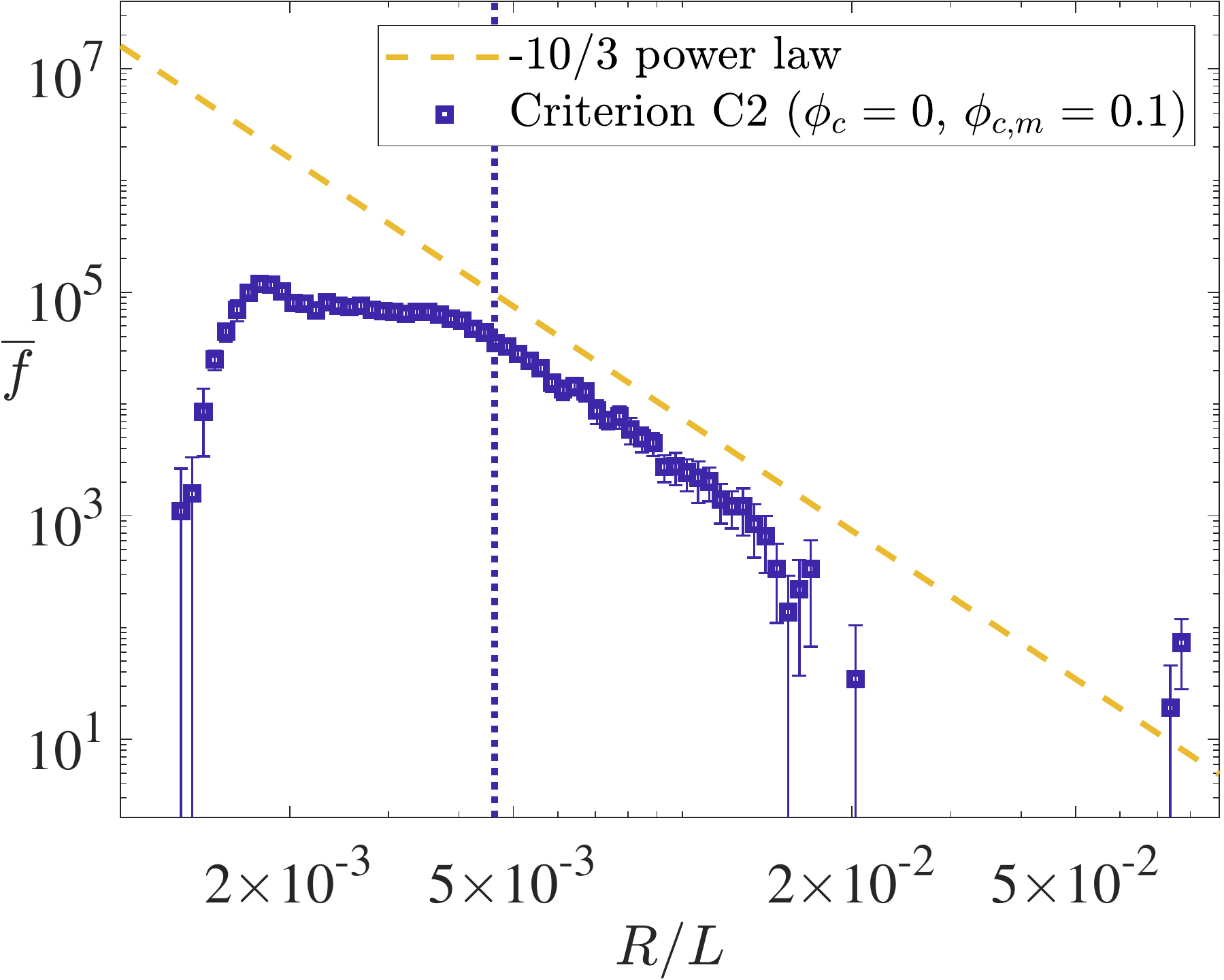}
}
  \caption{The bubble size distribution, $\ov{f}$, averaged over 30 statistically independent realizations of the breaking-wave simulation referenced in \S~\ref{sec:ident} and \S~\ref{sec:ident-test-wave}, using (a) Criterion A, (b) Criterion B1, (c) Criterion B2, (d) Criterion C1, and (e) Criterion C2. Refer to Table \ref{tab:criteria} for descriptions of the criteria. The distribution is normalized such that $\sum_j \ov{f}_j \Delta(R_j/L) = N_b$, where $N_b$ is the ensemble-averaged number of bubbles in the characteristic wave volume $L^3$. Thus, the magnitude of this distribution is indicative of the number of bubbles present in a particular size range. The distribution was computed using histogram bins of equal logarithmic spacing, where the smallest employed bins are larger than the radius errors \eqref{eqn:nondimvolerror} incurred by all the criteria considered here. The error bars denote twice the standard error over the 30 realizations. The distributions were computed for a flow snapshot occurring about 1 wave period after the wave was initialized. The dotted vertical line denotes the mesh resolution. The dashed sloped line exhibits a $R^{-10/3}$ power law, which is an idealized scaling expected in breaking waves~\citep{Garrett1,Chan7}.}
\label{fig:wavedist}
\end{figure}

In this demonstration case, bubble size distributions from an ensemble of breaking-wave simulations, performed using the solver referenced in \S~\ref{sec:ident-test-single}, are used to visualize what happens when the various grouping criteria are applied to a turbulent flow and wisps are present. Simulation parameters were discussed in \S~\ref{sec:ident} and are further detailed in Refs.~\citep{Chan3,Chan4,Chan5}. In particular, each ensemble realization is initialized using the interface and velocity field of a periodic third-order Stokes wave with a fundamental steepness of $0.55$ in a cubic domain whose side length is equal to the wavelength of the fundamental mode. The ensemble nominally resolves the Hinze scale, and mesh insensitivity in the bubble statistics was observed over a subrange of super-Hinze-scale bubble sizes (not shown here). Fig.~\ref{fig:wave} depicts a rendering of the wave surface from one of the ensemble realizations shortly after the wave has broken. Fig.~\ref{fig:wavedist} depicts the size distribution from another flow snapshot during the active air entrainment phase for various criteria. Table~\ref{tab:largestbubble} further lists the nondimensional equivalent diameter of the largest bubble in the system due to the various criteria. This bubble comprises the atmosphere above the wave surface and other subsurface air pockets connected to it, and is not included in the distributions in Fig.~\ref{fig:wavedist}. See the next paragraph for a discussion of this table.

The traditional criterion (Criterion A) yields a distribution of significantly lower magnitude compared to those from the other criteria, likely because wisps cause spurious connections of resolved bubbles. In particular, many bubbles reside near the convoluted wave surface, as evidenced in Fig.~\ref{fig:wave}, and reconnections with the atmosphere may spuriously occur through wisps, as suggested in the preamble of this section. As the criterion threshold for both clipping (Criterion B) and the proposed procedure (Criterion C) is increased, the magnitude of the size distribution is seen to correspondingly increase, and the equivalent diameter of the largest bubble is seen to decrease. Both observations suggest that connections to the atmosphere are reduced by Criteria B and C. Note that these observations may be attributed both to the removal of spurious connections with the atmosphere due to wisps, as well as the reduction of mergers between structures with a physically small separation, the latter of which was discussed at the end of \S~\ref{sec:ident} and \S~\ref{sec:ident-test-dist}. One may directly compare Table~\ref{tab:dropvoidnum}, which indicated that Criteria A and C yield global packing fractions that are identical to three significant figures for a wispless multi-drop system, and Table~\ref{tab:largestbubble}, which suggests that more subsurface air is connected to the atmosphere due to Criterion C2 than to Criterion C1. This comparison suggests that the difference in the magnitudes of the distributions of Criteria A and C is due to the removal of spurious connections between well-separated structures caused by wisps, while the corresponding difference between Criteria C1 and C2 is due to the suppression of mergers between closely spaced structures. Recalling the statement made at the end of \S~\ref{sec:ident} and reiterated in \S~\ref{sec:ident-test-dist}, the suppression or promotion of these mergers is not indicative of a more accurate algorithm, and merely reflects the choice of the algorithm in dealing with the finite numerical resolution of a discrete volume fraction field. Equivalent statements on the relative importance of the two mechanisms at play cannot be made with equal confidence for Criterion B, because Tables~\ref{tab:errtrans1a}--\ref{tab:dropvoidnum} and Fig.~\ref{fig:comp} demonstrated that errors in the bubble volumes and thus the size distribution are incurred by clipping. Fig.~\ref{fig:wavedist} does, however, suggest that Criterion B likely reduces connections more aggressively than Criterion C, owing to general differences in the size distribution magnitudes. In particular, the marked difference of the large-bubble distribution of Criterion B1 from the distributions of Criteria B2/C1/C2 suggests that Criterion B1 is especially aggressive at reducing connections and disconnecting moderately sized bubbles from the atmosphere. This reduction is also borne out in the marked decrease in the equivalent diameter of the largest bubble of Criterion B1 in Table~\ref{tab:largestbubble} relative to the other criteria, although the decrease may be attributed to both the reduction of connections and the reduction in bubble volumes due to clipping.

In addition, the traditional criterion identifies a significant number of subgrid bubbles, likely also as a result of connected wisps. The decrease in the number of subgrid bubbles with an increase in the criterion threshold for Criteria B and C supports this hypothesis. The traditional criterion also yields a distribution with large standard error relative to the distributions from the other criteria at a given bubble size. Taken together, these observations suggest that the traditional criterion incurs significant errors due to contamination from wisps and should generally be avoided.

One may show that clipping may also incur significant errors in the volumes of large, nonspherical bubbles with large surface-area-to-volume ratios due to its relatively large error coefficient $M$. In particular, the fractional volume error, $\Delta \mathcal{V}_\text{err}/(\pi D^3 / 6)$, may be expressed as a product of $M \Delta x$ and the dimensional surface-to-volume ratio. Thus, while Criteria B and C may yield visually similar size distributions notwithstanding differences from the mergers of closely spaced structures, Criterion C is generally recommended since it incurs demonstrably smaller errors in the quantities of interest than Criterion B and is quantitatively more reliable. As discussed at the end of \S~\ref{sec:ident-test-dist}, the results of Criterion C are not expected to be sensitive to the choice of $\phi_{c,m}$, except when the mergers of closely spaced structures result in phenomena of interest. Since these mergers are fundamentally a consequence of underresolved flow structures, they should strictly be mitigated by a higher mesh resolution in the original simulation.
 
\begin{table}
\begin{center}
\begin{tabular} { c || >{\centering\arraybackslash} m{0.35\textwidth}}
Criterion & Nondimensional equivalent diameter of largest bubble\\
&\\
A & $(9.8550\pm0.0001) \times 10^{-1}$\\
B1 & $(9.829\pm0.001) \times 10^{-1}$\\
B2 & $(9.841\pm0.001) \times 10^{-1}$\\
C1 & $(9.8395\pm0.0005) \times 10^{-1}$\\
C2 & $(9.850\pm0.002) \times 10^{-1}$\\
\end{tabular}
\end{center}
\caption{The equivalent diameter of the largest bubble, or the diameter of the sphere with an equivalent volume, as identified by various criteria, averaged over 30 statistically independent realizations of the breaking-wave simulation referenced in \S~\ref{sec:ident} and \S~\ref{sec:ident-test-wave}, and nondimensionalized by the fundamental wavelength of the initial waveform. Refer to Table \ref{tab:criteria} for descriptions of the criteria. The uncertainties denote twice the standard error over the 30 realizations. Note that this large bubble corresponds to the atmosphere above the wave surface, as well as any subsurface air pockets connected to it.}
\label{tab:largestbubble}
\end{table}

\section{Description of the tracking algorithm}\label{sec:track}

The identification algorithm in \S~\ref{sec:ident} yields the size distribution from a single simulation snapshot. As mentioned at the outset, 	important physical and modeling insights may be obtained by analyzing the evolution of the size distribution through breakup and coalescence statistics, including the breakup and coalescence frequencies, as well as the probability distributions of child and parent bubble and drop sizes in breakup and coalescence events, respectively. These statistics may be obtained using a tracking algorithm that records the history of every bubble and drop from its creation to its destruction, and traces the lineage of each bubble and drop. As in \S~\ref{sec:ident}, only drops will be referenced hereafter without loss of generality and for brevity. To construct these lineages, lists of drops with their sizes and locations are compared between successive simulation snapshots. The continuation of drops between these snapshots, as well as the occurrence of breakup and coalescence events, may be determined via a constraint based on mass conservation. For a VoF-based scheme, access to the conservative volume fraction field, $\phi$, allows reliance on the principle of mass conservation, obviating the need to rely on probabilistic techniques typically used in cell and particle tracking. It turns out that the construction of these lineages using consecutive time-steps may result in the detection of repeated confounding events involving slowly fragmenting and slowly coalescing drops (e.g., Ref.~\citep{Rubel1}), as discussed in \S~\ref{sec:intro}. In order for the algorithm to effectively post-process datasets where consecutive snapshots might not come from consecutive time-steps, this constraint will need to be supplemented with the knowledge that the simulation satisfies the Courant-Friedrichs-Lewy (CFL) condition, as well as a good understanding of the errors incurred by the identification algorithm that were discussed in \S~\ref{sec:ident-test}. These constraints will now be discussed.

\subsection{Constraints that should be satisfied during breakup and coalescence}\label{sec:track-cons}

As a drop is advected, its mass\textemdash and volume in an incompressible setting\textemdash remains constant between snapshots in the absence of phase change even if it deforms. Even if the drop breaks up into two, or if two drops coalesce into one, the principle of mass conservation is necessarily satisfied before and after the change in topology. Suppose drop \#0 breaks up into drops \#1 and \#2, or \#1 and \#2 coalesce to form \#0. Then, their volumes $\mathcal{V}_i$ exactly satisfy
\begin{equation}
\mathcal{V}_0 = \mathcal{V}_1 + \mathcal{V}_2.
\label{eqn:consmassC}
\end{equation}
The centroids $\bs{x}_i$ of the drops also instantaneously satisfy the following constraint
\begin{equation}
\bs{x}_0(t_E^\mp) \mathcal{V}_0 = \bs{x}_1(t_E^\pm) \mathcal{V}_1 + \bs{x}_2(t_E^\pm) \mathcal{V}_2
\label{eqn:centroidC}
\end{equation}
immediately before ($t_E^-$) and immediately after ($t_E^+$) the moment of breakup or coalescence $t_E$.

Now, consider a continuing drop, i.e., a drop that does not break up or coalesce between two snapshots $n$ and $n+1$. Because of the volume error $\Delta \mathcal{V}_\text{err}$ associated with the identification algorithm, the computed volume of this drop varies between snapshots. Denote the volume of a drop $i$ in snapshot $j$ as $\mathcal{V}_i^j$. Then, $\mathcal{V}_i^j$ satisfies
\begin{equation}
\left| \mathcal{V}_i^n - \mathcal{V}_i^{n+1} \right| < \Delta \mathcal{V}_\text{err}.
\label{eqn:consmassD}
\end{equation}
If the simulation satisfies the CFL condition, then each fluid--fluid interface cannot traverse more than a single cell width in a single time-step, multiplied by the maximum permissible Courant number for the advection scheme adopted in the simulation. It then follows that the drop centroid remains stationary insofar as the permissible distance error is the product of the local grid spacing $\Delta x$, the number of time-steps $N_t$ between the snapshots, and the Courant number $C$ adopted in the simulation. Denote the centroid of a drop $i$ in snapshot $j$ as $\bs{x}_i^j$. Then, $\bs{x}_i^j$ satisfies
\begin{equation}
\left|\left| \bs{x}_i^n - \bs{x}_i^{n+1} \right|\right| < C N_t \Delta x.
\label{eqn:centroidD}
\end{equation}
Note that the identification algorithm also generates spatial errors in the centroid, as discussed in \S~\ref{sec:ident-test-single}. However, these errors are typically much smaller than the grid spacing and will be neglected here. 

One may now use \eqref{eqn:consmassD} and \eqref{eqn:centroidD} to construct the discrete equivalents of \eqref{eqn:consmassC} and \eqref{eqn:centroidC}. Consider, first, the constraint arising from the principle of mass conservation. In the case of breakup, one may write
\begin{equation}
\left| \mathcal{V}_0^n - \left[ \mathcal{V}_1^{n+1} + \mathcal{V}_2^{n+1} \right] \right| < \Delta \mathcal{V}_\text{err},
\label{eqn:consmassDB}
\end{equation}
while in the case of coalescence, one may write
\begin{equation}
\left| \mathcal{V}_0^{n+1} - \left[ \mathcal{V}_1^n + \mathcal{V}_2^n \right] \right| < \Delta \mathcal{V}_\text{err}.
\label{eqn:consmassDC}
\end{equation}
Now, consider the constraint arising from the satisfaction of the CFL condition. In the absence of volume errors, one may write, for breakup, the following exact bound
\begin{equation}
\left|\left| \bs{x}_0^n - \f{ \bs{x}_1^{n+1} \mathcal{V}_1^{n+1} + \bs{x}_2^{n+1} \mathcal{V}_2^{n+1} }{\mathcal{V}_1^{n+1} + \mathcal{V}_2^{n+1}} \right|\right| < C N_t \Delta x,
\label{eqn:centroidDB}
\end{equation}
while in the case of coalescence, one may write
\begin{equation}
\left|\left| \bs{x}_0^{n+1} - \f{ \bs{x}_1^n \mathcal{V}_1^n + \bs{x}_2^n \mathcal{V}_2^n }{\mathcal{V}_1^n + \mathcal{V}_2^n} \right|\right| < C N_t \Delta x.
\label{eqn:centroidDC}
\end{equation}
These bounds are approximate when $\Delta \mathcal{V}_\text{err}$ in \eqref{eqn:consmassDB} and \eqref{eqn:consmassDC} are nonzero. The constraints \eqref{eqn:consmassD}--\eqref{eqn:centroidDC} are sufficient for the identification of breakup and coalescence events, as well as continuing drops, using the drop volumes and centroids. These constraints assume that all breakup and coalescence events are binary, i.e., a drop breaks up into a maximum of two children drops, and a coalescence event involves a maximum of two parent drops. It may be argued that with the exception of collections of breakup events involving the same parent drop where every event in a collection occurs at exactly the same physical time instant, ternary and polyadic breakup events are in principle series of two or more binary events happening in quick succession. An analogous argument holds for coalescence. These arguments imply that the aforementioned constraints are also relevant to complex scenarios, such as entrainment and degassing in breaking waves, which may respectively be decomposed into series of successive breakup and coalescence events. Note that \eqref{eqn:consmassDB} and \eqref{eqn:consmassDC} imply a critical size ratio $r$ above which events involving a small drop and a large drop cannot be distinguished from fluctuations in the large-drop volume between snapshots. This size ratio was discussed in \S~\ref{sec:ident-test-pair}. Choosing an identification scheme with a lower volume error increases this critical size ratio keeping the resolution of the smallest drop constant, allowing more relevant events to be captured accurately. The identification of these events will now be discussed.

\subsection{Identifying breakup and coalescence events}\label{sec:track-algo}

\begin{figure}
\begin{center}
\includegraphics[width=0.5\textwidth]{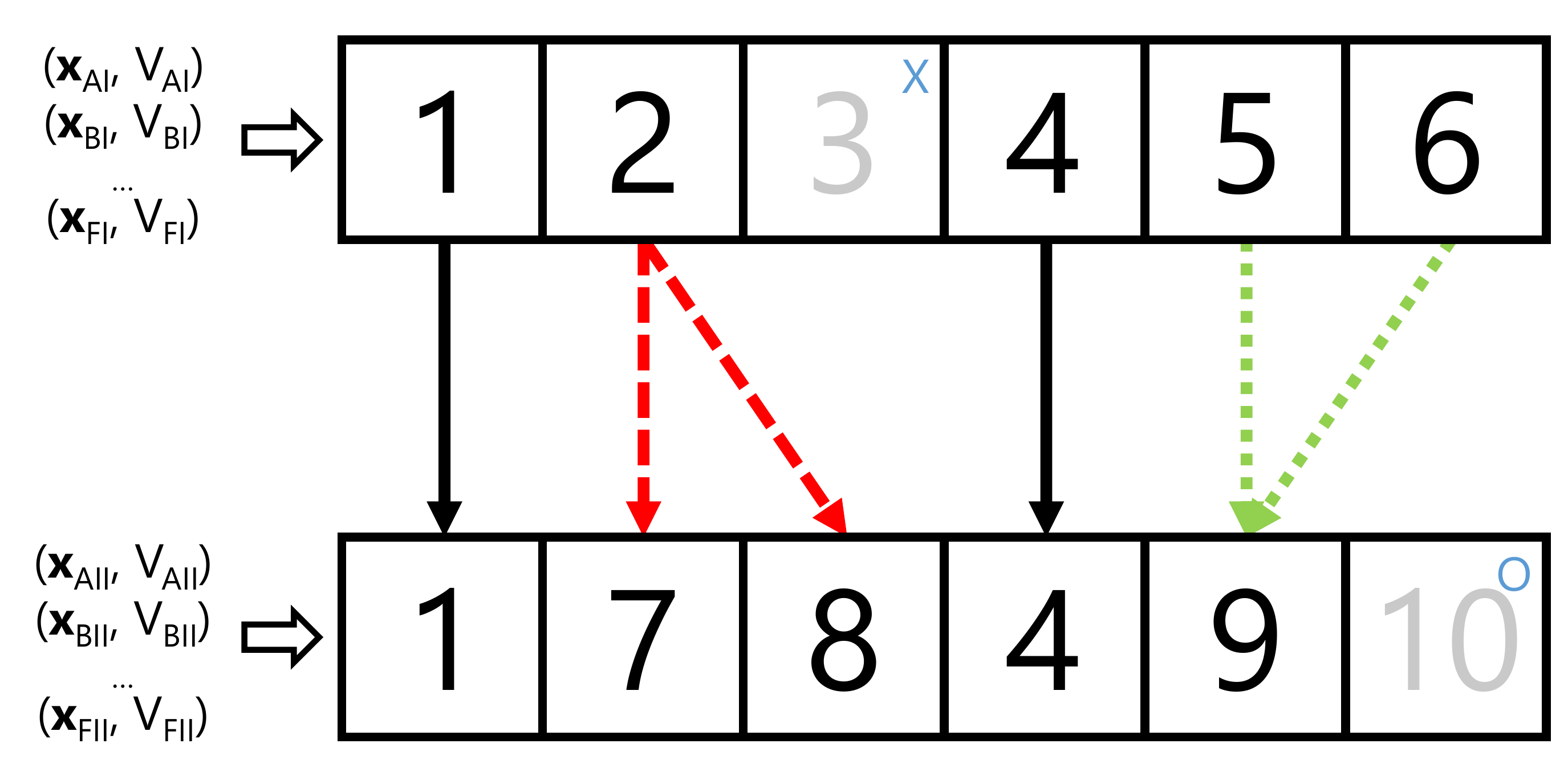}
\caption{An illustration of the tracking algorithm described in \S~\ref{sec:track-algo}. As depicted on the left, the algorithm requires two lists of drops: the list in the top row corresponds to the drops present in some flow snapshot (I), while the list in the bottom row corresponds to the drops in the succeeding snapshot (II). In this example, each list contains six drops (labeled A\textendash F) in both snapshots. No relation between the two sets of drops is assumed \emph{a priori}. The algorithm takes the centroid and volume of each drop as inputs. The two rows of cells on the right depict the same lists after the algorithm has been executed, and the numbers in each list entry correspond to the tags assigned to each drop. Suppose the drops in the first snapshot were assigned the tags \#1\textendash\#6. The solid arrows indicate that drops \#1 and \#4 are continuing. The dashed arrows indicate that drop \#2 split up into drops \#7 and \#8. The dotted arrows indicate that drops \#5 and \#6 coalesced to form drop \#9. Because drop \#3 could not be associated with a drop in the second list, its disappearance should be associated with a death event ($\times$), and because drop \#10 could not be associated with a drop in the first list, its formation should be associated with a birth event ($\circ$).
\label{fig:breakcoalealgo}}
\end{center}
\end{figure}

In order to identify breakup and coalescence events, comparisons are performed between lists of drop volumes and centroids from successive snapshots separated by a time interval $\Delta_{t,s} = N_t \Delta t$, where $\Delta t$ is the simulation time-step. This process is repeated from the first to the penultimate snapshot to obtain a family tree of drops spanning the flow evolution. First, the two lists are traversed in order to locate continuing drops. Continuing drops are identified through the satisfaction of \eqref{eqn:consmassD} and \eqref{eqn:centroidD}. This search for continuing drops incurs a computational cost $O\left(N_d^2\right)$, where $N_d$ is the higher number of drops in both lists. Then, the remaining noncontinuing drops are tested for breakup and coalescence events using the constraints \eqref{eqn:consmassDB}\textendash\eqref{eqn:centroidDC}. This search among noncontinuing drops incurs a computational cost $O\left(n_d^3\right)$, where $n_d \lesssim N_d$ is the higher number of noncontinuing drops in both lists. To avoid excessive computational costs per pair of snapshots, a sufficiently small $N_t$ should be chosen so that $n_d \ll N_d$. This is typically satisfied if the snapshot interval, $\Delta_{t,s}$, is less than the characteristic mean breakup and coalescence times of most of the drops. Finally, in order to account for the edge case of large-size-ratio breakup and coalescence events involving one large drop and one small drop, where the large drop may have been identified as continuing in the first traversal, the lists are looped over one final time. These events are identified using the constraints \eqref{eqn:consmassDB}\textendash\eqref{eqn:centroidDC} under the following restrictions: either the large drop is identified as drops \#0 and \#2 and the small drop is identified as drop \#1, or the large drop is identified as drops \#0 and \#1 and the small drop is identified as drop \#2. If such an event is determined to have occurred, then the large drop is no longer classified as continuing, and the three participating drops are reclassified as being part of a breakup or coalescence event.  This search, which requires one loop over the set of continuing drops and one loop over the set of noncontinuing drops, incurs a computational cost $O(N_d n_d)$. All remaining drops are then associated with birth (new drop appears) or death (existing drop disappears) events. It is emphasized that the tracking algorithm takes lists of drop volumes and centroids as inputs, and does not require direct knowledge of the underlying flow conditions. As such, the performance of the algorithm is not directly parameterized by material parameters of the constituent fluids.

One has to be more precise about the definitions of birth and death, since a breakup event also produces a new drop, and a coalescence event also consumes an existing drop. Here, a birth event is defined as the production of a drop independent of the presence of any existing identified drops, and a death event is correspondingly defined as the destruction of an existing drop without subsequent influence on any of the other identified drops. Such events may physically occur if there is a large reservoir of the dispersed phase with which individual structures may interact. For example, the entrainment of a bubble from the atmosphere by a breaking wave in the ocean, as well as the separation of a drop from the core jet through the rupture of a ligament during atomization in a combustor, may be associated with birth events. The popping of a bubble at the ocean surface and the reconnection of a stray drop with the core jet may correspondingly be associated with death events. Physical birth and death events may also occur when there are inlet and outlet boundaries in the computational domain allowing drops and bubbles to physically enter and leave the simulated system. Notwithstanding the presence of such a reservoir, or such inlet and outlet boundaries, if the total volume of all identified dispersed-phase structures is supposed to remain constant, then all detected birth and death events become reflective of errors in the tracking algorithm, which may be reduced by mesh refinement in tandem with a more accurate identification algorithm, and/or a more appropriate choice of the snapshot interval $\Delta_{t,s}$ given a fixed mesh resolution. Considerations in the selection of this interval are discussed in the context of the test cases in \S~\ref{sec:track-test}. In particular, the test case of drops with breakup timers in \S~\ref{sec:track-test-timer} addresses the intertwined roles of the snapshot interval and the mesh resolution. The tracking algorithm is illustrated in Fig. \ref{fig:breakcoalealgo} and summarized in Algorithm~\ref{alg:track}. Note that the birth and death events yielded at the end of each pass of the algorithm may be directly used to gain further insights into the dispersed-phase dynamics, if these events are ascertained to be physical and the aforementioned errors have been minimized. In particular, the final search loop of the algorithm is designed to maximize the probability that the resulting birth and death events are physical. For example, in a breaking wave, birth events near the wave surface can be attributed to entrainment, while death events near the wave surface can be attributed to degassing. This attribution can be performed through a geometrical test that references the interfacial boundary computational nodes or cells of the air mass constituting the atmosphere, and compares the distance of the centroid or boundary nodes/cells of the bubble in question to the boundary nodes/cells of the atmosphere. A similar procedure may be used to attribute birth events near an inlet boundary to entering structures and death events near an outlet boundary to leaving structures.

\begin{algorithm}
\SetAlgoLined
\KwResult{Bubble and drop lineages spanning the flow evolution over $N_s$ snapshots}
\For{$i_s\gets1$ \KwTo $N_s-1$ \KwBy $1$}{
 Prepare lists of bubbles and drops from snapshots $i_s$ and $i_s + 1$ using the algorithm in \S~\ref{sec:ident}\;
 \eIf{$i_s > 1$}{
	Retrieve bubble and drop tags for snapshot $i_s$\;
 }{
	Generate bubble and drop tags for snapshot $i_s$\;
 }
 Locate and tag continuing bubbles and drops using \eqref{eqn:consmassD} and \eqref{eqn:centroidD}\;
 Locate and tag breakup events using \eqref{eqn:consmassDB} and \eqref{eqn:centroidDB}\;
 Locate and tag coalescence events using \eqref{eqn:consmassDC} and \eqref{eqn:centroidDC}\;
 \For{continuing bubbles and drops $i$}{
  \For{remaining bubbles and drops $j$ not yet associated with breakup, coalescence, or continuation}{
   \If{bubbles or drops $i$ and $j$ satisfy the breakup conditions \eqref{eqn:consmassDB} and \eqref{eqn:centroidDB}}{
	Remove bubble or drop $i$ from the list of continuing bubbles or drops\;
	Add bubbles or drops $i$ and $j$ to the list of breakup events\;
   }
   \If{bubbles or drops $i$ and $j$ satisfy the coalescence conditions \eqref{eqn:consmassDC} and \eqref{eqn:centroidDC}}{
	Remove bubble or drop $i$ from the list of continuing bubbles or drops\;
	Add bubbles or drops $i$ and $j$ to the list of coalescence events\;
   }
  }
 }
 Associate remaining bubbles and drops with birth or death events\;
 Save bubble and drop tags for snapshot $i_s + 1$\;
}
 \caption{The bubble and drop tracking algorithm}
 \label{alg:track}
\end{algorithm}

\section{Test and demonstration cases of the tracking algorithm}\label{sec:track-test}

The performance of the tracking algorithm described in \S~\ref{sec:track} is now illustrated using a number of test and demonstration cases of varying complexities. In \S~\ref{sec:track-test-TG}, the breakup of a drop in a Taylor-Green vortex is used to introduce the concept of lineage construction, and to highlight that the ratio of the snapshot interval to the physical breakup/coalescence time is an important parameter at play. In \S~\ref{sec:track-test-glance}, the glancing approach of two drops is used to illustrate the presence of repeated confounding breakup/coalescence events that occur when this ratio of the snapshot interval to the physical breakup/coalescence time is exceedingly small. In \S~\ref{sec:track-test-timer}, the performance of the algorithm is demonstrated in a complex scenario with $O(10^3)$ drops and a scale separation of $O(10^2)$, under the assumptions that the algorithm has the ability to construct the necessary drop lineages, and that the ratio of the snapshot interval to the physical breakup time may be quantified and used to select an appropriate snapshot interval.

\subsection{Drop breakup in a Taylor-Green vortex}\label{sec:track-test-TG}

\begin{figure}
\begin{center}
  \centerline{
(a)
\includegraphics[width=0.425\linewidth,valign=t]{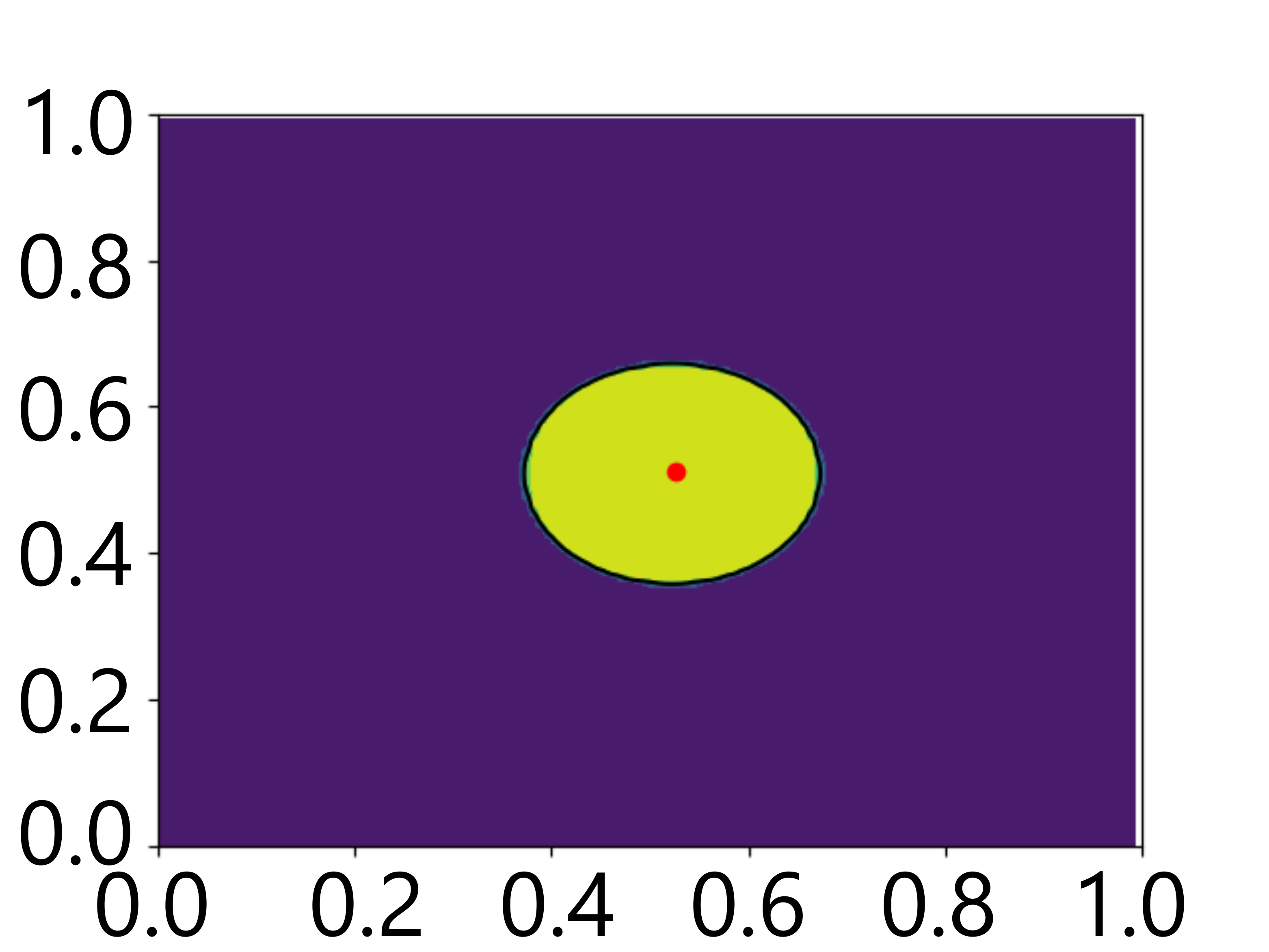}
\quad
(b)
\includegraphics[width=0.425\linewidth,valign=t]{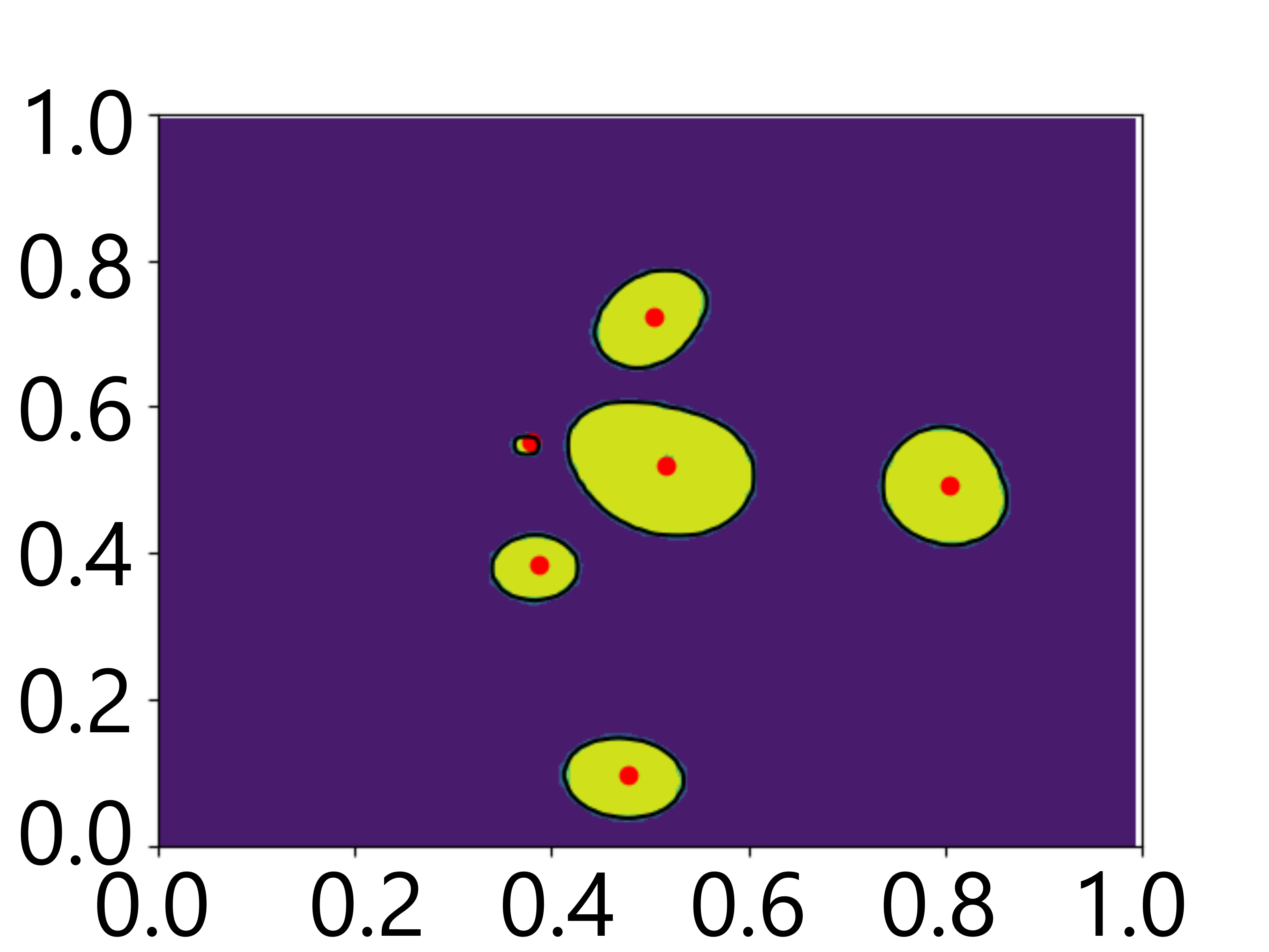}
}
\caption{Snapshots of drop breakup in a two-dimensional Taylor-Green vortex, (a) before and (b) after breakup. The contrasting colors denote different phases, and the small dots indicate the centroids of each drop. The snapshots are separated by $13.2$ characteristic breakup times, where the characteristic breakup time, $t_b$, is the ratio of the original drop radius to the initial maximum speed of the vortex.}
\label{fig:dropTGstill}
\end{center}
\end{figure}

\begin{figure}
\begin{center}
  \centerline{
(a)
\includegraphics[width=0.425\linewidth,valign=t]{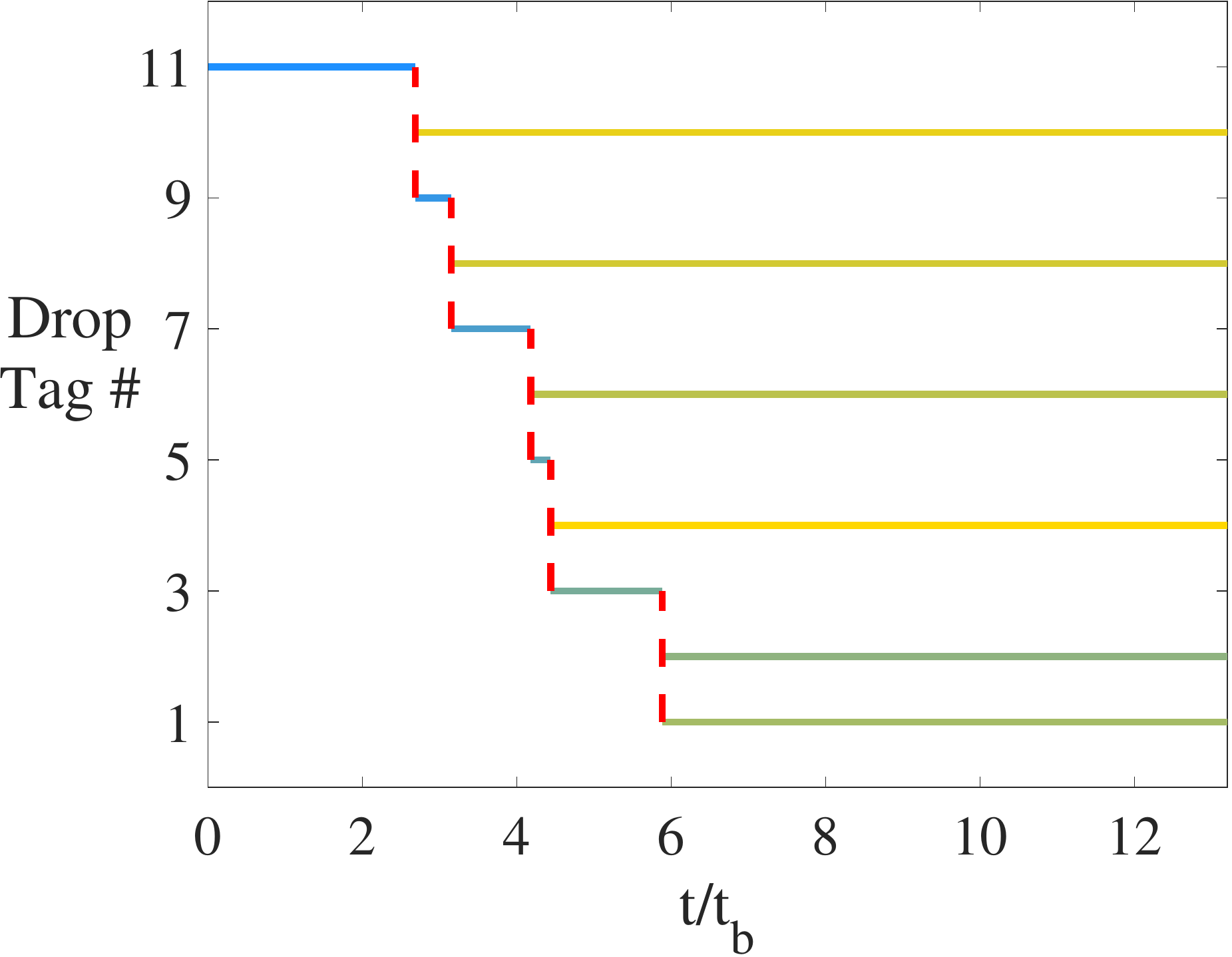}
\quad
(b)
\includegraphics[width=0.425\linewidth,valign=t]{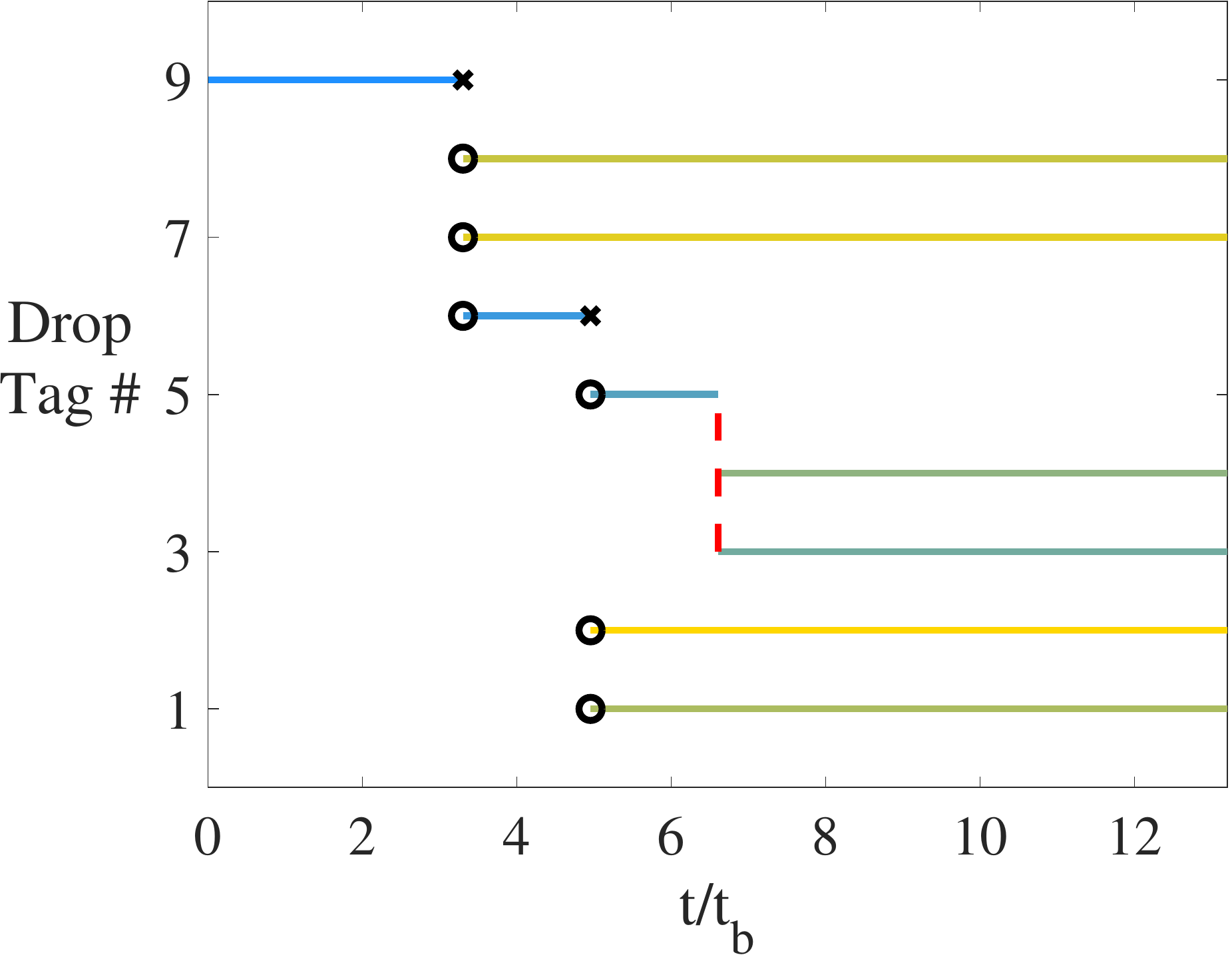}
}
\caption{Family trees of drops constructed by the algorithm in \S~\ref{sec:track-algo} for the breakup depicted in Fig. \ref{fig:dropTGstill} where the nondimensional time interval between snapshots, $\tau$, is (a) $5.2\times10^{-2}$ and (b) $8.3\times10^{-1}$. The drop tags are arbitrary, and time is nondimensionalized by the characteristic breakup time, $t_b$, defined in the main text and the caption of Fig.~\ref{fig:dropTGstill}. Each horizontal line depicts a drop, and adopts a color between blue (largest drop, darkest in grayscale) and yellow (smallest drop, lightest in grayscale). The colors of these lines are not essential to the interpretation of the trees. The dashed vertical lines depict breakup, similar to the usage in Fig.~\ref{fig:breakcoalealgo}, where the parent drop is to the left and the children drops are to the right of each vertical line. The circles denote birth events and the crosses denote death events, also consistent with the symbols introduced in Fig. \ref{fig:breakcoalealgo}.}
\label{fig:dropTGstats}
\end{center}
\end{figure}

\begin{figure}
\begin{center}
  \centerline{
(a)
\includegraphics[width=0.425\linewidth,valign=t]{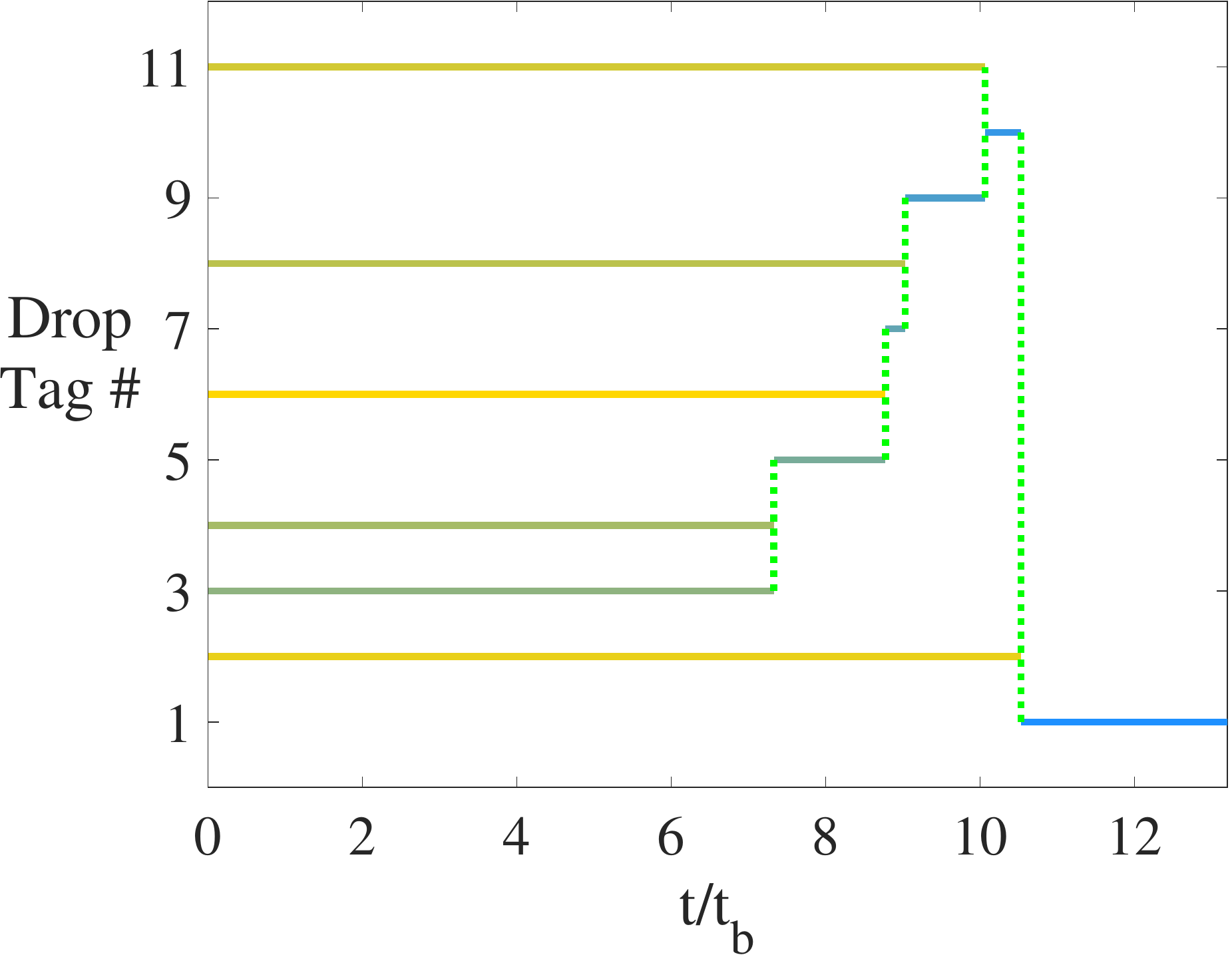}
\quad
(b)
\includegraphics[width=0.425\linewidth,valign=t]{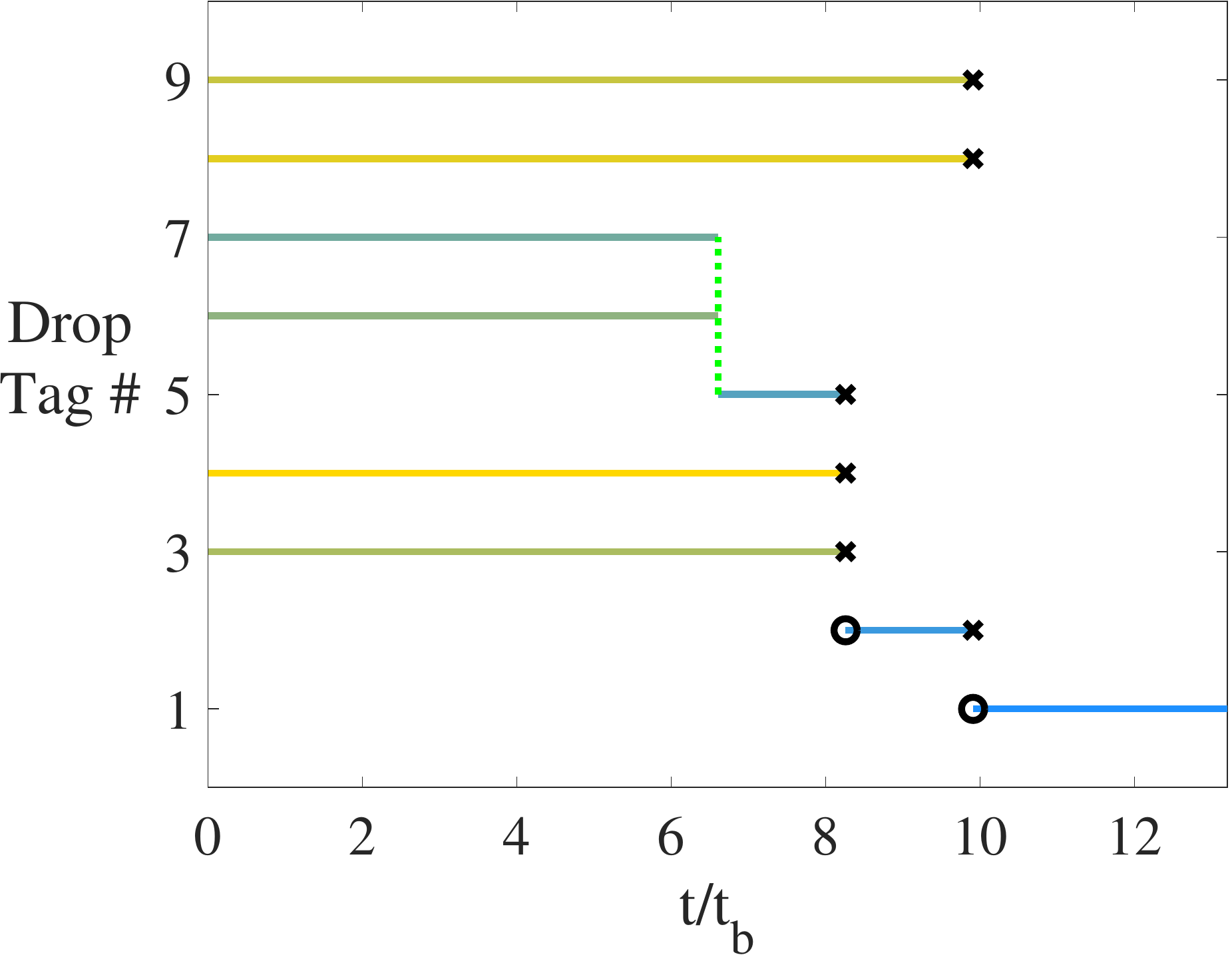}
}
\caption{Family trees of drops constructed by the algorithm in \S~\ref{sec:track-algo} by reversing the order of the snapshots considered in Fig.~\ref{fig:dropTGstats}. For a description of the snapshot intervals, drop tags, nondimensionalization of time, horizontal lines, and other symbols, refer to the caption of Fig.~\ref{fig:dropTGstats}. The dotted vertical lines depict coalescence, similar to the usage in Fig.~\ref{fig:breakcoalealgo}, where the parent drops are to the left and the child drop is to the right of each vertical line.}
\label{fig:dropTGreverse}
\end{center}
\end{figure}

In this test case, the tracking algorithm presented in \S~\ref{sec:track} is applied to the breakup of a drop in a two-dimensional freely decaying Taylor-Green vortex with $\nu_d/\nu_c = 5$ and $\rho_d/\rho_c = 5$, where $\nu$ and $\rho$ respectively denote the kinematic viscosity and density, and $d$ and $c$ respectively denote the dispersed and carrier phases. The flow was simulated using the two-phase VoF-based solver described in Refs.~\citep{Baraldi1,Dodd1}. The drop Reynolds and Weber numbers based on the initial drop diameter $D$ and carrier fluid conditions are $\RR_D = 300$ and $\We_D = 20$, respectively. It is reiterated that the tracking algorithm does not require direct knowledge of the flow conditions. As such, the performance of the algorithm is not directly parameterized by specific material parameters, and the dimensionless parameters above are provided only for reference. Fig.~\ref{fig:dropTGstill} depicts snapshots of the drop in the vortex before and after breakup, while Fig.~\ref{fig:dropTGstats} shows the breakup events detected using different dimensionless time intervals between snapshots, $\tau = \Delta_{t,s}/t_b$. These snapshot intervals are nondimensionalized by the characteristic breakup time, $t_b$, which is defined here as the ratio of the original drop radius to the initial maximum speed of the vortex. In the case of the shorter snapshot interval, all the breakup events are recovered; in the case of the longer snapshot interval, some breakup events are missed since more than one breakup event involving the same drop occurs in some of the intervals between snapshots. Note that the remaining event is captured in an unaltered fashion, indicating that a longer snapshot interval only misses events and does not distort the nature of the remaining detectable events. Fig.~\ref{fig:dropTGreverse} shows the coalescence events detected by running the algorithm on the snapshots in reverse order for the same time intervals. Analogous events are detected in the reverse sequences. These results underscore the need for the snapshot interval to resolve the mean breakup and coalescence times of interest, i.e., for $\tau$ to be sufficiently small. In particular, $\tau$ should be no more than $O(10^{-1})$.

\subsection{Glancing approach of two drops}\label{sec:track-test-glance}

\begin{figure}
\begin{center}
  \centerline{
(a)
\includegraphics[width=0.425\linewidth,valign=t]{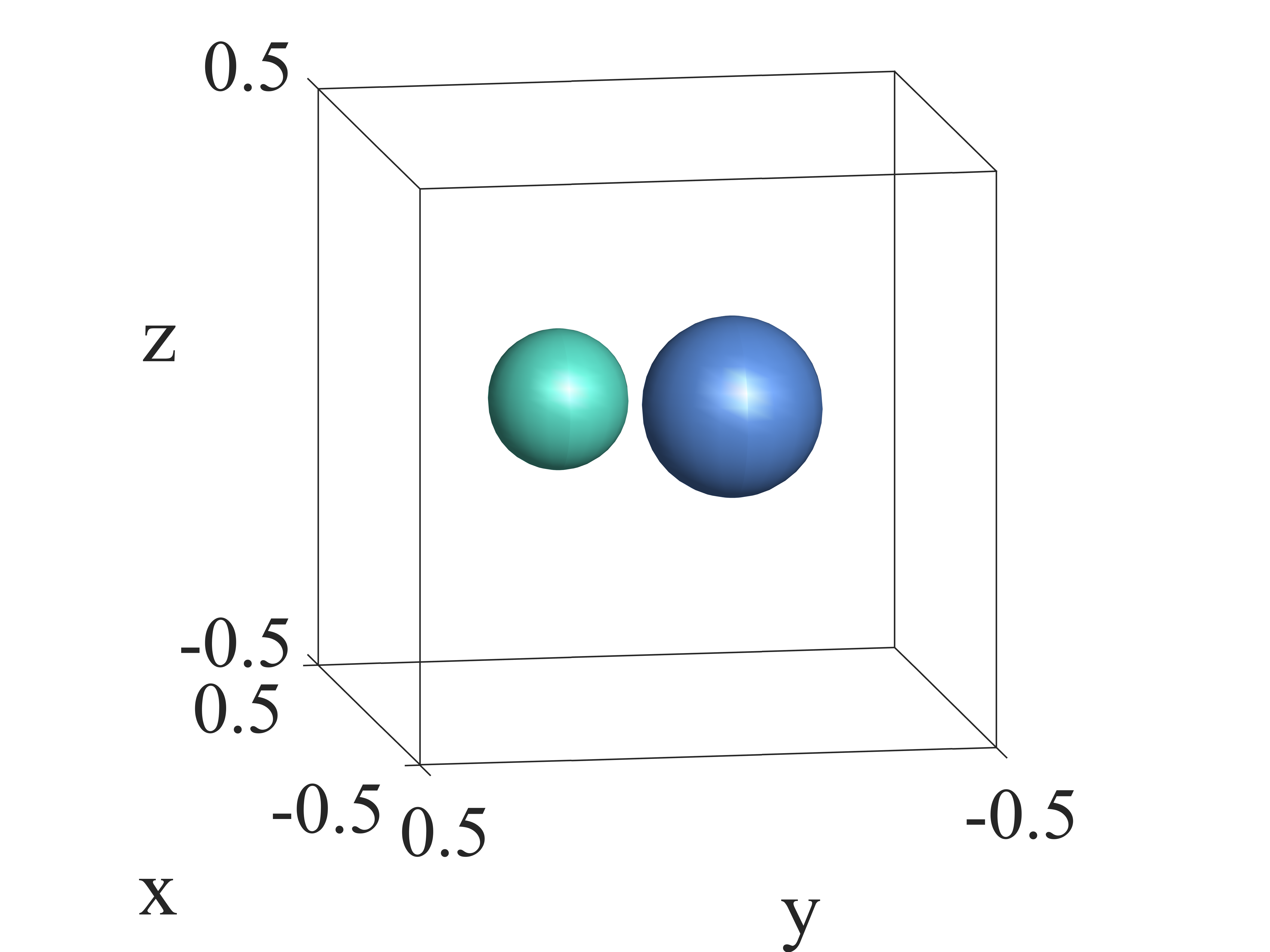}
\quad
(b)
\includegraphics[width=0.425\linewidth,valign=t]{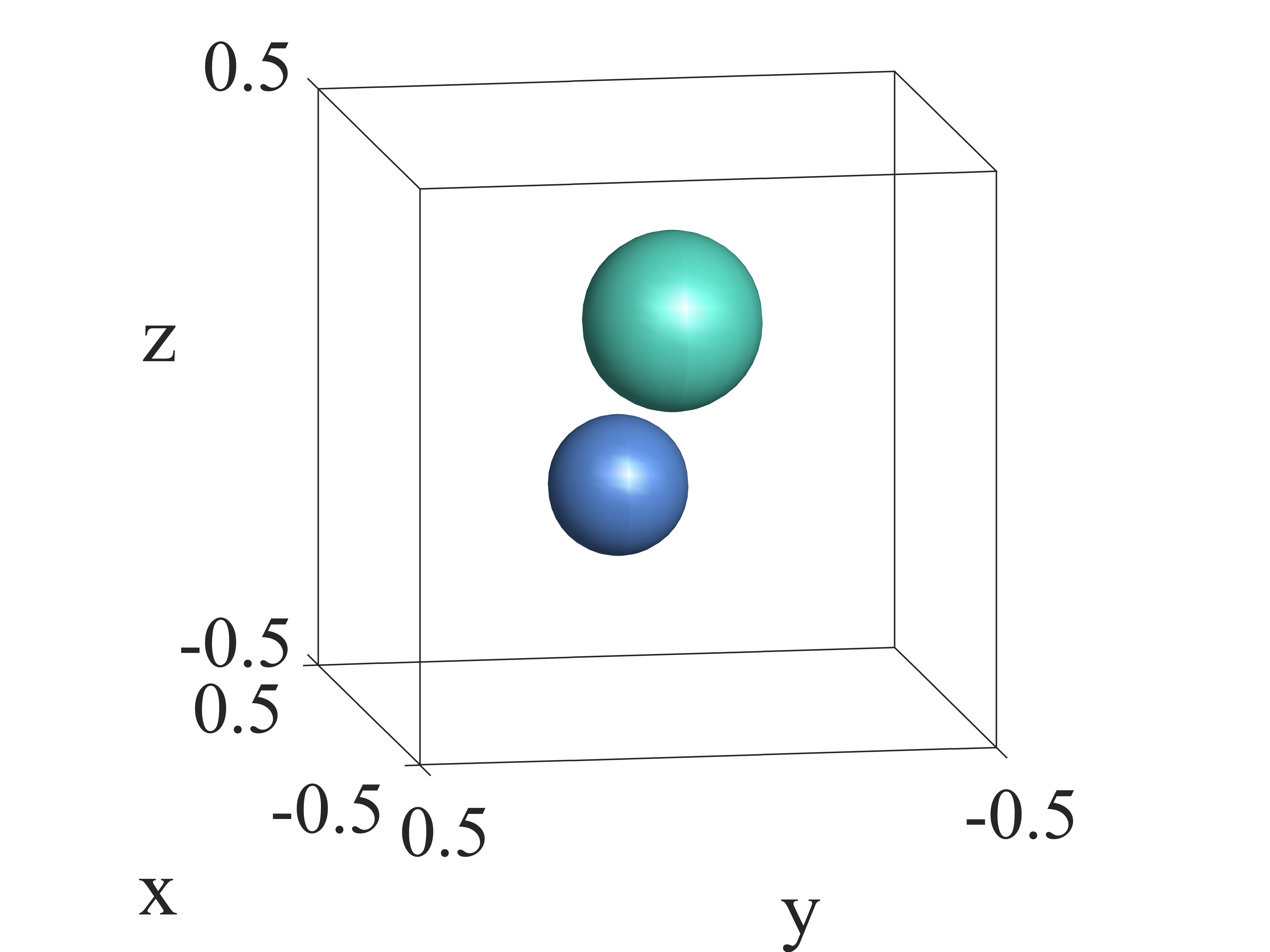}
}
\caption{Snapshots of the glancing approach of two drops (a) before and (b) after the approach. The drop on the right in (a), which is also the drop at the bottom in (b), is larger. The snapshots are separated by $0.95$ characteristic interaction times, where the characteristic interaction time, $t_b$, is the ratio of the average drop radius to the average drop speed.}
\label{fig:dropsglance}
\end{center}
\end{figure}

\begin{figure}
\begin{center}
  \centerline{
(a)
\includegraphics[width=0.425\linewidth,valign=t]{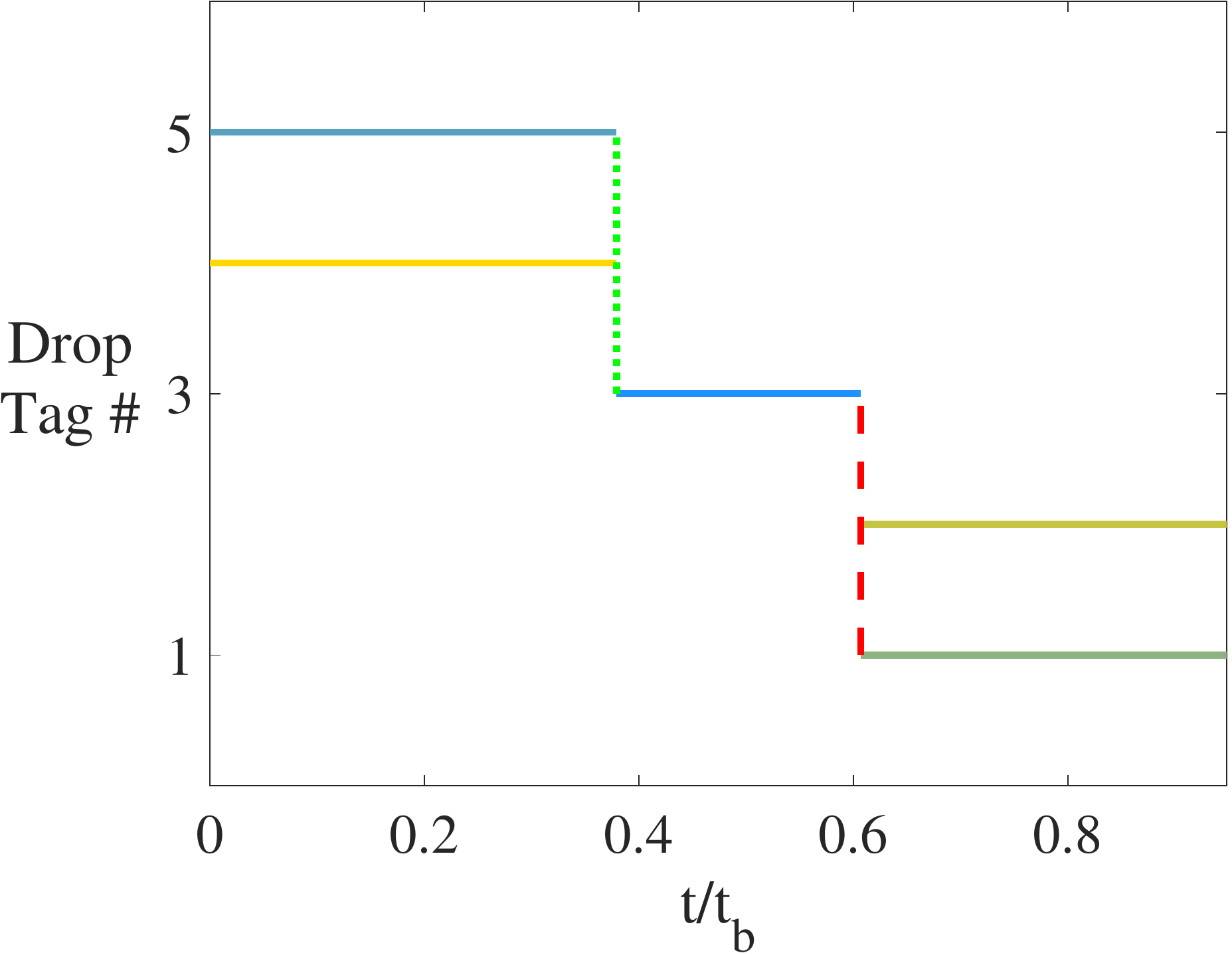}
\quad
(b)
\includegraphics[width=0.425\linewidth,valign=t]{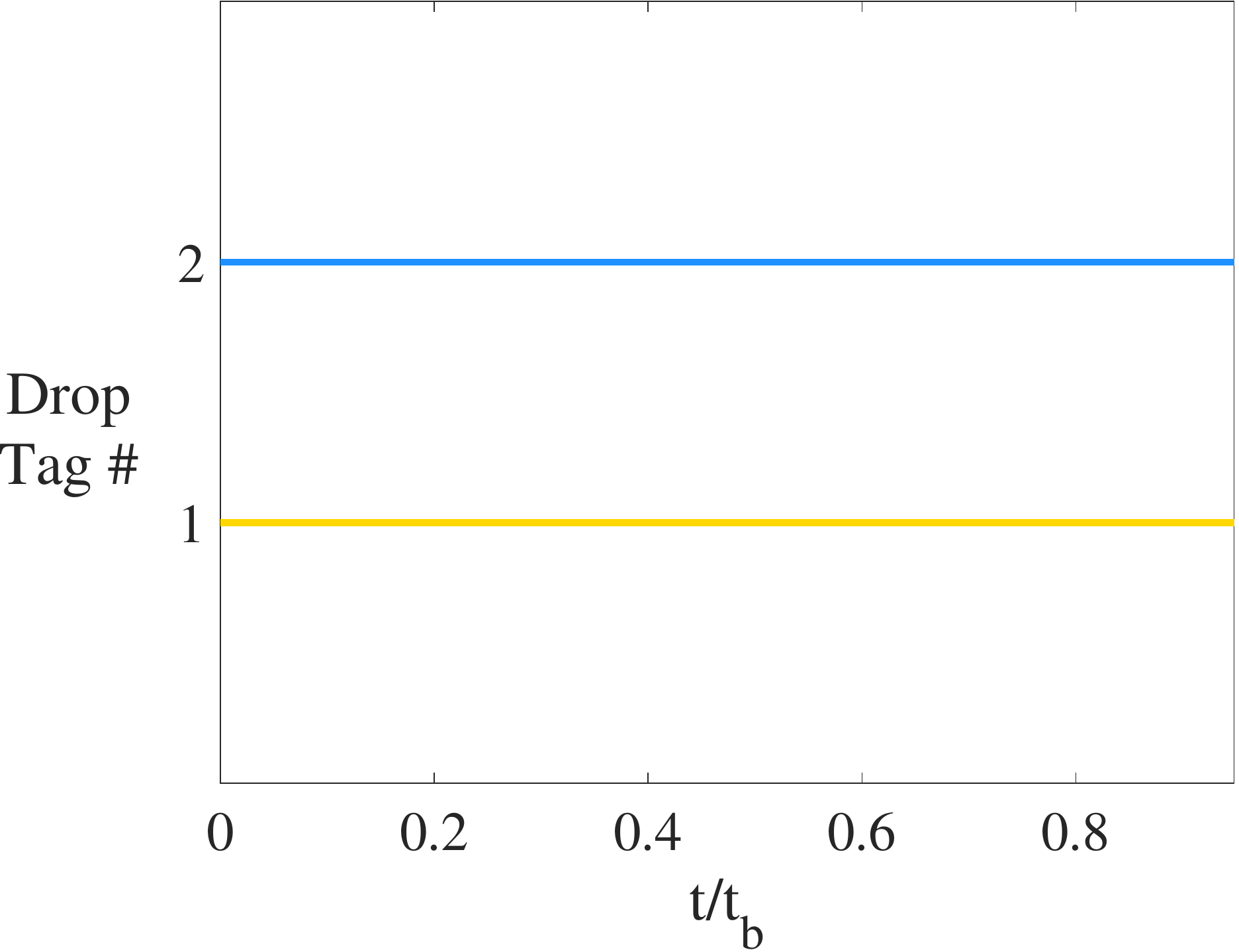}
}
\caption{Family trees of drops constructed by the algorithm in \S~\ref{sec:track-algo} for the glancing approach depicted in Fig. \ref{fig:dropsglance} where the nondimensional time interval between snapshots, $\tau$, is (a) $4.0\times10^{-2}$ and (b) $3.2\times10^{-1}$. Time is nondimensionalized by the characteristic interaction time, $t_b$, defined in the main text and the caption of Fig.~\ref{fig:dropsglance}. For a description of the drop tags and lines, refer to the captions of Figs.~\ref{fig:dropTGstats} and \ref{fig:dropTGreverse}.}
\label{fig:dropsglancestats}
\end{center}
\end{figure}

In this test case, the detection of confounding breakup and coalescence events due to the close proximity of drop pairs is artificially reconstructed by advecting two drops in the solver referenced in \S~\ref{sec:ident-test-single} with prescribed velocities such that they momentarily come within half a grid cell of each other. This mimics the scenario of slowly fragmenting and slowly coalescing drops, where repeated breakup and coalescence events may be detected if the fragmenting or coalescing drops remain in close proximity of each other for a sustained period of time. This phenomenon was also observed in other event detection algorithms~\citep{Rubel1}, and may be an inherent issue in event detection in turbulent flows with significant scale separation. The presence of these confounding events is especially challenging for complex processes like entrainment and degassing in breaking waves, and remains a pacing item in the development of drop and bubble tracking algorithms. Fig.~\ref{fig:dropsglance} depicts snapshots of the drops before and after their glancing approach, while Fig.~\ref{fig:dropsglancestats} shows the events detected using different nondimensional snapshot intervals, $\tau = \Delta_{t,s}/t_b$. Here, $t_b$ is defined as the characteristic interaction time, or the ratio of the average drop radius to the average drop speed, and should be interpreted as a proxy for the characteristic breakup/coalescence time for slowly fragmenting/coalescing drops. In the case of the shorter snapshot interval, confounding breakup and coalescence events are detected because the separation between the drops is momentarily not well resolved by the underlying mesh, and the corresponding VoF field momentarily reflects the presence of a single drop; in the case of the longer snapshot interval, these confounding events are skipped over, and the two drops are identified as merely passing each other. Thus, a sufficiently large $\tau$ is required to avoid the identification of these confounding events. In particular, $\tau$ should be no less than $O(10^{-1})$.

The results of \S~\ref{sec:track-test-TG} and \S~\ref{sec:track-test-glance} suggest that there exists an ideal order of magnitude for the snapshot interval such that the interval resolves the mean physical breakup and coalescence times of most of the drops in the flow, while being long enough to skip over confounding breakup and coalescence events of higher frequencies. Specifically, the results suggest that $\tau \sim 10^{-1}$. In a turbulent bubbly flow, for example, one would desire a snapshot interval that is about an order of magnitude shorter than the mean breakup time of bubbles of the size of the Hinze scale, even though the time-step required to resolve all pertinent flow dynamics may be much shorter.

\subsection{Drops with breakup timers}\label{sec:track-test-timer}

In this demonstration case, the efficacy of the algorithm in recovering the distribution of breakup events of a drop population is tested by seeding a domain with drops and randomly breaking them up with the passage of time, where the mean breakup time of each drop is a function of only its size. In order to control this breakup frequency, the system is evolved by maintaining a list of all drops with their sizes and centroids at every time-step independent of any surrounding flow, and artificially breaking the drops up by assigning a random countdown timer to each drop. One may interpret this as a Monte Carlo simulation of the corresponding population balance equation. In order to mimic the size and centroid variations that would be present in an actual flow solver as discussed in \S~\ref{sec:ident-test-single}, some noise is added to these sizes and centroids. 

\begin{figure}
\begin{center}
  \centerline{
(a)
\includegraphics[width=0.425\linewidth,valign=t]{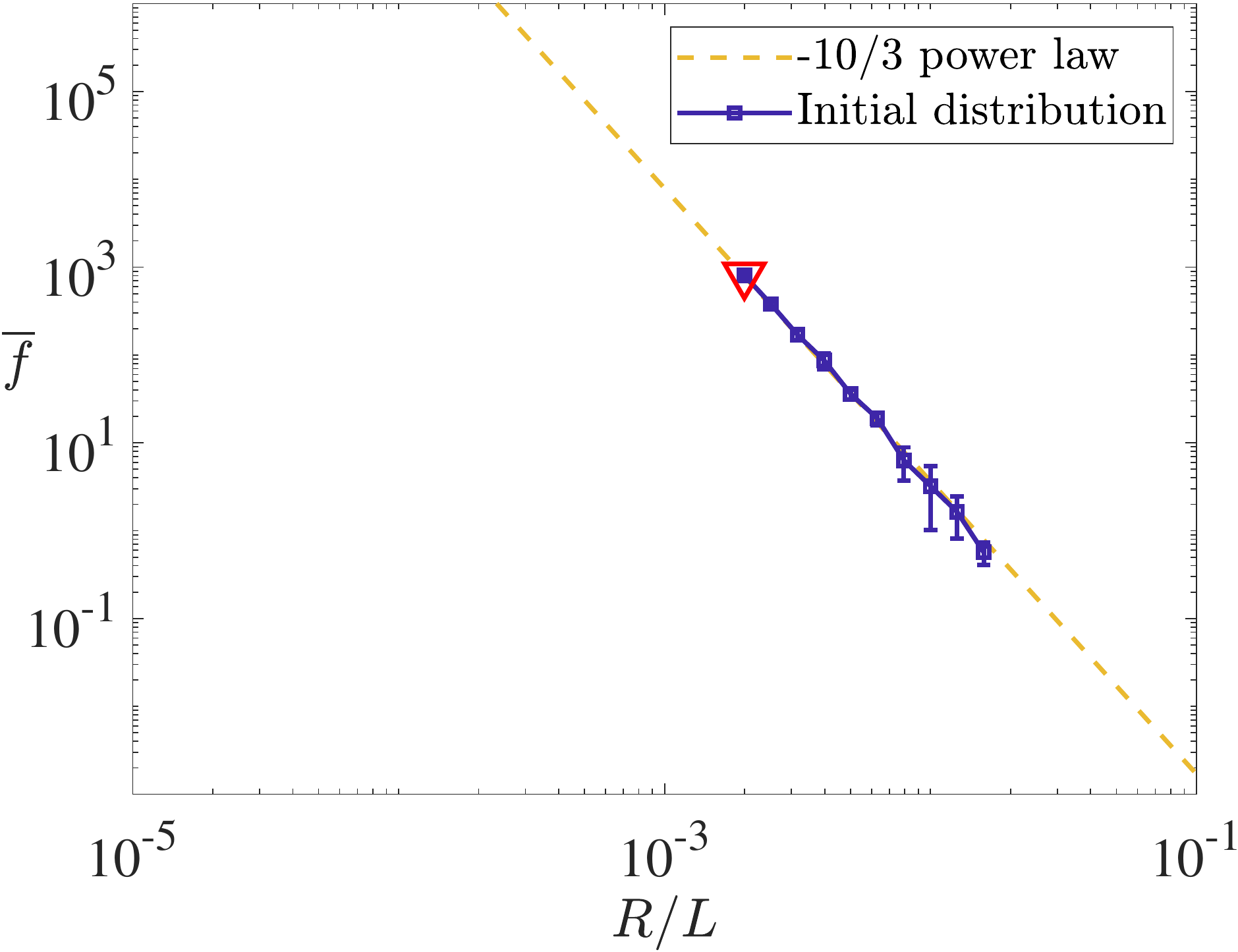}
\quad
(b)
\includegraphics[width=0.425\linewidth,valign=t]{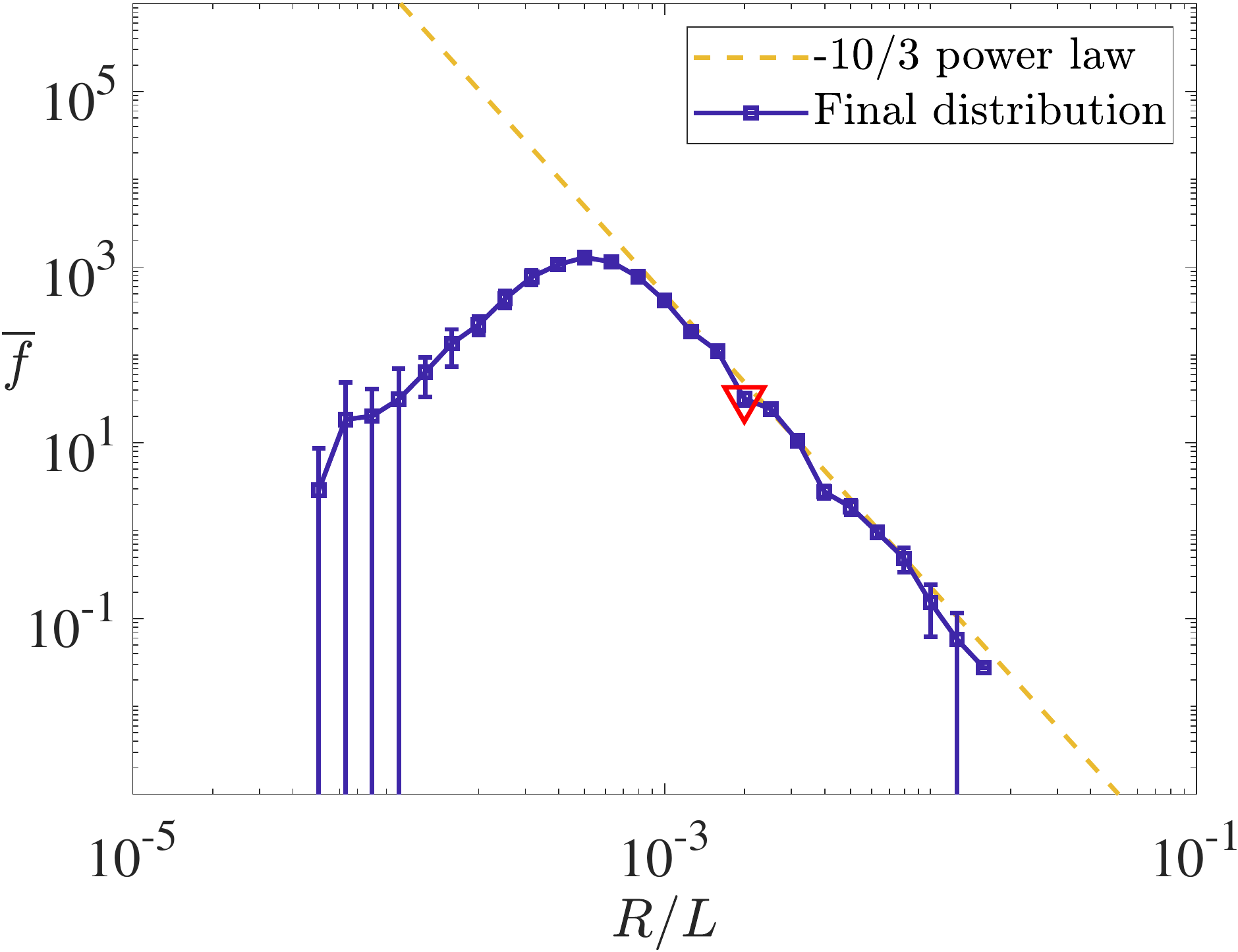}
}
\caption{The drop size distribution, $\ov{f}$, averaged over 3 statistically independent realizations of the 1{,}000-drop system, (a) initially and (b) after 400 time-steps, or $4t_b$. The distribution was computed using histogram bins of equal logarithmic spacing, For a description of the normalization and error bars, refer to the caption of Fig.~\ref{fig:dist}. The triangle marks the drop size with a mean breakup time of 100 time-steps, which is used in this case as the reference breakup time $t_b$ to define the nondimensional snapshot interval $\tau = \Delta_{t,s}/t_b$.}
\label{fig:timersizedist}
\end{center}
\end{figure}

\begin{figure}
\begin{center}
  \centerline{
(a)
\includegraphics[width=0.425\linewidth,valign=t]{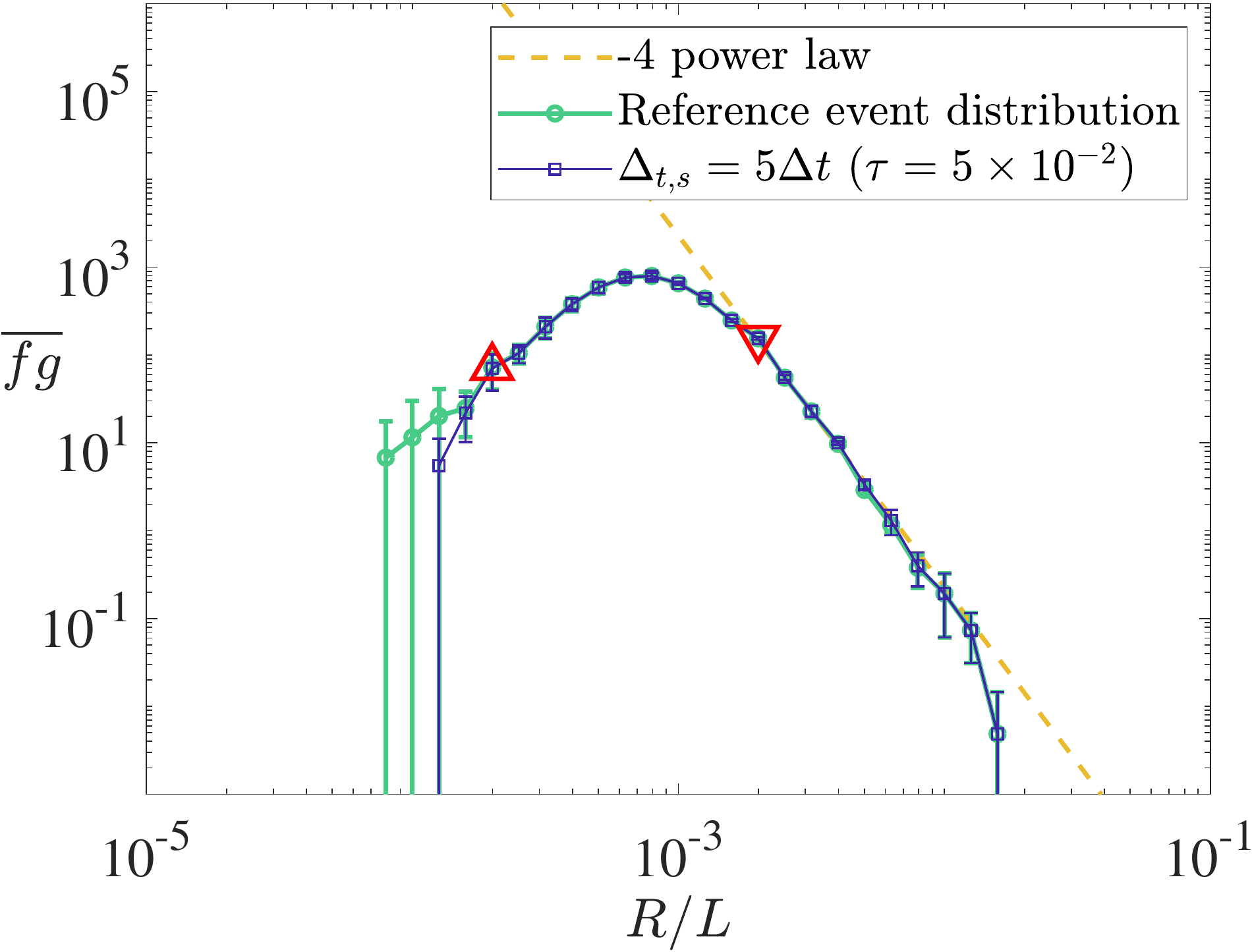}
\quad
(b)
\includegraphics[width=0.425\linewidth,valign=t]{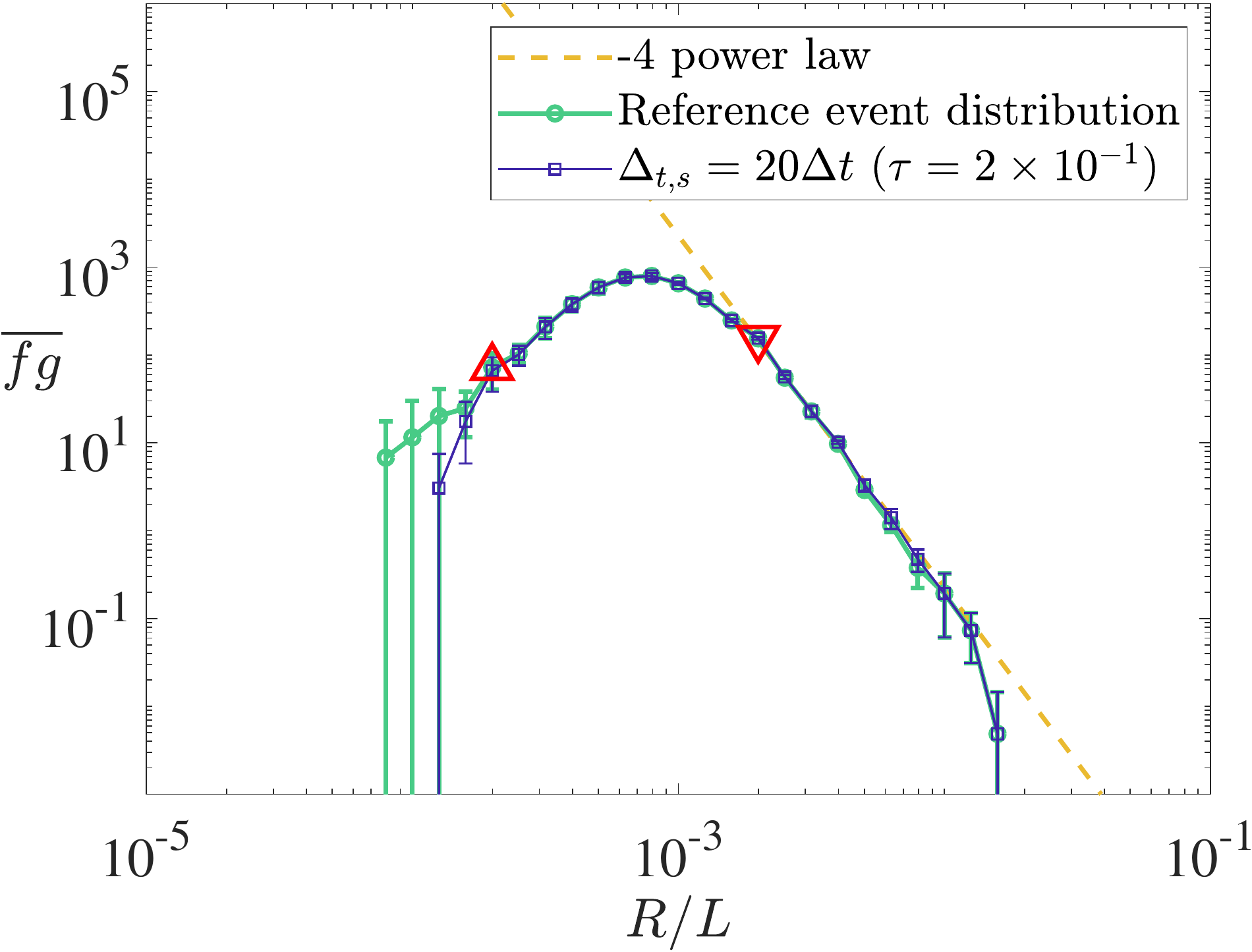}
}
\caption{The distribution of breakup events, $\ov{fg}$, averaged over 3 statistically independent realizations of the 1{,}000-drop system, where the time interval between snapshots is (a) 5 time-steps $(\tau = 5\times10^{-2})$ and (b) 20 time-steps $(\tau = 2\times10^{-1})$. For a description of the histogram bins, normalization, and error bars, refer to the caption of Fig.~\ref{fig:timersizedist}. The upward- and downward-pointing triangles mark the drop sizes with mean breakup times of 22 and 100 time-steps, respectively, or $0.22t_b$ and $t_b$. The dashed sloped line denotes the $R^{-4}$ power-law scaling expected for a drop population with a size distribution that scales as $R^{-10/3}$ and a breakup frequency that scales as $R^{-2/3}$.}
\label{fig:timerfreqdist}
\end{center}
\end{figure}

Here, a drop population obeying a $R^{-10/3}$ power-law size distribution is subject to a breakup frequency that scales as $R^{-2/3}$. This incidentally corresponds to the behavior of bubbles in a turbulent breaking wave with quasi-steady air entrainment~\citep{Garrett1,Kolmogorov3,Hinze1,MartinezBazan1}. 1{,}000 drops are randomly seeded in a similar fashion to the test case in \S~\ref{sec:ident-test-dist}, except that their radii $R \in [0.002L, 0.04L]$ span a smaller range of sizes so that their initial size distribution is more faithful to the desired $R^{-10/3}$ scaling. Note that the minimum drop radius is 500 times smaller than the box size, leading to a scale separation comparable to that of the breaking-wave simulations discussed in \S~\ref{sec:ident-test-wave}. These drops are initially separated from one another by at least $0.002L$, and their initial size distribution is plotted in Fig.~\ref{fig:timersizedist}(a). Timers with integer time-steps are assigned to these drops, as well as all ensuing children drops, such that the mean breakup time of the smallest initial drop is 100 time-steps. This will be used as the reference breakup time, $t_b$, for this test case in order to define the nondimensional snapshot interval $\tau = \Delta_{t,s}/t_b$. When a breakup is slated to occur, the parent drop with volume $\mathcal{V}_i$ is removed, and two children drops of volumes between $0.3\mathcal{V}_i$ and $0.7\mathcal{V}_i$ are added such that their centroid and total volume are respectively coincident with the parent-drop centroid and volume. The radius and each centroid coordinate of each drop are allowed to vary by up to $10^{-7}L$ every time-step, mimicking the values $M \sim 10^{-4}$ and $\Delta x/L \sim 10^{-3}$ for Criterion C1 in \S~\ref{sec:ident-test-single} and \S~\ref{sec:ident-test-pair}. The size distribution after 400 time-steps, or $4t_b$, is plotted in Fig.~\ref{fig:timersizedist}(b). The corresponding distribution of breakup events throughout this process is plotted for different snapshot intervals in Fig.~\ref{fig:timerfreqdist}. The statistics of breakup events whose mean breakup times are well resolved by the snapshot interval are accurately captured, corroborating the findings of \S~\ref{sec:track-test-TG} and \S~\ref{sec:track-test-glance} that $\tau = O(10^{-1})$ appropriately recovers the statistics of interest. Here, $\tau$ was defined using the shortest mean breakup time expected in the system from the initial conditions. This guideline for $\tau$ may be established with confidence because the demonstration case was constructed such that a reference ground truth exists for the statistics of interest, i.e., the distribution of breakup events as a function of bubble size. The recovery of the distribution appears to be comparable for the two values of $\tau$ adopted, suggesting that the statistics of interest are insensitive to the choice of $\tau$ between the values $\tau = 5\times10^{-2}$ and $\tau = 2\times10^{-1}$, even in the presence of a broad range of bubble sizes and thus correspondingly disparate time-scales. Note the divergence of the measured and reference event distributions at small drop sizes. It turns out that the effective $M$ and $\Delta x/L$ correspond to a critical size ratio of about $r \sim 30$--$40$, as discussed in \S~\ref{sec:ident-test-pair} and \S~\ref{sec:track-cons}. The size ratio between the largest and smallest drops at late times exceeds this critical ratio, resulting in an inability of the algorithm to capture the associated breakup events even with a sufficiently small snapshot interval. This reinforces the need to have an accurate identification algorithm to maximize $r$ in order to yield accurate breakup and coalescence statistics for a drop population spanning a large size range given a fixed mesh resolution. Note that $r$ may be further increased by improving the mesh resolution. Notwithstanding this constraint, the results indicate that bubble and drop lineages may be constructed using snapshots not necessarily from consecutive time-steps as long as the snapshot interval resolves all relevant physical time-scales, i.e., $\tau \sim 10^{-1}$.

\section{Conclusions}\label{sec:concl}

This paper introduces a toolbox for obtaining dispersed-phase statistics related to breakup and coalescence in two-phase flow simulations. These statistics are of practical importance in maritime, climate, and turbulent combustion studies, among others. Two steps are required to obtain accurate breakup and coalescence statistics. First, bubble or drop sizes need to be identified with sufficient accuracy. Traditional algorithms used for the identification of connected regions do not account for the aggregation of spurious structures due either to errors in the interface-advection scheme or to energetic collisions that pinch off small underresolved structures. A modification to these algorithms is proposed to eliminate these spurious structures without undue impact on the volume accuracy of the remaining identified structures. The volume preservation property of this modification is demonstrated for the computation of the volume of a single drop, as well as the size distribution of a drop population. The latter test case reveals tradeoffs between the volume accuracy of individual drops, the shape of the resulting size distribution, and the merger of closely spaced drops. However, sensitivity to the threshold employed by the proposed algorithm modification, $\phi_{c,m}$, is shown to be minimal. The proposed modification has also been applied to recent numerical studies of bubble breakup and coalescence in breaking waves~\citep{Chan3,Chan4,Chan5}. A crucial performance metric for the identification algorithm is its ability to distinguish a large drop--small drop pair from fluctuations in the large-drop volume. This significantly influences the accuracy of the construction of bubble and drop lineages and the detection of breakup and coalescence events. The principle of mass conservation and the CFL condition are used to track bubbles and drops between successive simulation snapshots. The selection of the snapshot interval for a given mesh resolution is crucial to the performance of the tracking algorithm: an excessively long interval misses events, while an excessively short interval picks up confounding events that are the result of slowly fragmenting and slowly coalescing bubbles and drops, which momentarily result in thin underresolved features. The proposed algorithm affords the flexibility to select snapshots that need not arise from consecutive time-steps in order to optimize the choice of this interval. In particular, the ratio of the snapshot interval to the characteristic breakup or coalescence time of the system, $\tau$, should be chosen to be on the order of $10^{-1}$. This algorithm is tested on configurations of varying complexities, and the importance of an accurate identification algorithm is underscored. A more accurate identification scheme enables the identification of large-size-ratio bubble pairs and drop pairs, which is crucial to maintaining the accuracy of the tracking algorithm for bubble and drop populations spanning a wide range of sizes. Note that events may also be missed due to the occurrence of ternary/polyadic breakup and/or coalescence. As remarked in \S~\ref{sec:track-cons}, ternary and polyadic events are in principle series of binary events occurring in quick succession, and only appear when the simulation time-step and/or snapshot interval is larger than the characteristic time-scale associated with these successive events. Recent experiments have observed that most resolvable breakup events in turbulent bubbly flows are binary~\citep{Qi1}, while recent theoretical investigations indicate that nonbinary events do not significantly influence the size locality of a breakup process~\citep{Chan7}, suggesting that the assumption of binarity may be generally sufficient. Nevertheless, polyadic events can be captured by modifying the constraints in \S~\ref{sec:track-cons} and lengthening the search procedure in \S~\ref{sec:track-algo}, and are a potential target for further investigation. The computation of breakup and coalescence statistics enabled by this toolbox promotes further insights into the breakup and coalescence mechanisms underlying a variety of flows of practical interest. In particular, it allows deeper analysis of model kernels in population balance equations with an eye towards improved models for dispersed-phase dynamics. The test and demonstration cases of this work indicate that many quantities and statistics of interest are insensitive to the threshold choices required by the algorithms, i.e., $\phi_{c,m}$ and $\tau$. To put it another way, the algorithms in the proposed toolbox have been designed such that sensitivities to these thresholds are kept to a minimum. Thus, these algorithms do not necessitate precise tuning of parameters, and are generally applicable to a wide range of two-phase flow simulations.

\section*{Acknowledgments}
This investigation was funded by the Office of Naval Research, Grant \#N00014-15-1-2726, and is also supported by the Advanced Simulation and Computing program of the U.S. Department of Energy’s National Nuclear Security Administration via the PSAAP-II Center at Stanford University, Grant \#DE-NA0002373. W.~H.~R. Chan is also funded by a National Science Scholarship from the Agency of Science, Technology and Research in Singapore. The authors acknowledge computational resources from the U.S. Department of Energy's INCITE Program. The authors are grateful to J. Urzay for insightful discussions on the algorithms, and S.~S. Jain for his comments on an early version of this manuscript and for useful discussions on the diffuse-interface method.

\bibliographystyle{model1-num-names}
\bibliography{bubbletoolboxA_R1}

\end{document}